# Identification of oscillatory brain networks with Hidden Gaussian Graphical Spectral models of EEG/MEG

**Deirel Paz-Linares[1,2], Eduardo Gonzalez-Moreira[1,3,6], Ariosky Areces-Gonzalez[1,4], Ying Wang[1], Min Li[1], Eduardo Martinez-Montes[2], Jorge Bosch-Bayard[5,2], Maria L. Bringas-Vega[1,2], Mitchel J. Valdes-Sosa and Pedro A. Valdes-Sosa[1,2,*]**

[1] The Clinical Hospital of Chengdu Brain Science Institute, MOE Key Lab for Neuroinformation, University of Electronic Science and Technology of China, Chengdu, China. [2] Neuroinformatic Department, Cuban Neuroscience Center, Havana, Cuba. [3] Center for Research in Informatics, Central University “Marta Abreu” of Las Villas, Santa Clara, Cuba. [4] Faculty of Technical Sciences, University of Pinar del Río “Hermanos Saiz Montes de Oca”, Pinar del Rio, Cuba. [5] McGill Centre for Integrative Neurosciences MCIN, Ludmer Centre for Mental Health, Montreal Neurological Institute, McGill University, Montreal, Canada. [6] Research Unit in Neurodevelopment, Institute of Neurobiology, Autonomous University of Mexico, Querétaro, México

[*] corresponding author: Pedro A. Valdes-Sosa (pedro.valdes@neuroinformatics-collaboratory.org)

## Abstract

**Identifying the functional networks underpinning indirectly observed processes poses an inverse problem for neurosciences or other fields. A solution of such inverse problems estimates as a first step the activity emerging within functional networks from EEG or MEG data. These EEG or MEG estimates are a direct reflect functional brain network activity with a temporal resolution that no other in vivo neuroimage may provide. A second step estimating functional connectivity from such activity pseudodata unveil the oscillatory brain networks that strongly correlate with all cognition and behavior. Simulations of such MEG or EEG inverse problem also reveal estimation errors of the functional connectivity determined by any of the state-of-the-art inverse solutions. We disclose a significant cause of estimation errors originating from misspecification of the functional network model incorporated into either inverse solution steps. We introduce the Bayesian identification of a Hidden Gaussian Graphical Spectral (HIGGS) model specifying such oscillatory brain networks model. In human EEG alpha rhythm simulations. Estimation errors measured as ROC performance do not surpass 2% in our HIGGS inverse solution and reach 20% in state-of-the-art methods. Macaque simultaneous EEG/ECoG recordings provide experimental confirmation for our inverse-solution with 1/3 more congruence according to Riemannian distances than state-of-the-art methods.**

## 1 Introduction

In vivo identification of functional brain networks would greatly benefit from the inverse solutions for MEG/EEG electromagnetic observation modalities (Larson-Prior et al. 2013; Lopes da Silva 2013). A direct relationship to the activity of postsynaptic potentials (PSPs) is attributed to the latent brain variables generating the MEG/EEG (Paul L. Nunez 1974; Freeman 1975; Valdes et al. 1999). Where the actual latent brain variables are local currents, vector process $\boldsymbol{\iota}(t)$ in time-domain $t$ with components representing locally summated PSP myriads within a spatial scale of millimeters (Paul L Nunez and Srinivasan 2006). See appendices *I. Notation for variables and mathematical operations* and *II. Nomenclature for theoretical quantities.*

MEG/EEG observations $\boldsymbol{v}(t)$ with millisecond temporal resolution are due to these latent brain variables (or activity) $\boldsymbol{\iota}(t)$ converted according to a linear and stationary forward-operator $\mathbf{L}_{v\iota}$ or Lead Field summarizing an *Electromagnetic Forward Model* (EFM) (Jörge Javier Riera and Fuentes 1998; Baillet, Mosher, and Leahy 2001; Grech et al. 2008). An inverse solution from data $\boldsymbol{v}(t)$ provides estimates $\hat{\boldsymbol{\iota}}(t)$ for this latent brain activity within a natural timescale (Hämäläinen and Ilmoniemi 1994). Estimates $\hat{\boldsymbol{\iota}}(t)$ could therefore help identify functional brain networks in association with the actual PSP activity and within very small timescales, avoiding time-domain distortions (He et al. 2019; Reid et al. 2019).

The target of all identification is estimating functional connectivity within a given functional network model interpreted as the set of parameters $\boldsymbol{\Theta}_{\iota}$ governing coactivation in $\boldsymbol{\iota}(t)$, which is without loss of generality represented by a multivariate probability $p(\boldsymbol{\iota}(t)|\boldsymbol{\Theta}_{\iota})$ (K. Friston 2011; Valdes-Sosa et al. 2011). Therefore, inverse solutions estimating $\hat{\boldsymbol{\iota}}(t)$ are only a first step to subsequently estimate latent functional connectivity $\boldsymbol{\Theta}_{\iota}$ (a **multistep** procedure). Estimation $\widehat{\boldsymbol{\Theta}}_{\iota}$ commonly requires a second-step inverse solution from such brain activity pseudodata $\hat{\boldsymbol{\iota}}(t)$ given the functional network model $p(\hat{\boldsymbol{\iota}}(t)|\boldsymbol{\Theta}_{\iota})$ (Razi and Friston 2016; Lee, Friston, and Horwitz 2006; Valdés-Sosa et al. 2005).

Functional brain networks synchronized at specific frequencies or bands of frequency (Fig. 1a) are thought to underpin the periodic components or oscillations observed in the MEG/EEG (Niedermeyer and da Silva 2005). These oscillations are a spectral vector process $\boldsymbol{v}(t,f)$ at a frequency $f$ obtained from the narrow band filtered, or alternatively Hilbert transform of the narrow band filtered vector process $\boldsymbol{v}(t)$. Oscillations $\boldsymbol{v}(t,f)$ then reflect analogous latent brain oscillations $\boldsymbol{\iota}(t,f)$ (Bruns 2004; Nolte et al. 2020; Hipp et al. 2012) according to the forward operator $\mathbf{L}_{v\iota}$ of the EFM spectral equivalent. These so-called oscillatory brain networks are described by frequency-specific functional connectivity $\boldsymbol{\Theta}_{\iota}(f)$, which produces brain oscillations $\boldsymbol{\iota}(t,f)$ even within a millisecond timescale $t$(Moffett et al. 2017). Specific oscillatory network patterns exhibit a transient behavior which may be consider stationary in a timescale of approximately seconds (Engel, Fries, and Singer 2001; Varela et al. 2001; Vidaurre, Abeysuriya, et al. 2018; Vidaurre, Hunt, et al. 2018; Tewarie et al. 2019).

Important properties of oscillatory brain networks are characterized by *Gaussian Graphical Spectral* (GGS) models. A GGS model defines the multivariate probabilities governing such latent brain oscillations $\boldsymbol{\iota}(t,f)$ under specific stationarity and mixing conditions in the frequency domain (Brillinger 1975). Details on such conditions and limits for the applicability of GGS models are provided in Materials and Methods section 2.1. The GGS model probabilities $p\big(\boldsymbol{\iota}(t,f)|\boldsymbol{\Theta}_{\boldsymbol{u}}(f)\big)$ are based upon a frequency-specific hermitian precision-matrix $\boldsymbol{\Theta}_{\boldsymbol{u}}(f)$ or inverse of the covariance-matrix denominated cross-spectrum $\boldsymbol{\Sigma}_{\boldsymbol{u}}(f)$ ($\boldsymbol{\Theta}_{\boldsymbol{u}}(f)=\boldsymbol{\Sigma}_{\boldsymbol{u}}^{-1}(f)$): where the hermitian precision-matrix $\boldsymbol{\Theta}_{\boldsymbol{u}}(f)$ off-diagonal entries define functional connectivity parameters as hermitian graph elements, known as undirected graph elements within similar real-valued functional network models (Qiao, Guo, and James 2019; Andersen et al. 1995; Jordan 1998; Jaakkola and Jordan 2000; Attias 2000; Koller and Friedman 2009; J. Friedman, Hastie, and Tibshirani 2008; Ghahramani and Beal 2001; Zhang and Zou 2014). Identifying these networks may then be regarded as equivalent to estimating the hermitian precision matrix $\widehat{\boldsymbol{\Theta}}_{\boldsymbol{u}}(f)$, which would in turn provide all functional connectivity proxies under the same GGS model assumption (Vidaurre, Hunt, et al. 2018; Hipp et al. 2012; M. J. Brookes, Woolrich, and Barnes 2012).
The identification of oscillatory brain networks is potentially viable with **multistep** MEG/EEG inverse solutions (Fig. 1b), obtaining functional connectivity proxies $\widehat{\boldsymbol{\Theta}}_{\boldsymbol{u}}(f)$ from brain-oscillation pseudodata $\hat{\boldsymbol{\iota}}(f,t)$ with strong correlation to cognition and behavior in normal or abnormal brain conditions (Matthew J. Brookes, Hale, et al. 2011; Matthew J. Brookes, Woolrich, et al. 2011a; G. L. Colclough et al. 2015; Engel, Fries, and Singer 2001; Varela et al. 2001; Vidaurre, Abeysuriya, et al. 2018; Vidaurre, Hunt, et al. 2018; Tewarie et al. 2019; Hipp et al. 2012; Wens et al. 2015; Mahjoory et al. 2017; Marzetti, Del Gratta, and Nolte 2008; Nolte et al. 2020; K. J. Friston et al. 2012; Faes and Nollo 2011; Faes, Erla, and Nollo 2012; Faes, Stramaglia, and Marinazzo, n.d.; Le Van Quyen and Bragin 2007; Valdés et al. 1992; Valdes-Sosa et al. 2009).
State-of-the-art practice targets the precision matrix $\widehat{\boldsymbol{\Theta}}_{\boldsymbol{u}}(f)$ by a second-step inverse solution (G. L. Colclough et al. 2015; Paz-Linares et al. 2018; Wodeyar 2019). In addition to this second-step inverse solution, postprocessing may also target corrections to spatial distortions or leakage in $\widehat{\boldsymbol{\Theta}}_{\boldsymbol{u}}(f)$ (Wens et al. 2015; M. J. Brookes, Woolrich, and Barnes 2012; G. L. Colclough et al. 2015; Palva et al. 2018). Leakage emerges in brain-oscillation pseudodata $\hat{\boldsymbol{\iota}}(f,t)$ due to the first-step inverse solutions (Van Veen et al. 1997; Pascual-Marqui et al. 2006), which is regarded as a major cause of distortions in precision-matrix estimates $\widehat{\boldsymbol{\Theta}}_{\boldsymbol{u}}(f)$ or any functional connectivity proxy.
Nevertheless, either inverse solutions (first step for $\hat{\boldsymbol{\iota}}(f,t)$ and second step for $\widehat{\boldsymbol{\Theta}}_{\boldsymbol{u}}(f)$) add estimation errors to these distortions in $\widehat{\boldsymbol{\Theta}}_{\boldsymbol{u}}(f)$, which stems from misspecifying the actual GGS model $p\big(\boldsymbol{\iota}(t,f)|\boldsymbol{\Theta}_{\boldsymbol{u}}(f)\big)$ governing the latent brain oscillations $\boldsymbol{\iota}(t,f)$. It is therefore the purpose of this manuscript to explain such estimation errors and theoretically correct estimation in terms of Bayesian inverse solutions.

**1.1 Bayesian inverse solution as a maximum a posteriori probability (MAP)**

An inverse solution (equation 1-1) estimating the latent variables (categories $\mathcal{X}$ and $\mathcal{Z}$) from their data (cathegory $\mathcal{Y}$) is without loss of generality a Bayesian *maximum a posteriori probability* (MAP) (Bishop 1995). A MAP $\widehat{\mathcal{X}}$ is the optimum value or estimate for an a posteriori probability $p(\mathcal{X}|\mathcal{Y})$ computed for a given likelihood $p(\mathcal{Y}|\mathcal{X})$ explaining $\mathcal{Y}$ upon $\mathcal{X}$ and positing an a priori probability $p(\mathcal{X})$ upon $\mathcal{X}$. The nominal Bayesian procedure (Mackay 1999) is a *first-type MAP* (MAP1) involving only a given first-type likelihood $p(\mathcal{Y}|\mathcal{X})$. The full Bayesian procedure (MacKay 2003) is a *second-type MAP* (MAP2) involving a second-type likelihood $p(\mathcal{Y}|\mathcal{X})$, which must be determined marginalizing a given first-type likelihood $p(\mathcal{Y}|\mathcal{Z})$ explaining $\mathcal{Y}$ upon (parameters) $\mathcal{Z}$. An a priori probability $p(\mathcal{Z}|\mathcal{X})$ explains these parameters upon (hyperparameters) $\mathcal{X}$.

$$\begin{gathered}\widehat{\mathcal{X}}=argmax_{\mathcal{X}}\{p(\mathcal{X}|\mathcal{Y})\}\\ p(\mathcal{X}|\mathcal{Y})=p(\mathcal{Y}|\mathcal{X})p(\mathcal{X})/p(\mathcal{Y})\\ p(\mathcal{Y}|\mathcal{X})=\int p(\mathcal{Y}|\mathcal{Z})p(\mathcal{Z}|\mathcal{X})d\mathcal{Z}\end{gathered}\qquad\text{(I-1)}$$

Where such likelihood model $p(\mathcal{Y}|\mathcal{X})$ (first-type or second-type) has an ill-conditioned nature and estimating $\widehat{\mathcal{X}}$ (equation 1-1) poses an inverse problem, standing for no unique entries for $\mathcal{X}$ explain a given data in $\mathcal{Y}$ (Tarantola 2005). Then, an MAP posits an a priori probability $p(\mathcal{X})$ regularizing such ill conditioning and preferably pursuing sparse selection of those variables within $\mathcal{X}$, actually explaining the given data in $\mathcal{Y}$. The term regularization by sparse variable selection is valid for the type of ill conditioning, such as that dealth with inverse solutions (equation 1-1) in the first step for latent vectors $\boldsymbol{\iota}(t,f)$ (Tibshirani 1996; Tipping 2001) and in the second step for latent precision matrices $\boldsymbol{\Theta}_{\boldsymbol{u}}(f)$ (J. Friedman, Hastie, and Tibshirani 2008; H. Wang 2012).

**1.2 First step MAP1 inverse solution for brain oscillations and estimation errors**

The first step MAP1 (equation 1-1) is for the brain oscillations (Fig. 1b) $\boldsymbol{\iota}(f,t)$ upon the MEG/EEG oscillations (data) $\boldsymbol{v}(t,f)$ (equation 1-2): where the EFM spectral equivalent yields a Gaussian first-type likelihood $p\big(\boldsymbol{v}(t,f)|\boldsymbol{\iota}(f,t)\big)$. Here, likelihood ill-conditioning is most severe in the sense of Hadamard (Hadamard and Morse 1953), caused by a large number of variables describing latent brain oscillations $\boldsymbol{\iota}(f,t)$ compared to the available data $\boldsymbol{v}(t,f)$ (Hämäläinen and Ilmoniemi 1994).
An a priori probability $p\big(\boldsymbol{\iota}(f,t)|\boldsymbol{\Lambda}(f)\big)$ (equation 1-2) regularizes this ill-conditioning also incorporating some information $\boldsymbol{\Lambda}(f)$, preferably with a Gaussian model upon covariance-matrix $\boldsymbol{\Lambda}(f)$ (sometimes a diagonal matrix) somehow specified from MEG/EEG data or structural information (Mattout et al. 2006; K. Friston et al. 2008; Trujillo-Barreto, Aubert-Vázquez, and Valdés-Sosa 2004; D. Wipf and Nagarajan 2009; Paz-Linares et al. 2017; Tipping 2001; D. P. Wipf et al. 2006; Casella et al. 2010).

Fig. 1| Illustration of the ontological levels involved and basic multistep identification of connectivity. a. Generation by the electromagnetic forward model of the periodic components or oscillations in MEG/EEG observations $\boldsymbol{v}(t,f)$ from brain oscillations $\boldsymbol{\iota}(t,f)$ emerging from functional graph-elements or functional connectivity $\boldsymbol{\Theta}_{\boldsymbol{u}}(f)$ defining a cortical oscillatory network at a given frequency $f$. b. Functional connectivity distortions due to multistep inverse-solutions of the optimal cortical oscillations $\hat{\boldsymbol{\iota}}(t,f)$ explaining the observations and functional connectivity $\widehat{\boldsymbol{\Theta}}_{\boldsymbol{u}}(f)$, from pseudodata $\hat{\boldsymbol{\iota}}(t,f)$. Even with with perfect spatial localization estimates $\hat{\boldsymbol{\iota}}(t,f)$ by any first step inverse-solution produce however false positives and false negatives functional graph-elements in the second-step inverse-solution $\widehat{\boldsymbol{\Theta}}_{\boldsymbol{u}}(f)$.

State-of-the-art practice choses this Gaussian leading to the type of linear and stationary inverse solution to avoid introducing time-domain distortions in brain oscillations pseudodata $\hat{\boldsymbol{\iota}}(f,t)$, which may further affect functional connectivity estimates $\widehat{\boldsymbol{\Theta}}_{\boldsymbol{u}}(f)$ (M. J. Brookes, Woolrich, and Barnes 2012; G. L. Colclough et al. 2015; Marinazzo et al. 2019; Nolte et al. 2020). The most common examples are *Exact Low Resolution Electromagnetic Tomographic Analysis* (**eLORETA**) (Pascual-Marqui et al. 2006) and Linearly Constrained Minimum Variance (**LCMV**) (Van Veen et al. 1997). This MAP1 may include, in addition to $\boldsymbol{\Lambda}(f)$, a sensor noise covariance $\mathbf{B}(f)$ within a Gaussian likelihood $p\big(\boldsymbol{v}(t,f)|\boldsymbol{\iota}(f,t),\mathbf{B}(f)\big)$. Henceforth, it is not essential to the main exposition to specify likelihood or a priori probability model formulas for this MAP1 as well as $\boldsymbol{\Lambda}(f)$ or $\mathbf{B}(f)$, which are determined by other methods.

$$\hat{\boldsymbol{\iota}}(f,t) = argmax_{\boldsymbol{\iota}(f,t)}\{p\big(\boldsymbol{\iota}(f,t)|\boldsymbol{v}(t,f),\boldsymbol{\Lambda}(f)\big)\}$$
$$p\big(\boldsymbol{\iota}(f,t)|\boldsymbol{v}(t,f),\boldsymbol{\Lambda}(f)\big) \propto p\big(\boldsymbol{v}(t,f)|\boldsymbol{\iota}(f,t)\big)p\big(\boldsymbol{\iota}(f,t)|\boldsymbol{\Lambda}(f)\big) \quad \text{(I-2)}$$

This first-step MAP1 (equations 1-2) carries on estimation errors (Fig. 1b) when posits ad hoc any a priori model $p\big(\boldsymbol{\iota}(f,t)|\boldsymbol{\Lambda}(f)\big)$ and with covariance-matrix $\boldsymbol{\Lambda}(f)$ having no relation whatsoever with the functional connectivity, whereas a GGS model $p\big(\boldsymbol{\iota}(f,t)|\boldsymbol{\Theta}_{\boldsymbol{u}}(f)\big)$ actually governs brain oscillations $\boldsymbol{\iota}(f,t)$. Then, the **multistep** inverse solution causes errors from the first step MAP1 estimate $\hat{\boldsymbol{\iota}}(f,t)$ (equations 1-2) to blow up in any second-step estimate $\widehat{\boldsymbol{\Theta}}_{\boldsymbol{u}}(f)$.

At the same time, a first step MAP1 faces an obvious circularity when posits the actual GGS model $p\big(\boldsymbol{\iota}(f,t)|\boldsymbol{\Theta}_{\boldsymbol{u}}(f)\big)$, where the precision-matrix $\boldsymbol{\Theta}_{\boldsymbol{u}}(f)$ is unknown and must be the target of the second step MAP1. A similar circularity is dealt with estimating a covariance-matrix in a *Covariance Component Model* (CCM) (C. Liu and Rubin 1994; Galka, Yamashita, and Ozaki 2004; K. Friston et al. 2006, 2007) or autoregressive-coefficient in a *State Space Model* (SSM) (Jorge J. Riera et al. 2004; Faes, Stramaglia, and Marinazzo, n.d.; Galka et al. 2004). Here, this circularity is caused by a *Hidden GGS* (HIGGS) model for latent brain oscillations $\boldsymbol{\iota}(f,t)$, similar to a CCM, but upon precision matrices and with a graphical sparse a priori model. The inverse solution in **onestep** baypassing this circularity is a MAP2 (equation 1-1) (MacKay 2003).

**1.3 Second step MAP1 inverse solution for functional connectivity of the GGS model and estimation errors**

The second step MAP1 is for the GGS precision matrix (Fig. 1 b) $\boldsymbol{\Theta}_{\boldsymbol{u}}(f)$ upon sampled covariance matrix (pseudodata) $\widehat{\boldsymbol{\Sigma}}_{\boldsymbol{u}}(f)$ (equations 1-3) obtained from brain oscillation estimates $\hat{\boldsymbol{\iota}}(f,t)$ in the first step MAP1 (equations 1-2). The GGS model $p\big(\hat{\boldsymbol{\iota}}(f,t)|\boldsymbol{\Theta}_{\boldsymbol{u}}(f)\big)$ yields a Wishart first-type likelihood $p\left(\widehat{\boldsymbol{\Sigma}}_{\boldsymbol{u}}(f)\middle|\boldsymbol{\Theta}_{\boldsymbol{u}}(f)\right)$ for a sampled covariance-matrix $\widehat{\boldsymbol{\Sigma}}_{\boldsymbol{u}}(f)$. Likelihood ill-conditioning is here in the sense of low condition order for any sampled covariance-matrix $\widehat{\boldsymbol{\Sigma}}_{\boldsymbol{u}}(f)$ caused by a limited number of samples $\hat{\boldsymbol{\iota}}(f,t)$ in time-domain $t$ (Ledoit and Wolf 2004). A graphical a priori probability $p\big(\boldsymbol{\Theta}_{\boldsymbol{u}}(f)\big)$ regularizes this ill-conditioning with sparse selection of undirected graph elements and amplitude (Fig. 2 a) of the hermitian precision-matrix off-diagonal entries. The graphical a priori probability $p\big(\boldsymbol{\Theta}_{\boldsymbol{u}}(f)\big)$ base sparse selection upon an extension of the *Least Absolute Shrinkage and Selection Operator* (LASSO) (Tibshirani 1996) denominated *graphical LASSO* (**gLASSO**) in real-valued multivariate statistics (J. Friedman, Hastie, and Tibshirani 2008).

Fig. 2| Connectivity distortions in the real-valued approximation of a hermitian GGS model. a. Undirected network as defined by the amplitude of hermitian graph-elements or entries of the hermitian GGS precision-matrix. b. Binary GGS-precision-matrix amplitudes $\boldsymbol{\Theta}_{\boldsymbol{u}}(f)$ are perfectly retrievd by estimation $\widehat{\boldsymbol{\Theta}}_{\boldsymbol{u}}(f)$ based on a hermitian GGS model with hermitian graphical LASSO prior and distorted by estimation based on a real-valued GGS model with graphical LASSO prior.

**a. Undirected network defined as precision-matrix amplitudes**

**b. Estimation of amplitudes via hermitian or real-valued model**

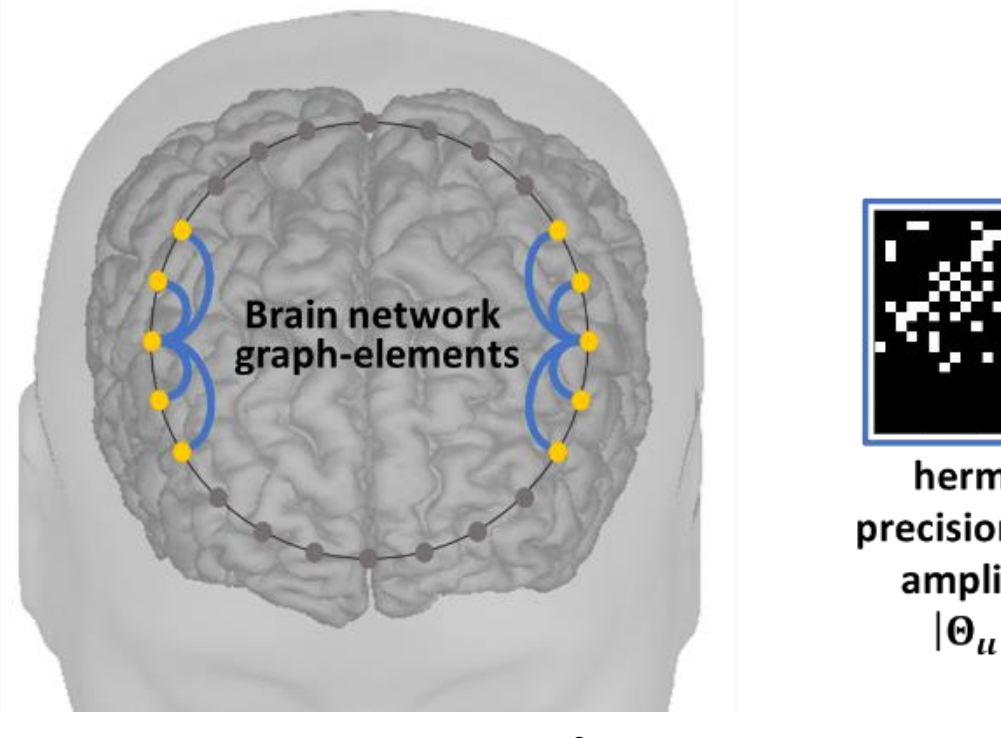

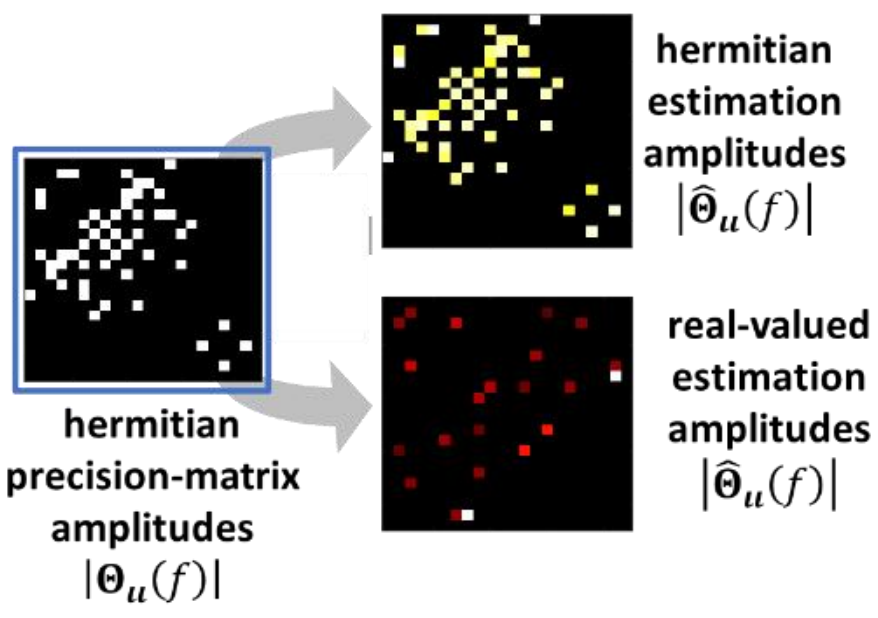

$$\widehat{\boldsymbol{\Theta}}_{\boldsymbol{u}}(f) = argmax_{\boldsymbol{\Theta}_{\boldsymbol{u}}(f)} \left\{ p\left( \boldsymbol{\Theta}_{\boldsymbol{u}}(f) | \widehat{\boldsymbol{\Sigma}}_{\boldsymbol{u}}(f) \right) \right\}$$
$$p\left( \boldsymbol{\Theta}_{\boldsymbol{u}}(f) | \widehat{\boldsymbol{\Sigma}}_{\boldsymbol{u}}(f) \right) \propto p\left( \widehat{\boldsymbol{\Sigma}}_{\boldsymbol{u}}(f) \middle| \boldsymbol{\Theta}_{\boldsymbol{u}}(f) \right) p\left(\boldsymbol{\Theta}_{\boldsymbol{u}}(f)\right) \tag{I-3}$$

The second step MAP1 (equations 1-3) carries on estimation errors when posits a GGS model $p\left(\boldsymbol{\iota}(f,t)|\boldsymbol{\Theta}_{\boldsymbol{u}}(f)\right)$ adhoc defined as real-valued (Fig. 2 b), whereas a complex-valued hermitian actually governs brain oscillations in the frequency domain (Faes and Nollo 2011; Nolte et al. 2020; Marzetti, Del Gratta, and Nolte 2008; K. J. Friston et al. 2012). Within such models functional connectivity parameters are hermitian precision-matrix entries $\boldsymbol{\Theta}_{\boldsymbol{u}}(f)$ or hermitian graph-elements encoding phase and amplitude (Schreier and Scharf 2010; Andersen et al. 1995; Tugnait 2019). Indeed, estimation errors in the precision-matrix $\widehat{\boldsymbol{\Theta}}_{\boldsymbol{u}}(f)$ even reflect wrong sparse selection of the amplitudes for the real-valued compared to the hermitian one. Furthermore, this MAP1 in either case may also carry on estimation errors due to sparse bias (Janková and van de Geer 2018) or instability in large-scale dimensions (C. Hsieh 2014).

We present a novel MAP1 algorithm determining the precision matrix $\widehat{\boldsymbol{\Theta}}_{\boldsymbol{u}}(f)$ (equations 1-2, 1-5) with Bayesian hierarchical representation of the sparse hermitian graphical a priori model or *Hermitian Graphical LASSO* (**hgLASSO**). We demonstrate in GGS model simulations with large-scale and hermitian precision-matrices the unbiased performance of this **hgLASSO** algorithm, which provides estimates following the theoretical Rayleigh distribution (Janková and van de Geer 2018).

### 1.4 Successive approximations to onestep MAP2 inverse solution for functional connectivity via multistep MAP1 inverse solution

A **onestep** MAP2 (equation 1-1) is for the HIGGS precision-matrix (Fig. 3 a) $\boldsymbol{\Theta}_{\boldsymbol{u}}(f)$ upon the MEG/EEG oscillations sampled covariance-matrix (data) $\widehat{\boldsymbol{\Sigma}}_{\boldsymbol{vv}}(f)$. The second-type likelihood $p\left(\widehat{\boldsymbol{\Sigma}}_{\boldsymbol{vv}}(f)\middle|\boldsymbol{\Theta}_{\boldsymbol{u}}(f)\right)$ (equations 1-4) is obtained marginalizing $\boldsymbol{\iota}(f,t)$ in a first-type likelihood $p\left(\boldsymbol{v}(t,f)|\boldsymbol{\iota}(f,t)\right)$ under an a priori HIGGS model $p\left(\boldsymbol{\iota}(f,t)\middle|\boldsymbol{\Theta}_{\boldsymbol{u}}(f)\right)$. A MAP2 for similar models may be impractical and must commonly be approached by A*pproximated Bayesian Computation* (ABC) (Csilléry et al. 2010). ABC employs successive approximations to the second-type likelihood $p\left(\widehat{\boldsymbol{\Sigma}}_{\boldsymbol{vv}}(f)\middle|\boldsymbol{\Theta}_{\boldsymbol{u}}(f)\right)$ (equations 1-4) (Cabras, Nueda, and Ruli 2015) with Gibbs sampling of the a posteriori probability $p\left(\boldsymbol{\Theta}_{\boldsymbol{u}}(f)|\widehat{\boldsymbol{\Sigma}}_{\boldsymbol{vv}}(f)\right)$ (Fearnhead and Prangle 2012). Gibbs sampling algorithms might be impractical even for an a posteriori probability of the second-step MAP1 $p\left(\boldsymbol{\Theta}_{\boldsymbol{u}}(f)|\widehat{\boldsymbol{\Sigma}}_{\boldsymbol{u}}(f)\right)$ (equation 1-2) (H. Wang 2012) requiring alternative approaches (C. Hsieh 2014; C.-J. Hsieh et al. 2013).

$$\widehat{\boldsymbol{\Theta}}_{\boldsymbol{u}}(f) = argmax_{\boldsymbol{\Theta}_{\boldsymbol{u}}(f)} \left\{ p\left( \boldsymbol{\Theta}_{\boldsymbol{u}}(f) | \widehat{\boldsymbol{\Sigma}}_{\boldsymbol{vv}}(f) \right) \right\}$$
$$p\left( \boldsymbol{\Theta}_{\boldsymbol{u}}(f) | \widehat{\boldsymbol{\Sigma}}_{\boldsymbol{vv}}(f) \right) \propto p\left( \widehat{\boldsymbol{\Sigma}}_{\boldsymbol{vv}}(f) \middle| \boldsymbol{\Theta}_{\boldsymbol{u}}(f) \right) p\left(\boldsymbol{\Theta}_{\boldsymbol{u}}(f)\right) \tag{I-4}$$

These successive approximations to a second-type likelihood (Fig. 3 b) may be implemented through the *Expectation Maximization* (EM) algorithm as in a CCM (C. Liu and Rubin 1994; Galka, Yamashita, and Ozaki 2004; K. Friston et al. 2006, 2007). EM is interpretable as first-step (equation 1-2) and second-step (equation 1-3) MAP1 estimators obtaining a MAP2 local optima (McLachlan and Krishnan 2007): where the EM maximization stage is a second-step MAP1 in a loop $(k)$ with $p\left(\boldsymbol{\Theta}_{\boldsymbol{u}}(f)|\widehat{\boldsymbol{\Psi}}_{\boldsymbol{u}}^{(k)}(f)\right)$ (equations 1-5) computed from an equivalent first-type likelihood $p\left(\widehat{\boldsymbol{\Psi}}_{\boldsymbol{u}}^{(k)}(f)\middle|\boldsymbol{\Theta}_{\boldsymbol{u}}(f)\right)$ (equations 1-3). This first-type likelihood is interpreted as a GGS model explaining an expected covariance-matrix $\widehat{\boldsymbol{\Psi}}_{\boldsymbol{u}}^{(k)}(f)$ upon the latent precision-matrix $\boldsymbol{\Theta}_{\boldsymbol{u}}(f)$ and the a priori probability $p\left(\boldsymbol{\Theta}_{\boldsymbol{u}}(f)\right)$.

$$\widehat{\boldsymbol{\Theta}}_{\boldsymbol{u}}^{(k+1)}(f) = argmax_{\boldsymbol{\Theta}_{\boldsymbol{u}}(f)} \left\{ p\left( \boldsymbol{\Theta}_{\boldsymbol{u}}(f) | \widehat{\boldsymbol{\Psi}}_{\boldsymbol{u}}^{(k)}(f) \right) \right\}$$
$$p\left( \boldsymbol{\Theta}_{\boldsymbol{u}}(f) | \widehat{\boldsymbol{\Psi}}_{\boldsymbol{u}}^{(k)}(f) \right) \propto p\left( \widehat{\boldsymbol{\Psi}}_{\boldsymbol{u}}^{(k)}(f) \middle| \boldsymbol{\Theta}_{\boldsymbol{u}}(f) \right) p\left(\boldsymbol{\Theta}_{\boldsymbol{u}}(f)\right) \tag{I-5}$$

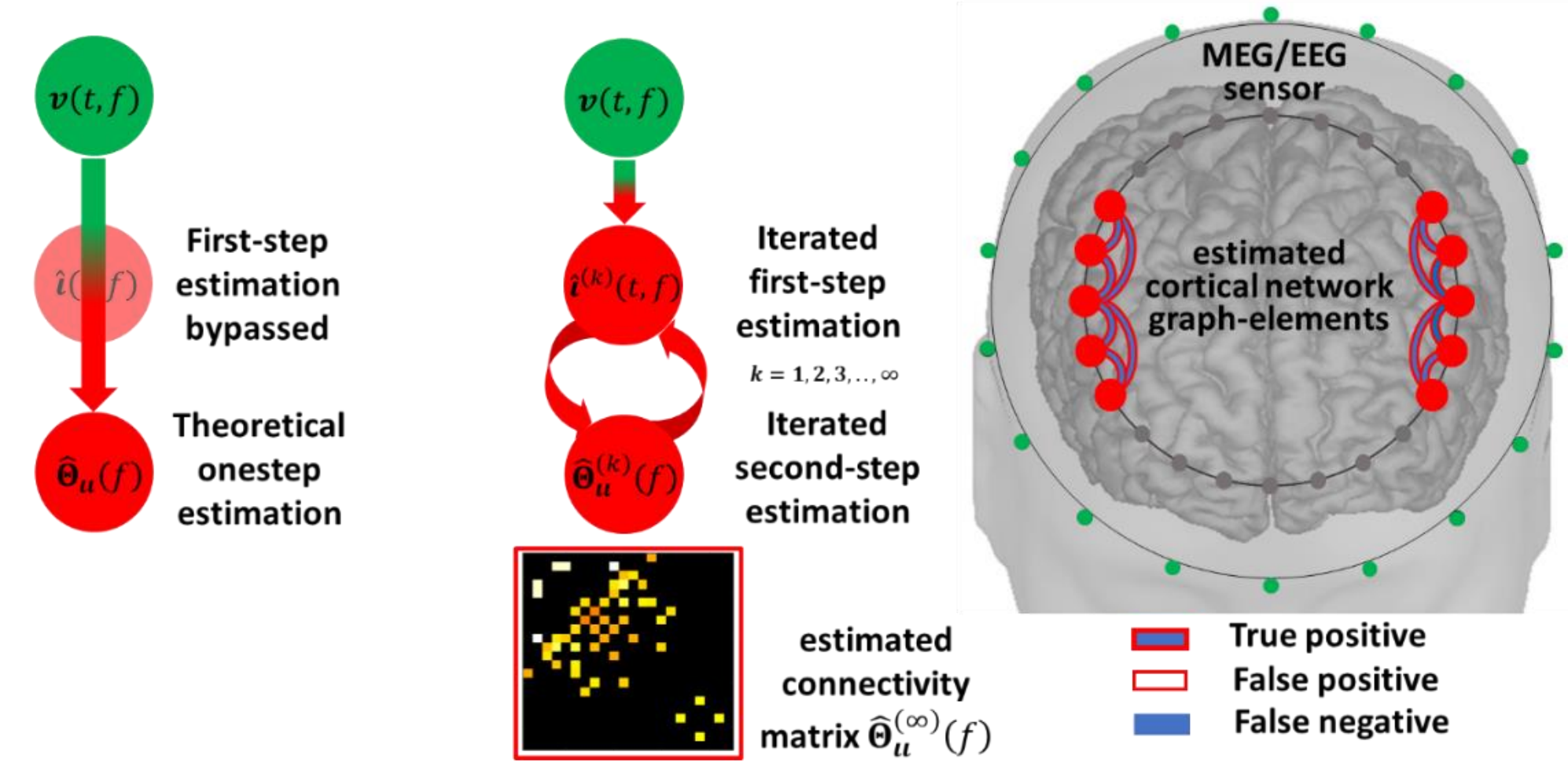

Fig. 3| Theoretically correct procedure for the estimation of connectivity and to eliminate distortions. a. The ideal but impractical solution is to bypass intermediate estimators $\hat{\boldsymbol{\iota}}(t,f)$ onestep search of optimal connectivity $\widehat{\boldsymbol{\Theta}}_{\boldsymbol{u}}(f)$ explaining the observations. b. Our implementation of efficient sequential approximations to the onestep procedure by means of Expectation Maximization iterations of $\hat{\boldsymbol{\iota}}^{(k)}(t,f)$ and $\widehat{\boldsymbol{\Theta}}_{\boldsymbol{u}}^{(k)}(f)$ produces perfect functional graph-elements due to statistical guaranties.

The expected covariance matrix $\widehat{\boldsymbol{\Psi}}_{\boldsymbol{u}}^{(k)}(f)$ (equations 1-5) is determined from an ensemble covariance matrix $\widehat{\boldsymbol{\Pi}}_{\boldsymbol{u}}^{(k)}(f)$, and its sampled estimator $\widehat{\boldsymbol{\Sigma}}_{\boldsymbol{u}}^{(k)}(f)$ in the expectation stage MAP1 (Dempster, Laird, and Rubin 1977) is also common for a first-step MAP1 (Mattout et al. 2006; K. Friston et al. 2008; Trujillo-Barreto, Aubert-Vázquez, and Valdés-Sosa 2004; D. Wipf and Nagarajan 2009; Paz-Linares et al. 2017; Tipping 2001; D. P. Wipf et al. 2006; Casella et al. 2010).

$$\widehat{\boldsymbol{\Psi}}_{\boldsymbol{u}}^{(k)}(f) = \widehat{\boldsymbol{\Sigma}}_{\boldsymbol{u}}^{(k)}(f) + \widehat{\boldsymbol{\Pi}}_{\boldsymbol{u}}^{(k)}(f) \tag{I-6}$$

The EM expectation stage obtaining $\widehat{\boldsymbol{\Psi}}_{\boldsymbol{u}}^{(k)}(f)$ is a first-step MAP1 in a loop $(k)$ with $p\left(\boldsymbol{\iota}(f,t)|\boldsymbol{v}(t,f),\widehat{\boldsymbol{\Theta}}_{\boldsymbol{u}}^{(k)}(f)\right)$ (equations 1-7) computed from an equivalent first-type likelihood $p\big(\boldsymbol{v}(t,f)|\boldsymbol{\iota}(f,t)\big)$ (equations 1-2) and positing an a priori GGS model $p\left(\boldsymbol{\iota}(f,t)|\widehat{\boldsymbol{\Theta}}_{\boldsymbol{u}}^{(k)}(f)\right)$ upon the precision-matrix $\widehat{\boldsymbol{\Theta}}_{\boldsymbol{u}}^{(k)}(f)$, which is determined in the previous maximization stage. The MAP2 circularity is then solved by successive approximations in a multistep loop $(k)$ with maximization (equations 1-5) and expectation (equations 1-6) specifying the actual HIGGS model.

$$\begin{gathered}\hat{\boldsymbol{\iota}}^{(k)}(f,t) = argmax_{\boldsymbol{\iota}(f,t)}\left\{p\left(\boldsymbol{\iota}(f,t)|\boldsymbol{v}(t,f),\widehat{\boldsymbol{\Theta}}_{\boldsymbol{u}}^{(k)}(f)\right)\right\} \\ p\left(\boldsymbol{\iota}(f,t)|\boldsymbol{v}(t,f),\widehat{\boldsymbol{\Theta}}_{\boldsymbol{u}}^{(k)}(f)\right) \propto p\big(\boldsymbol{v}(t,f)|\boldsymbol{\iota}(f,t)\big)p\left(\boldsymbol{\iota}(f,t)|\widehat{\boldsymbol{\Theta}}_{\boldsymbol{u}}^{(k)}(f)\right)\end{gathered} \tag{I-7}$$

A proof of concept for the HIGGS model in EEG simulations reveals that **multistep** inverse solutions (equations 1-2, 1-3) may produce a very distorted precision matrix $\widehat{\boldsymbol{\Theta}}_{\boldsymbol{u}}(f)$ in a second step MAP1 (equations 1-3). When even no ill-conditioning holds (ruling out leakage) in a first step MAP1 (equation 1-2) determining the brain oscillations $\hat{\boldsymbol{\iota}}(f,t)$. In the same situation, the **onestep** inverse solution (equations 1-5, 1-6, 1-7) determines $\widehat{\boldsymbol{\Theta}}_{\boldsymbol{u}}(f)$ exactly. We employ the sparse hermitian graphical a priori model to determine $\widehat{\boldsymbol{\Theta}}_{\boldsymbol{u}}(f)$ via the unbiased **hgLASSO** algorithm for both **multistep** (equations 1-3) and **onestep** (equations 1-5) inverse solutions. Implementations of **onestep** inverse solutions that violate such unbiasedness conditions with other hermitian graphical a priori models also outperform **multistep hgLASSO** inverse solutions. The **multistep** procedure is then the systemic cause of distortions irremediable even with an exact second-step inverse solution for $\widehat{\boldsymbol{\Theta}}_{\boldsymbol{u}}(f)$. This proof of concept also reveals that **multistep** inverse solutions employing a sparse real-valued graphical a priori model or *Graphical LASSO* (**gLASSO**) that approximates the HIGGS model further distorts $\widehat{\boldsymbol{\Theta}}_{\boldsymbol{u}}(f)$. We confirm this difference in performance between **multistep** and **onestep** procedures by comparing EEG inverse solutions against their fine-grained ECoG reference inverse solutions in macaque simultaneous recordings.

# 2 Materials and Methods

## 2.1 Gaussian Graphical Spectral (GGS) model and MAP1 inverse solution with Hermitian Graphical LASSO (hgLASSO)

### 2.1.1 GGS model and interpretations of functional connectivity

The PSP processes myriad fulfills some frequency-domain mixing conditions and consequently Gaussianity in the complex-valued follows for brain oscillations $\boldsymbol{\iota}(t,f)$ (Brillinger 1975; Rosenblatt 1956; Roweis and Ghahramani 1999; Freeman 1975). Then, without loss of generality, the *Gaussian Graphical Spectral* (GGS) model (equation 2-1) with complex-valued hermitian precision matrix $\boldsymbol{\Theta}_{\boldsymbol{u}}(f)$ is the functional network model for brain oscillations. This GGS model is valid at any frequency $f$ within the human spectrum $(f \in \mathbb{F})$ from 0.5 to 140 hertz (Niedermeyer and da Silva 2005; Le Van Quyen and Bragin 2007; Frauscher et al. 2018) and within a resting state or task transient ($\forall t \in \mathbb{T}$) from a block design of approximately 2 seconds (Engel, Fries, and Singer 2001; Varela et al. 2001; Vidaurre, Abeysuriya, et al. 2018; Vidaurre, Hunt, et al. 2018; Tewarie et al. 2019).

$$p(\boldsymbol{\iota}(t,f)|\boldsymbol{\Theta}_{\boldsymbol{u}}(f)) = N^{\mathbb{C}}\big(\boldsymbol{\iota}(t,f)\big|\mathbf{0},\boldsymbol{\Theta}_{\boldsymbol{u}}^{-1}(f)\big) \tag{II-1}$$

where the complex-valued precision matrix $\boldsymbol{\Theta}_{\boldsymbol{u}}(f)$, or inverse of the cross-spectrum $\boldsymbol{\Sigma}_{\boldsymbol{u}}(f)$ ($\boldsymbol{\Theta}_{\boldsymbol{u}}(f) = \boldsymbol{\Sigma}_{\boldsymbol{u}}^{-1}(f)$)[a], defines within off-diagonal functional connectivity parameters as hermitian graph elements (Schreier and Scharf 2010; Andersen et al. 1995; Tugnait 2019). Functional connectivity proxies are then a function of the hermitan tensor $\{\boldsymbol{\Theta}_{\boldsymbol{u}}(f): \forall f \in \mathbb{F}\}$ at all frequencies (Valdés et al. 1992; Faes and Nollo 2011; Nolte et al. 2020; Marzetti, Del Gratta, and Nolte 2008; K. J. Friston et al. 2012).

A first interpretation of these hermitian graph elements is amplitude $\{|\boldsymbol{\Theta}_{\boldsymbol{u}}(g,g';f)|\}$ encoding undirected edges $\{g \leftrightarrow g'\}$ at a given frequency ($f \in \mathbb{F}$). This interpretation is also common for a real-valued and symmetric approximation for $\boldsymbol{\Theta}_{\boldsymbol{u}}(f)$ due to a GGS model describing $\boldsymbol{\iota}(t,f)$ as a narrow band filtered process.

A second interpretation is phase lag across frequencies in $\{\boldsymbol{\Theta}_{\boldsymbol{u}}(g,g';f): \forall f \in \mathbb{F}\}$ encoding directed edges $\{\dots, g \leftarrow g', \dots\}$, which are determined by other functional connectivity proxies such as coherences and the phase slope (Carter 1987; Bastos and Schoffelen 2016). A real-valued and symmetric approximation for $\boldsymbol{\Theta}_{\boldsymbol{u}}(f)$ misses this phase information and is limited interpreted to encode the degree of correlation or anticorrelation interpreted with the sign in $\{\boldsymbol{\Theta}_{\boldsymbol{u}}(g,g';f)\}$ (Matthew J. Brookes, Woolrich, et al. 2011a; Mantini et al. 2007; Hipp et al. 2012).

A third interpretation of the hermitian graph elements ($\boldsymbol{\Theta}_{\boldsymbol{u}}(f)$;$\forall f \in \mathbb{F}$) employed here in GGS model simulations here follows from transient second-order stochastic stationarity due to time-domain mixing conditions. The time domain $\boldsymbol{\iota}(t)$, as expressed by the stochastic integral (equation 2-2), is driven by the convolution kernel $\mathbf{K}_{\boldsymbol{u}}(\tau)$ ($\forall \tau, t \in \mathbb{T}$) and the perturbative process $\boldsymbol{\zeta}(t)$.

$$\boldsymbol{\iota}(t) = \int_0^t \mathbf{K}_{\boldsymbol{u}}(\tau)\boldsymbol{\iota}(t-\tau)d\tau + \boldsymbol{\zeta}(t) \tag{II-2}$$

where the convolution kernel $\mathbf{K}_{\boldsymbol{u}}(\tau)$ encodes mutltiply time lagged lagged causal relations ($\forall \tau \in \mathbb{T}$) physically interpreted as synaptic conductance values and axonal delays (Valdes-Sosa et al. 2009; David and Friston 2003; Jirsa and Haken 1997; Faes and Nollo 2011; Galka et al. 2004; Baccalá and Sameshima 2001; Babiloni et al. 2005).

The relation between the precision-matrix $\boldsymbol{\Theta}_{\boldsymbol{u}}(f)$ in (equation 1-1) and the convolution kernel $\mathbf{K}_{\boldsymbol{u}}(\tau)$ follows from analogous GGS model conditions held by the frequency-domain perturbative process $\boldsymbol{\zeta}(t,f)$ with precision-matrix $\boldsymbol{\Theta}_{\zeta\zeta}(f)$ in (equation 2-3).

$$\begin{gathered}\boldsymbol{\iota}(t,f) = \mathbf{K}_{\boldsymbol{u}}(f)\boldsymbol{\iota}(t,f) + \boldsymbol{\zeta}(t,f)\\ p\big(\boldsymbol{\zeta}(t,f)\big|\boldsymbol{\Theta}_{\zeta\zeta}(f)\big) = N^{\mathbb{C}}\Big(\boldsymbol{\zeta}(t,f)\Big|\mathbf{0},\boldsymbol{\Theta}_{\zeta\zeta}^{-1}(f)\Big)\end{gathered} \tag{II-3}$$

where the frequency-domain kernel $\mathbf{K}_{\boldsymbol{u}}(f)$ and precision-matrix $\boldsymbol{\Theta}_{\zeta\zeta}(f)$ are spectral factors for $\boldsymbol{\Theta}_{\boldsymbol{u}}(f)$ (equations 2-4) (Sayed and Kailath 2001; Ephremidze, Janashia, and Lagvilava 2007; Janashia, Lagvilava, and Ephremidze 2011; Jafarian and McWhirter 2012). Then, from the kernel or directed transfer function $\mathbf{K}_{\boldsymbol{u}}(f)$ and precision-matrix $\boldsymbol{\Theta}_{\zeta\zeta}(f)$, other functional connectivity proxies such as directed partial coherence or granger causality index encode directed edges $\{\dots, g \leftarrow g', \dots\}$ (K. Friston 2011; Faes, Stramaglia, and Marinazzo, n.d.; Blinowska 2011).

$$\begin{gathered}p(\boldsymbol{\iota}(t,f)|\boldsymbol{\Theta}_{\boldsymbol{u}}(f)) = N^{\mathbb{C}}\big(\boldsymbol{\iota}(t,f)\big|\mathbf{0},\boldsymbol{\Theta}_{\boldsymbol{u}}^{-1}(f)\big)\\ \boldsymbol{\Theta}_{\boldsymbol{u}}(f) = \big(\mathbf{I}-\mathbf{K}_{\boldsymbol{u}}(f)\big)^{\dagger}\boldsymbol{\Theta}_{\zeta\zeta}(f)\big(\mathbf{I}-\mathbf{K}_{\boldsymbol{u}}(f)\big)\end{gathered} \tag{II-4}$$

**2.1.2 GGS model MAP1 inverse solution for functional connectivity and hermitian graphical a priori probabilities**

Assume, for the moment, directly observed cortical oscillations $\boldsymbol{\iota}(t,f)$ explained by the GGS model (equation 2-1). Estimating the precision-matrix $\boldsymbol{\Theta}_{\boldsymbol{u}}(f)$ in this GGS model (equation 2-1) may be regarded as a pseudoinverse for the sampled covariance-matrix $\widehat{\boldsymbol{\Sigma}}_{\boldsymbol{u}}(f)$ (equation 2-5) for several samples ($\forall t \in \mathbb{T}$), where the sample size $T = |\mathbb{T}|$.

$$\widehat{\boldsymbol{\Sigma}}_{\boldsymbol{u}}(f) = \frac{1}{T}\sum_{t\in\mathbb{T}} \boldsymbol{\iota}(t,f)\boldsymbol{\iota}(t,f)^{\dagger} \tag{II-5}$$

The first-type likelihood explaining the sampled covariance-matrix $\widehat{\boldsymbol{\Sigma}}_{\boldsymbol{u}}(f)$ (equation 2-3) is a complex-valued Wishart (equation 2-6), with degree of freedom $T$ and scale matrix $T^{-1}\boldsymbol{\Theta}_{\boldsymbol{u}}^{-1}(f)$. The a priori probability (J. Friedman, Hastie, and Tibshirani 2008) is a Gibbs form (equations 2-6) upon a given scalar penalty function P for $\boldsymbol{\Theta}_{\boldsymbol{u}}(f)$, a scale parameter $\alpha_{\iota}$ and a selection matrix $\mathbf{A}_{\boldsymbol{u}}$.

$$\begin{gathered}p\left(\widehat{\boldsymbol{\Sigma}}_{\boldsymbol{u}}(f)\middle|\boldsymbol{\Theta}_{\boldsymbol{u}}(f)\right) = W^{\mathbb{C}}\big(\widehat{\boldsymbol{\Sigma}}_{\boldsymbol{u}}(f)\big|T^{-1}\boldsymbol{\Theta}_{\boldsymbol{u}}^{-1}(f),T\big)\\ p\big(\boldsymbol{\Theta}_{\boldsymbol{u}}(f)\big) = exp\big(\mathrm{P}\big(\mathbf{A}_{\boldsymbol{u}} \odot \boldsymbol{\Theta}_{\boldsymbol{u}}(f)\big)\big|\alpha_{\iota}T\big)\end{gathered} \tag{II-6}$$

where $\mathbf{A}_{\boldsymbol{u}}$ shall be a matrix of ones that places priors only the off-diagonal elements $\boldsymbol{\Theta}_{\boldsymbol{u}}(\nu)$, i.e., $\mathbf{A}_{\boldsymbol{u}}(i,i) = 0$ for all $i$ and $\mathbf{A}_{\boldsymbol{u}}(i,j) = 1$ for $i \neq j$. In subsequent work, this matrix will be used to reflect a priori connectivity information, for example, from anatomical data (Wodeyar 2019). Then, with $\widehat{\boldsymbol{\Sigma}}_{\boldsymbol{u}}(f)$ directly observed or determined from a first step (equation 1-2), MAP1 $\widehat{\boldsymbol{\Theta}}_{\boldsymbol{u}}(f)$ (equation 1-3) based on a posteriori probability $p\left(\boldsymbol{\Theta}_{\boldsymbol{u}}(f)|\widehat{\boldsymbol{\Sigma}}_{\boldsymbol{u}}(f)\right)$ from the GGS model likelihood and a priori probability (equation 2-6).

[a] This matrix may also be represent a time-frequency connectivity behaivor $\boldsymbol{\Theta}_{\boldsymbol{u}}(t_0,f)$ for a transient $\mathbb{T} = (t_0, t_0 + \Delta t)$ in a sliding in a slower timescale $t_0$. This timescale $t_0$ is the domain for evolving network configurations as represented across task or resting state transients.

$$\widehat{\boldsymbol{\Theta}}_{\boldsymbol{u}}(f) = argmax_{\boldsymbol{\Theta}_{\boldsymbol{u}}(f)}\left\{p\left(\boldsymbol{\Theta}_{\boldsymbol{u}}(f)|\widehat{\boldsymbol{\Sigma}}_{\boldsymbol{u}}(f)\right)\right\}$$
$$p\left(\boldsymbol{\Theta}_{\boldsymbol{u}}(f)|\widehat{\boldsymbol{\Sigma}}_{\boldsymbol{u}}(f)\right) \propto W^{\mathbb{C}}\left(\widehat{\boldsymbol{\Sigma}}_{\boldsymbol{u}}(f)\middle|T^{-1}\boldsymbol{\Theta}_{\boldsymbol{u}}^{-1}(f),T\right) exp\left(\mathrm{P}\left(\mathbf{A}_{\boldsymbol{u}} \odot \boldsymbol{\Theta}_{\boldsymbol{u}}(f)\right)\middle|\alpha_\iota T\right) \tag{II-7}$$

The equivalent problem of maximizing this a posteriori probability in MAP1 (equations 2-7) is commonly expressed in the literature as minimizing the $-log$ transformation (equations 2-8) as a penalized cost function $\mathcal{L}\left(\boldsymbol{\Theta}(f)\right) = -log\ p\left(\boldsymbol{\Theta}_{\boldsymbol{u}}(f)\middle|\widehat{\boldsymbol{\Sigma}}_{\boldsymbol{u}}(f)\right)$.

$$\widehat{\boldsymbol{\Theta}}_{\boldsymbol{u}}(f) = argmin_{\boldsymbol{\Theta}_{\boldsymbol{u}}(f)}\left\{\mathcal{L}\left(\boldsymbol{\Theta}_{\boldsymbol{u}}(f)\right)\right\}$$
$$\mathcal{L}\left(\boldsymbol{\Theta}_{\boldsymbol{u}}(f)\right) = -log|\boldsymbol{\Theta}_{\boldsymbol{u}}(f)| + tr\left(\widehat{\boldsymbol{\Sigma}}_{\boldsymbol{u}}(f)\boldsymbol{\Theta}_{\boldsymbol{u}}(f)\right) + \alpha_\iota \mathrm{P}\left(\mathbf{A}_{\boldsymbol{u}} \odot \boldsymbol{\Theta}_{\boldsymbol{u}}(f)\right) \tag{II-8}$$

The *Hermitian Graphical Naïve* (**hgNaïve**) estimator (a priori free $\mathrm{P} = 0$ in equation 2-8) is the inverse of the sample cross-spectral matrix $\widehat{\boldsymbol{\Theta}}_{\boldsymbol{u}}(f) \leftarrow \widehat{\boldsymbol{\Sigma}}_{\boldsymbol{u}}^{-1}(f)$ and usually yields a quite dense matrix with many spurious connectivities, which suggests the use of different penalizations. Then, we introduce (equations 2-8) the L2 norm $\mathrm{P} = \|\cdot\|_2^2$, the *Hermitian Graphical Ridge* (**hgRidge**) (Honorio and Jaakkola 2013; van Wieringen 2019), the L1 norm $\mathrm{P} = \|\cdot\|_1$, the *Hermitian Graphical LASSO* (**hgLASSO**) (Tugnait 2019), known as *Graphical LASSO* (**gLASSO**), in the real variable (J. Friedman, Hastie, and Tibshirani 2008).

In this paper, we shall emphasize that **hgLASSO** is the a priori model proposed to produce unbiased sparse estimates with the optimum of the target function (equations 2-5), where **hgNaïve** and **hgRidge** as well as **gLASSO** are hereinafter model violations. The critical issues are stable and scalable **hgLASSO** calculations in ultrahigh matrix dimensions preserving unbiasedness. We introduce a novel algorithm reformulating MAP1 (equations 2-7) with Bayesian hierarchical **hgLASSO**. This algorithm is plugged into multistep, and **onestep** methods are deferred to avoid interrupting the flow of exposition *2.3 hermitian graphical LASSO (hgLASSO)*.

### 2.2 Hidden Gaussian Graphical Spectral (HIGGS) model and inverse solutions via multistep MAP1 and onestep MAP2

#### 2.2.1 HIGGS model underneath the Gaussian EFM spectral equivalent

As with latent brain activity $\boldsymbol{\iota}(t)$, the conversion of brain oscillations $\boldsymbol{\iota}(t,f)$ is given according to the same Electromagnetic Forward Model (EFM) (Jörge Javier Riera and Fuentes 1998; Baillet, Mosher, and Leahy 2001; Grech et al. 2008). The EFM forward operator $\mathbf{L}_{\boldsymbol{v\iota}}$ (equation 2-11) describes in the time domain ($\forall t \in \mathbb{T}$) and spectral domain ($\forall f \in \mathbb{F}$) a purely linear and stationary measurement process. These measurements possess no relation whatsoever with any other biological mechanism [b] and are only perturbed by a spectral noise process $\boldsymbol{\xi}(t,f)$ at the sensors.

$$\boldsymbol{v}(t,f) = \mathbf{L}_{\boldsymbol{v\iota}}\boldsymbol{\iota}(t,f) + \boldsymbol{\xi}(t,f) \tag{II-9}$$

Hereinafter, EFM represents cortical activity due to the high sensitivity of MEG/EEG to the activity of pyramidal layers within the cortical columnar organization (P. L. Nunez et al. 1994; Paul L Nunez and Srinivasan 2006). Estimating activity for noncortical structures encounters another problem (Krishnaswamy et al. 2017) that must be addressed by some compensation measures for their bias to zero (Haufe et al. 2011).

Due to similar mixing conditions, the spectral noise process $\boldsymbol{\xi}(t,f)$ is asymptotically Gaussian with precision matrix $\boldsymbol{\Theta}_{\boldsymbol{\xi\xi}}(f)$, which is valid at any frequency ($f \in \mathbb{F}$) and time-domain ($\forall t \in \mathbb{T}$) (Brillinger 1975; Nolte et al. 2020). This is not accurate for the time-domain noise process $\boldsymbol{\xi}(t)$ commonly assumed in an equivalent EFM of MEG/EEG data $\boldsymbol{v}(t)$ (equations 2-9).

Then, this EFM Gaussian spectral equivalent represents the first-type likelihood explaining MEG/EEG oscillations (data) $\boldsymbol{v}(t,f)$ upon the latent brain oscillation parameters $\boldsymbol{\iota}(t,f)$ (equations 2-10). The a priori probability for parameters $\boldsymbol{\iota}(t,f)$ (equation 2-1) specifies the Hidden GGS (HIGGS) explaining $\boldsymbol{\iota}(t,f)$ upon the precision-matrix $\boldsymbol{\Theta}_{\boldsymbol{u}}(f)$. In addition, the hyperparameters $\boldsymbol{\Omega}(f)$ specified as precision matrices $\boldsymbol{\Omega}(f) = \left\{\boldsymbol{\Theta}_{\boldsymbol{u}}(f), \boldsymbol{\Theta}_{\boldsymbol{\xi\xi}}(f)\right\}$ within this HIGGS model require some a priori probability model $p\left(\boldsymbol{\Omega}(f)\right)$.

$$p\left(\boldsymbol{v}(t,f)|\boldsymbol{\iota}(t,f), \boldsymbol{\Theta}_{\boldsymbol{\xi\xi}}(f)\right) = N^{\mathbb{C}}\left(\boldsymbol{v}(t,f)\middle|\mathbf{L}_{\boldsymbol{v\iota}}\boldsymbol{\iota}(t,f), \boldsymbol{\Theta}_{\boldsymbol{\xi\xi}}^{-1}(f)\right)$$
$$p\left(\boldsymbol{\iota}(t,f)|\boldsymbol{\Theta}_{\boldsymbol{u}}(f)\right) = N^{\mathbb{C}}\left(\boldsymbol{\iota}(t,f)\middle|\mathbf{0}, \boldsymbol{\Theta}_{\boldsymbol{u}}^{-1}(f)\right) \tag{II-10}$$
$$p\left(\boldsymbol{\Omega}(f)\right) = p\left(\boldsymbol{\Theta}_{\boldsymbol{u}}(f)\right)p\left(\boldsymbol{\Theta}_{\boldsymbol{\xi\xi}}(f)\right)$$

#### 2.2.2 HIGGS model MAP1 inverse solution for brain oscillations

We introduce the first step MAP1 based on the HIGGS model (equations 2-10), with first-type likelihood $N^{\mathbb{C}}\left(\boldsymbol{v}(t,f)\middle|\mathbf{L}_{\boldsymbol{v\iota}}\boldsymbol{\iota}(t,f), \boldsymbol{\Theta}_{\boldsymbol{\xi\xi}}^{-1}(f)\right)$ and a priori probability $N^{\mathbb{C}}\left(\boldsymbol{\iota}(t,f)\middle|\mathbf{0}, \boldsymbol{\Theta}_{\boldsymbol{u}}^{-1}(f)\right)$, upon hyperparameters $\boldsymbol{\Omega}(f) = \left\{\boldsymbol{\Theta}_{\boldsymbol{u}}(f), \boldsymbol{\Theta}_{\boldsymbol{\xi\xi}}(f)\right\}$ (equations 2-11). Here, $\boldsymbol{\Omega}(f)$ incorporates in $\boldsymbol{\Theta}_{\boldsymbol{u}}^{-1}(f)$ similar information to the covariance-matrix $\boldsymbol{\Lambda}(f)$ in a first step MAP1 (equation 1-2). In addition, $\boldsymbol{\Omega}(f)$ is incorporated into $\boldsymbol{\Theta}_{\boldsymbol{\xi\xi}}^{-1}(f)$ information about the sensor noise covariance-matrix $\mathbf{B}(f)$. This MAP1 represents a large class of inverse solutions distinguished by the methods determining $\boldsymbol{\Lambda}(f)$ and $\mathbf{B}(f)$ in the Bayesian literature (Mattout et al. 2006; K. Friston et al. 2008; Trujillo-Barreto, Aubert-Vázquez, and Valdés-Sosa 2004; D. Wipf and Nagarajan 2009; Paz-Linares et al. 2017; Tipping 2001; D. P. Wipf et al. 2006; Casella et al. 2010).

$$\hat{\boldsymbol{\iota}}(f,t) = argmax_{\boldsymbol{\iota}(f,t)}\left\{p\left(\boldsymbol{\iota}(f,t)|\boldsymbol{v}(t,f), \boldsymbol{\Omega}(f)\right)\right\}$$
$$p\left(\boldsymbol{\iota}(f,t)|\boldsymbol{v}(t,f), \boldsymbol{\Omega}(f)\right) \propto N^{\mathbb{C}}\left(\boldsymbol{v}(t,f)\middle|\mathbf{L}_{\boldsymbol{v\iota}}\boldsymbol{\iota}(t,f), \boldsymbol{\Theta}_{\boldsymbol{\xi\xi}}^{-1}(f)\right) N^{\mathbb{C}}\left(\boldsymbol{\iota}(t,f)\middle|\mathbf{0}, \boldsymbol{\Theta}_{\boldsymbol{u}}^{-1}(f)\right) \tag{II-11}$$

---

[b] The alternative, fMRI observations, suffer from temporal/spectral (also spatial) deformations due to the nonlinear metabolic-hemodynamic forward model of the Blood Oxygenation Level Depend (BOLD) signal which is acquired with a very poor temporal resolution.

A HIGGS MAP1 estimate $\hat{\boldsymbol{\iota}}(f,t)$ (equation 2-11) is from a Gaussian a posteriori probability for $\boldsymbol{\iota}(t,f)$ with mean value $\hat{\boldsymbol{\iota}}(f,t)$ and covariance-matrix $\boldsymbol{\Pi}_{\iota\iota}(f)$ (equation 2-12). This MAP1 is a consequence of conjugated Gaussian relations between likelihood and a priori probability (equation 2-11), where estimate or mean value $\hat{\boldsymbol{\iota}}(f,t)$ is obtained through a linear inverse operator $\mathbf{T}_{\iota v}(f)$ from data $\boldsymbol{v}(t,f)$.

$$\hat{\boldsymbol{\iota}}(f,t) = \mathbf{T}_{\iota v}(f)\boldsymbol{v}(t,f)$$
$$p\big(\boldsymbol{\iota}(f,t)|\boldsymbol{v}(t,f),\boldsymbol{\Omega}(f)\big) = N^{\mathbb{C}}\big(\boldsymbol{\iota}(t,f)\big|\mathbf{T}_{\iota v}(f)\boldsymbol{v}(t,f),\boldsymbol{\Pi}_{\iota\iota}(f)\big) \tag{II-12}$$

A linear inverse operator $\mathbf{T}_{\iota v}(f)$ (equation 2-10) is the pseudoinverse for $\mathbf{L}_{v\iota}$ (forward-operator), as expressed in relation to a well-conditioned inverse or covariance-matrix $\boldsymbol{\Pi}_{\iota\iota}(f)$. This pseudoinverse $\mathbf{T}_{\iota v}(f)$ is then a compact and efficient representation for estimating $\hat{\boldsymbol{\iota}}(f,t)$ (equations 2-9) from the data $\boldsymbol{v}(t,f)$, which requires only a definition for parameters $\boldsymbol{\Omega}(f)$.

$$\mathbf{T}_{\iota v}(f) = \boldsymbol{\Pi}_{\iota\iota}(f)\mathbf{L}_{\iota v}\boldsymbol{\Theta}_{\xi\xi}(f)$$
$$\boldsymbol{\Pi}_{\iota\iota}(f) = \Big(\mathbf{L}_{\iota v}\boldsymbol{\Theta}_{\xi\xi}(f)\mathbf{L}_{v\iota} + \boldsymbol{\Theta}_{\iota\iota}(f)\Big)^{-1} \tag{II-13}$$

This practice based on $\mathbf{T}_{\iota v}(f)$ (equations 2-12 and 2-13) renders linear and stationary a whole process targeting brain oscillations $\boldsymbol{\iota}(f,t)$, which avoids time-domain distortions (M. J. Brookes, Woolrich, and Barnes 2012; G. L. Colclough et al. 2015; Marinazzo et al. 2019; Nolte et al. 2020). While a theoretical $\boldsymbol{\iota}(f,t)$ is converted by the forward-operator $\mathbf{L}_{v\iota}$ into data $\boldsymbol{v}(t,f)$, these data are in turn converted by the pseudoinverse $\mathbf{T}_{\iota v}(f)$ into estimates $\hat{\boldsymbol{\iota}}(f,t)$. In other words, a MAP1 estimate (equation 2-12) is equivalent to estimating the sampled covariance matrix of brain oscillations $\hat{\boldsymbol{\Sigma}}_{\iota\iota}(f)$ directly by a linear matrix transformation $\mathbf{T}_{\iota v}(f)$ (equation 2-14) pon the data sampled covariance matrix $\hat{\boldsymbol{\Sigma}}_{vv}(f)$: where $\hat{\boldsymbol{\Sigma}}_{vv}(f)$ is determined for the oscillation data $\boldsymbol{v}(t,f)$ at a specific frequency $f$ and within a transient $\forall t \in \mathbb{T}$ with sample size $T = |\mathbb{T}|$.

$$\hat{\boldsymbol{\Sigma}}_{\iota\iota}(f) = \mathbf{T}_{\iota v}(f)\hat{\boldsymbol{\Sigma}}_{vv}(f)\mathbf{T}_{v\iota}(f)$$
$$\hat{\boldsymbol{\Sigma}}_{vv}(f) = \tfrac{1}{T}\textstyle\sum_{t\in\mathbb{T}}\boldsymbol{v}(t,f)\boldsymbol{v}(t,f)^{\dagger} \tag{II-14}$$

In state-of-the-art practice, such a definition for $\boldsymbol{\Omega}(f)$ is most commonly through two methods: *Exact Low Resolution Electromagnetic Tomographic Analysis* (**ELORETA**) (Pascual-Marqui et al. 2006) and *Linearly Constrained Minimum Variance* (**LCMV**) (Van Veen et al. 1997). These methods then determine $\boldsymbol{\Omega}(f)$ by some optimal criteria for brain oscillations $\hat{\boldsymbol{\iota}}(f,t)$ explaining the sampled covariance matrix $\hat{\boldsymbol{\Sigma}}_{vv}(f)$ (equation 2-14) (Nolte et al. 2020; Marzetti, Del Gratta, and Nolte 2008).

Elsewhere, this type of inverse operator $\mathbf{T}_{v\iota}(f)$ may also impose sparsity to identify nonzero copmponents in the vector $\boldsymbol{\iota}(t,f)$ and be obtained iteratively along with $\boldsymbol{\Omega}(f)$ through Bayesian hierarchical methods (Mattout et al. 2006; K. Friston et al. 2008; Trujillo-Barreto, Aubert-Vázquez, and Valdés-Sosa 2004; D. Wipf and Nagarajan 2009; Paz-Linares et al. 2017; Tipping 2001; D. P. Wipf et al. 2006; Casella et al. 2010) or proximal projection methods (Gramfort et al. 2013; Haufe et al. 2008, 2011; Vega-Hernández et al. 2008; Bunea et al. 2011; Fan Jianqing 2001; Zou and Hastie 2005; Hunter and Li 2005). Additionally, elsewhere, an inverse operator that is nonlinear $\mathbf{T}_{\iota v}\big(\boldsymbol{v}(t,f)\big)$, linear nonstationary $\mathbf{T}_{\iota v}(t,f)$, or both nonlinear and nonstationary $\mathbf{T}_{\iota v}\big(t,f,\boldsymbol{v}(t,f)\big)$ is preferable to identify the type of latent brain activity causing evoked potentials in MEG/EEG (Makeig 2002; Makeig et al. 2004).

**2.2.3 HIGGS model MAP2 inverse solution for functional connectivity via successive approximations of MAP1 inverse solution**

Assume a HIGGS model (equations 2-10) explaining the MEG/EEG oscillation data $\boldsymbol{v}(t,f)$ upon $\boldsymbol{\iota}(t,f)$. Then, marginalizing brain-oscillations $\boldsymbol{\iota}(t,f)$ for the likelihood $p\left(\boldsymbol{v}(t,f)|\boldsymbol{\iota}(t,f),\boldsymbol{\Theta}_{\xi\xi}(f)\right)$ and under the a priori GGS model $p\big(\boldsymbol{\iota}(t,f)|\boldsymbol{\Theta}_{\iota\iota}(f)\big)$ translates into another GGS model (equations 2-15). This GGS model explains oscillations data $\boldsymbol{v}(t,f)$ upon a precision-matrix $\boldsymbol{\Theta}_{vv}$ dependent on hyperparameters $\boldsymbol{\Omega}(f) = \{\boldsymbol{\Theta}_{\iota\iota}(f),\boldsymbol{\Theta}_{\xi\xi}(f)\}$.

$$p\big(\boldsymbol{v}(t,f)|\boldsymbol{\Omega}(f)\big) = N^{\mathbb{C}}\big(\boldsymbol{v}(t,f)\big|\mathbf{0},\boldsymbol{\Theta}_{vv}^{-1}(f)\big)$$
$$\boldsymbol{\Theta}_{vv}^{-1}(f) = \mathbf{L}_{v\iota}\boldsymbol{\Theta}_{\iota\iota}^{-1}(f)\mathbf{L}_{\iota v} + \boldsymbol{\Theta}_{\xi\xi}^{-1}(f) \tag{II-15}$$

where estimating a precision matrix $\boldsymbol{\Theta}_{vv}(f)$ in this GGS model (equations 2-15) may also be regarded as a pseudoinverse for the sampled covariance matrix $\hat{\boldsymbol{\Sigma}}_{vv}(f)$ (equations 2-16) for several samples ($\forall t \in \mathbb{T}$), where the sample size $T = |\mathbb{T}|$.

$$\hat{\boldsymbol{\Sigma}}_{vv}(f) = \tfrac{1}{T}\textstyle\sum_{t\in\mathbb{T}}\boldsymbol{v}(t,f)\boldsymbol{v}(t,f)^{\dagger} \tag{II-16}$$

However, estimating the precision matrix $\boldsymbol{\Theta}_{\iota\iota}(f)$ in this GGS model is twice a pseudoinverse, first for the sampled covariance matrix $\hat{\boldsymbol{\Sigma}}_{vv}(f)$ and second the forward operator $\mathbf{L}_{v\iota}$ (equations 2-15), also involving estimation of the precision matrix $\boldsymbol{\Theta}_{\xi\xi}(f)$. The second-type likelihood explaining the sampled covariance-matrix $\hat{\boldsymbol{\Sigma}}_{vv}(f)$ (equation 2-16) is a complex-valued Wishart (equation 2-17), with degree of freedom $T$ and scale matrix $T^{-1}\boldsymbol{\Theta}_{vv}^{-1}(f)$. With this second-type likelihood (equation 2-17), direct estimation of the precision-matrices or hyperparameters $\boldsymbol{\Omega}(f) = \{\boldsymbol{\Theta}_{\iota\iota}(f),\boldsymbol{\Theta}_{\xi\xi}(f)\}$ is impractical, requiring variational Bayes approximations or Gibbs sampling not extensible to high dimensions.

$$p\left(\hat{\boldsymbol{\Sigma}}_{vv}(f)\middle|\boldsymbol{\Omega}(f)\right) = W^{\mathbb{C}}\left(\hat{\boldsymbol{\Sigma}}_{\iota\iota}(f)\middle|T^{-1}\left(\mathbf{L}_{v\iota}\boldsymbol{\Theta}_{\iota\iota}^{-1}(f)\mathbf{L}_{\iota v} + \boldsymbol{\Theta}_{\xi\xi}^{-1}(f)\right),T\right)$$
$$p\big(\boldsymbol{\Omega}(f)\big) = p\big(\boldsymbol{\Theta}_{\iota\iota}(f)\big)p\left(\boldsymbol{\Theta}_{\xi\xi}(f)\right) \tag{II-17}$$

Therefore, we use the successive approximations to the second-type likelihood $p\left(\hat{\boldsymbol{\Sigma}}_{vv}(f)\middle|\boldsymbol{\Omega}(f)\right)$ (equations 2-17) by the Expectation Maximization (EM) algorithm (Dempster, Laird, and Rubin 1977; C. Liu and Rubin 1994; McLachlan and Krishnan 2007; K. Friston et al.

2008; K. J. Friston, Trujillo-Barreto, and Daunizeau 2008; Wills, Ninness, and Gibson 2009). These successive approximations at an EM $k$-th iteration are expressed in Gibbs exponential form $p^{(k)}\left(\widehat{\boldsymbol{\Sigma}}_{\boldsymbol{vv}}(f)\middle|\boldsymbol{\Omega}(f)\right)$ of an expected $-log$ second-type likelihood of the data sampled covariance-matrix $\widehat{\boldsymbol{\Sigma}}_{\boldsymbol{vv}}(f)$ implicit in the cost function $Q\left(\boldsymbol{\Omega}(f),\widehat{\boldsymbol{\Omega}}^{(k)}(f)\right)$ (equations 2-18).

$$p^{(k)}\left(\widehat{\boldsymbol{\Sigma}}_{\boldsymbol{vv}}(f)\middle|\boldsymbol{\Omega}(f)\right) = exp\left(Q\left(\boldsymbol{\Omega}(f),\widehat{\boldsymbol{\Omega}}^{(k)}(f)\right)\middle| 1\right) \quad \text{(II-18)}$$

where the cost function $Q\left(\boldsymbol{\Omega}(f),\widehat{\boldsymbol{\Omega}}^{(k)}(f)\right)$ is obtained from the expected $-log$ second-type likelihood (equation 2-19) of the data $\boldsymbol{v}(t,f)$ for several samples ($\forall t \in \mathbb{T}$) and upon hyperparameters $\boldsymbol{\Omega}(f)$ and hyperparameters from the previous EM iteration $\widehat{\boldsymbol{\Omega}}^{(k)}(f)$.

$$Q\left(\boldsymbol{\Omega}(f),\widehat{\boldsymbol{\Omega}}^{(k)}(f)\right) = -\sum_{t\in\mathbb{T}} E\left\{log\left(p\left(\boldsymbol{v}(t,f),\boldsymbol{\iota}(t,f)\middle|\boldsymbol{\Omega}(f)\right)\right)\middle|\widehat{\boldsymbol{\Omega}}^{(k)}(f)\right\} \quad \text{(II-19)}$$

This likelihood is obtained from expectation applied to the $-log$ joint probability $p\left(\boldsymbol{v}(t,f),\boldsymbol{\iota}(t,f)\middle|\boldsymbol{\Omega}(f)\right)$ over parameters (brain oscillations) $\boldsymbol{\iota}(t,f)$ (equation 2-20). The joint probability is composed of the HIGGS model first-type likelihood and a priori probability (equations 2-10), and the expectation is based on the a posteriori probability of the HIGGS MAP1 (equations 2-12).

$$E\left\{log\left(p\left(\boldsymbol{v}(t,f),\boldsymbol{\iota}(t,f)\middle|\boldsymbol{\Omega}(f)\right)\right)\middle|\widehat{\boldsymbol{\Omega}}^{(k)}(f)\right\} = \int log\left(p\left(\boldsymbol{v}(t,f),\boldsymbol{\iota}(t,f)\middle|\boldsymbol{\Omega}(f)\right)\right) p\left(\boldsymbol{\iota}(t,f)\middle|\boldsymbol{v}(t,f),\widehat{\boldsymbol{\Omega}}^{(k)}(f)\right) d\boldsymbol{\iota}(t,f)$$

$$p\left(\boldsymbol{v}(t,f),\boldsymbol{\iota}(t,f)\middle|\boldsymbol{\Omega}(f)\right) = N^{\mathbb{C}}\left(\boldsymbol{v}(t,f)\middle|\mathbf{L}_{\boldsymbol{v\iota}}\boldsymbol{\iota}(t,f),\boldsymbol{\Theta}_{\boldsymbol{\xi\xi}}^{-1}(f)\right) N^{\mathbb{C}}\left(\boldsymbol{\iota}(t,f)\middle|\mathbf{0},\boldsymbol{\Theta}_{\boldsymbol{\iota\iota}}^{-1}(f)\right) \quad \text{(II-20)}$$

$$p\left(\boldsymbol{\iota}(f,t)|\boldsymbol{v}(t,f),\widehat{\boldsymbol{\Omega}}^{(k)}(f)\right) = N^{\mathbb{C}}\left(\boldsymbol{\iota}(t,f)\middle|\mathbf{T}_{\boldsymbol{\iota v}}^{(k)}(f)\boldsymbol{v}(t,f),\boldsymbol{\Pi}_{\boldsymbol{\iota\iota}}^{(k)}(f)\right)$$

The function $Q\left(\boldsymbol{\Omega}(f),\widehat{\boldsymbol{\Omega}}^{(k)}(f)\right)$ possesses an additive form (equation 2-21) upon expected covariance-matrices for parameters (brain-oscillations) $\widehat{\boldsymbol{\Psi}}_{\boldsymbol{\iota\iota}}^{(k)}(f)$ and residuals (spectral sensor noise process) $\widehat{\boldsymbol{\Psi}}_{\boldsymbol{\xi\xi}}^{(k)}(f)$. These expected covariance matrices are dependent on the data sampled covariance $\widehat{\boldsymbol{\Sigma}}_{\boldsymbol{vv}}(f)$ hyperparameters $\widehat{\boldsymbol{\Omega}}^{(k)}(f)$ at the $k$-th EM expectation.

$$Q\left(\boldsymbol{\Omega}(f),\widehat{\boldsymbol{\Omega}}^{(k)}(f)\right) = Q\left(\boldsymbol{\Theta}_{\boldsymbol{\iota\iota}}(f),\widehat{\boldsymbol{\Psi}}_{\boldsymbol{\iota\iota}}^{(k)}(f)\right) + Q\left(\boldsymbol{\Theta}_{\boldsymbol{\xi\xi}}(f),\widehat{\boldsymbol{\Psi}}_{\boldsymbol{\xi\xi}}^{(k)}(f)\right)$$

$$Q\left(\boldsymbol{\Theta}_{\boldsymbol{\iota\iota}}(f),\widehat{\boldsymbol{\Psi}}_{\boldsymbol{\iota\iota}}^{(k)}(f)\right) = -T log\left|\boldsymbol{\Theta}_{\boldsymbol{\iota\iota}}(f)\right| + T tr\left(\widehat{\boldsymbol{\Psi}}_{\boldsymbol{\iota\iota}}^{(k)}(f)\boldsymbol{\Theta}_{\boldsymbol{\iota\iota}}(f)\right) \quad \text{(II-21)}$$

$$Q\left(\boldsymbol{\Theta}_{\boldsymbol{\xi\xi}}(f),\widehat{\boldsymbol{\Psi}}_{\boldsymbol{\xi\xi}}^{(k)}(f)\right) = -T log\left|\boldsymbol{\Theta}_{\boldsymbol{\xi\xi}}(f)\right| + T tr\left(\widehat{\boldsymbol{\Psi}}_{\boldsymbol{\xi\xi}}^{(k)}(f)\boldsymbol{\Theta}_{\boldsymbol{\xi\xi}}(f)\right)$$

The expected covariance-matrix $\widehat{\boldsymbol{\Psi}}_{\boldsymbol{\iota\iota}}^{(k)}(f)$ (equation 2-22) is determined from an ensemble covariance-matrix $\widehat{\boldsymbol{\Pi}}_{\boldsymbol{\iota\iota}}^{(k)}(f)$ and its sampled estimator $\widehat{\boldsymbol{\Sigma}}_{\boldsymbol{\iota\iota}}^{(k)}(f)$ due to the a posteriori probability $p\left(\boldsymbol{\iota}(f,t)|\boldsymbol{v}(t,f),\widehat{\boldsymbol{\Omega}}^{(k)}(f)\right)$ (equation 2-20) (Dempster, Laird, and Rubin 1977).

$$\widehat{\boldsymbol{\Psi}}_{\boldsymbol{\iota\iota}}^{(k)}(f) = \widehat{\boldsymbol{\Sigma}}_{\boldsymbol{\iota\iota}}^{(k)}(f) + \widehat{\boldsymbol{\Pi}}_{\boldsymbol{\iota\iota}}^{(k)}(f)$$

$$\widehat{\boldsymbol{\Sigma}}_{\boldsymbol{\iota\iota}}^{(k)}(f) = \mathbf{T}_{\boldsymbol{\iota v}}^{(k)}(f)\widehat{\boldsymbol{\Sigma}}_{\boldsymbol{vv}}(f)\mathbf{T}_{\boldsymbol{v\iota}}^{(k)}(f)$$

$$\mathbf{T}_{\boldsymbol{\iota v}}^{(k)}(f) = \widehat{\boldsymbol{\Pi}}_{\boldsymbol{\iota\iota}}^{(k)}(f)\mathbf{L}_{\boldsymbol{\iota v}}\widehat{\boldsymbol{\Theta}}_{\boldsymbol{\xi\xi}}^{(k)}(f) \quad \text{(II-22)}$$

$$\widehat{\boldsymbol{\Pi}}_{\boldsymbol{\iota\iota}}^{(k)}(f) = \left(\mathbf{L}_{\boldsymbol{\iota v}}\widehat{\boldsymbol{\Theta}}_{\boldsymbol{\xi\xi}}^{(k)}\mathbf{L}_{\boldsymbol{v\iota}} + \widehat{\boldsymbol{\Theta}}_{\boldsymbol{\iota\iota}}^{(k)}\right)^{-1}$$

Analogous formulas are obtained for an expected covariance-matrix $\widehat{\boldsymbol{\Psi}}_{\boldsymbol{\xi\xi}}^{(k)}(f)$ (equation 2-23) determined from an ensemble covariance-matrix $\widehat{\boldsymbol{\Pi}}_{\boldsymbol{\xi\xi}}^{(k)}(f)$ and its sampled estimator $\widehat{\boldsymbol{\Sigma}}_{\boldsymbol{\iota\iota}}^{(k)}(f)$ in this case due to the a posteriori probability for the sensor noise process $p\left(\boldsymbol{\xi}(f,t)|\boldsymbol{v}(t,f),\widehat{\boldsymbol{\Omega}}^{(k)}(f)\right)$ (Dempster, Laird, and Rubin 1977).

$$\widehat{\boldsymbol{\Psi}}_{\boldsymbol{\xi\xi}}^{(k)}(f) = \widehat{\boldsymbol{\Sigma}}_{\boldsymbol{\xi\xi}}^{(k)}(f) + \widehat{\boldsymbol{\Pi}}_{\boldsymbol{\xi\xi}}^{(k)}(f)$$

$$\widehat{\boldsymbol{\Sigma}}_{\boldsymbol{\xi\xi}}^{(k)}(f) = \mathbf{T}_{\boldsymbol{\xi v}}^{(k)}(f)\widehat{\boldsymbol{\Sigma}}_{\boldsymbol{vv}}(f)\mathbf{T}_{\boldsymbol{\xi\iota}}^{(k)}(f)$$

$$\mathbf{T}_{\boldsymbol{\xi v}}^{(k)}(f) = \mathbf{I} - \widehat{\boldsymbol{\Pi}}_{\boldsymbol{\xi\xi}}^{(k)}(f)\widehat{\boldsymbol{\Theta}}_{\boldsymbol{\xi\xi}}^{(k)}(f) \quad \text{(II-23)}$$

$$\widehat{\boldsymbol{\Pi}}_{\boldsymbol{\xi\xi}}^{(k)}(f) = \mathbf{L}_{\boldsymbol{v\iota}}\widehat{\boldsymbol{\Pi}}_{\boldsymbol{\iota\iota}}^{(k)}(f)\mathbf{L}_{\boldsymbol{\iota v}}$$

Then, due to the cost function additive form (equation 2-21), the Gibbs exponential form $p^{(k)}\left(\widehat{\boldsymbol{\Sigma}}_{\boldsymbol{vv}}(f)\middle|\boldsymbol{\Omega}(f)\right)$ (equation 2-18) admits a factorization into two marginal second-type Wishart ($W^{\mathbb{C}}$) likelihood models (equation 2-24). Se details the derivation for these successive approximations to the HIGGS second-type likelihood in appendix *III. Bimodal Wishart form of the HIGGS expected second-type likelihood (Lemma 1)*.

$$p^{(k)}\left(\widehat{\boldsymbol{\Sigma}}_{\boldsymbol{vv}}(f)\middle|\boldsymbol{\Omega}(f)\right) = p^{(k)}\left(\widehat{\boldsymbol{\Sigma}}_{\boldsymbol{vv}}(f)\middle|\boldsymbol{\Theta}_{\boldsymbol{\iota\iota}}(f)\right) p^{(k)}\left(\widehat{\boldsymbol{\Sigma}}_{\boldsymbol{vv}}(f)\middle|\boldsymbol{\Theta}_{\boldsymbol{\xi\xi}}(f)\right)$$

$$p^{(k)}\left(\widehat{\boldsymbol{\Sigma}}_{\boldsymbol{vv}}(f)\middle|\boldsymbol{\Theta}_{\boldsymbol{\iota\iota}}(f)\right) = W^{\mathbb{C}}\left(\widehat{\boldsymbol{\Psi}}_{\boldsymbol{\iota\iota}}^{(k)}(f)\middle|T^{-1}\boldsymbol{\Theta}_{\boldsymbol{\iota\iota}}^{-1}(f),T\right) \quad \text{(II-24)}$$

$$p^{(k)}\left(\widehat{\boldsymbol{\Sigma}}_{\boldsymbol{vv}}(f)\middle|\boldsymbol{\Theta}_{\boldsymbol{\xi\xi}}(f)\right) = W^{\mathbb{C}}\left(\widehat{\boldsymbol{\Psi}}_{\boldsymbol{\xi\xi}}^{(k)}(f)\middle|T^{-1}\boldsymbol{\Theta}_{\boldsymbol{\xi\xi}}^{-1}(f),T\right)$$

The EM approximated MAP2 (maximization stage) $\widehat{\boldsymbol{\Theta}}_{\boldsymbol{\iota\iota}}^{(k+1)}(f)$ and $\widehat{\boldsymbol{\Theta}}_{\boldsymbol{\xi\xi}}^{(k+1)}(f)$ (equation 2-25) may be trapped at local optima but is explicitly Wishart and scalable to high dimensions combined with regularization priors $p\left(\boldsymbol{\Theta}_{\boldsymbol{\iota\iota}}(f)\right)$ and $p\left(\boldsymbol{\Theta}_{\boldsymbol{\xi\xi}}(f)\right)$ that may facilitate the search closest to the global optimal (McLachlan and Krishnan 2007; Wills, Ninness, and Gibson 2009). As with the previous first-type likelihood for $\boldsymbol{\Theta}_{\boldsymbol{\iota\iota}}(f)$ (equations 2-6, 2-7, 2-8), any of the penalizations discussed in the previous section may be used here to regularize the EM procedure. See Lemma 1 in *IV. HIGGS second-type maximum a posteriori and priors (Corollary to Lemma 1)*.

$$
\begin{aligned}
\widehat{\mathbf{\Theta}}_{\boldsymbol{u}}^{(k+1)}(f) &= argmax_{\mathbf{\Theta}_{\boldsymbol{u}}(f)}\left\{p^{(k)}\left(\mathbf{\Theta}_{\boldsymbol{u}}(f)|\widehat{\mathbf{\Sigma}}_{\boldsymbol{vv}}(f)\right)\right\} \\
\widehat{\mathbf{\Theta}}_{\boldsymbol{\xi\xi}}^{(k+1)}(f) &= argmax_{\mathbf{\Theta}_{\boldsymbol{\xi\xi}}(f)}\left\{p^{(k)}\left(\mathbf{\Theta}_{\boldsymbol{\xi\xi}}(f)\middle|\widehat{\mathbf{\Sigma}}_{\boldsymbol{vv}}(f)\right)\right\} \\
p^{(k)}\left(\mathbf{\Theta}_{\boldsymbol{u}}(f)|\widehat{\mathbf{\Sigma}}_{\boldsymbol{vv}}(f)\right) &= W^{\mathbb{C}}\left(\widehat{\mathbf{\Psi}}_{\boldsymbol{u}}^{(k)}(f)\middle|T^{-1}\mathbf{\Theta}_{\boldsymbol{u}}^{-1}(f),T\right)p\left(\mathbf{\Theta}_{\boldsymbol{u}}(f)\right) \\
p^{(k)}\left(\mathbf{\Theta}_{\boldsymbol{\xi\xi}}(f)\middle|\widehat{\mathbf{\Sigma}}_{\boldsymbol{vv}}(f)\right) &= W^{\mathbb{C}}\left(\widehat{\mathbf{\Psi}}_{\boldsymbol{\xi\xi}}^{(k)}(f)\middle|T^{-1}\mathbf{\Theta}_{\boldsymbol{\xi\xi}}^{-1}(f),T\right)p\left(\mathbf{\Theta}_{\boldsymbol{\xi\xi}}(f)\right)
\end{aligned}
\qquad \text{(II-25)}
$$

The EM maximization stage (equations 2-25) poses the same formalism as for a pair of observed GGS models at every iteration based on the $-log$ transformation of this posterior distribution, an additive penalized cost function $\mathcal{L}^{(k)}\left(\mathbf{\Omega}(f)\right) = \mathcal{L}^{(k)}\left(\mathbf{\Theta}_{\boldsymbol{u}}(f)\right) + \mathcal{L}^{(k)}\left(\mathbf{\Theta}_{\boldsymbol{\xi\xi}}(f)\right)$. Note that for expected covariance-matrix $\widehat{\mathbf{\Psi}}_{\boldsymbol{u}}^{(k)}(f)$ (equation 2-26), which is not directly observable, this is the same problem as for the observed $\widehat{\mathbf{\Sigma}}_{\boldsymbol{u}}(f)$ (equation 2-6, 2-7, 2-8), employing the graphical a priori models $\mathrm{P}$. A novel algorithm exposed here computes this solution based on the Bayesian hierarchical **hgLASSO** a priori model _2.3 GGS model MAP1 via Hermitian Graphical LASSO (hgLASSO)_.

$$
\begin{aligned}
\widehat{\mathbf{\Theta}}_{\boldsymbol{u}}^{(k+1)}(f) &= argmin_{\mathbf{\Theta}_{\boldsymbol{u}}(f)}\left\{\mathcal{L}^{(k)}\left(\mathbf{\Theta}_{\boldsymbol{u}}(f)\right)\right\} \\
\mathcal{L}^{(k)}\left(\mathbf{\Theta}_{\boldsymbol{u}}(f)\right) &= -log|\mathbf{\Theta}_{\boldsymbol{u}}(f)| + tr\left(\breve{\mathbf{\Psi}}_{\boldsymbol{u}}^{(k)}(f)\mathbf{\Theta}_{\boldsymbol{u}}(f)\right) + \alpha_{\iota}\mathrm{P}\left(\mathbf{A}_{\boldsymbol{u}} \odot \mathbf{\Theta}_{\boldsymbol{u}}(f)\right)
\end{aligned}
\qquad \text{(II-26)}
$$

For the spectral sensor noise process, we assume that $\mathbf{\Theta}_{\boldsymbol{\xi\xi}}(f)$ is known but a scalar factor $\theta_{\xi}^{2}(f)$ to be estimated: $\mathbf{\Theta}_{\boldsymbol{\xi\xi}}(f) = \theta_{\xi}^{2}(f)\mathbf{A}_{\boldsymbol{\xi\xi}}$, for which we use an exponential prior with scale parameter $\alpha_{\xi}$ (equations 2-27), with p being the number of MEG/EEG sensors and $T$ being the sample number.

$$
p\left(\theta_{\xi}^{2}(f)\middle|\alpha_{\xi}\right) = exp\left(\theta_{\xi}^{2}(f)\middle|\alpha_{\xi}\mathrm{p}T\right) \qquad \text{(II-27)}
$$

In this paper, we define $\mathbf{A}_{\boldsymbol{\xi\xi}}$ as the identity matrix, but note that it might be used to encode spurious EEG sensor connectivity due to scalp leakage currents. $\alpha_{\xi}$ (equation 2-27) could be used to encode the instrumental noise inferior threshold that we define as 10% of the EEG signal (equation 2-28).

$$
\begin{aligned}
\hat{\theta}_{\xi}^{2\,(k+1)}(f) &= 1/\left(tr\left(\breve{\mathbf{\Psi}}_{\boldsymbol{\xi\xi}}^{(k)}(f)\mathbf{A}_{\boldsymbol{\xi\xi}}\right)/\mathrm{p} + \alpha_{\xi}\right) \\
\mathcal{L}^{(k)}\left(\theta_{\xi}^{2}(f)\right) &= -\mathrm{p}log\left(\theta_{\xi}^{2}(f)\right) + \theta_{\xi}^{2}(f)tr\left(\breve{\mathbf{\Psi}}_{\boldsymbol{\xi\xi}}^{(k)}(f)\mathbf{A}_{\boldsymbol{\xi\xi}}\right) + \alpha_{\xi}\mathrm{p}\theta_{\xi}^{2}
\end{aligned}
\qquad \text{(II-28)}
$$

## 2.3 GGS model MAP1 inverse solution via Hermitian Graphical LASSO (hgLASSO) algorithm

### 2.3.1 Unbiasedness of the MAP1 inverse solution with hgLASSO a priori probability

We leverage recent results on the distribution of high-dimensional estimators of precision matrices of Jankova and Van de Geer (JVDG) (Janková and van de Geer 2018) to provide statistical guarantees for the HIGGS inverse solution. First, we reduce the bias of $\widehat{\mathbf{\Theta}}_{\boldsymbol{u}}(f)$ at each iteration of EM maximization in the **onestep hgLASSO** inverse solution (equations 2-11) or second step in the multistep **hgLASSO** inverse solution (equations 2-6) by substituting $\breve{\mathbf{\Psi}}_{\boldsymbol{u}}^{(k)}(f)$ in $\widehat{\mathbf{\Sigma}}_{\boldsymbol{u}}(f)$, an unbiased estimator via “desparsification” (equation 2-14). This unbiased estimator is the complex-valued extension for the graphical LASSO (gLASSO) (J. Friedman, Hastie, and Tibshirani 2008) exposed in the JVDG theory (Janková and van de Geer 2018).

$$
unb\left(\widehat{\mathbf{\Theta}}_{\boldsymbol{u}}(f)\right) = 2\widehat{\mathbf{\Theta}}_{\boldsymbol{u}}(f) - \widehat{\mathbf{\Theta}}_{\boldsymbol{u}}(f)\widehat{\mathbf{\Sigma}}_{\boldsymbol{u}}(f)\widehat{\mathbf{\Theta}}_{\boldsymbol{u}}(f) \qquad \text{(II-29)}
$$

Debiasing $\widehat{\mathbf{\Theta}}_{\boldsymbol{u}}(f)$ is particularly relevant to increase the reliability of functional connectivity estimates and reduce distortions. This unbiased estimator $unb\left(\widehat{\mathbf{\Theta}}_{\boldsymbol{u}}(f)\right)$ follows the hermitian Gaussian distribution (equations 2-15), allowing us to carry the thresholding of the EM maximization in the **onestep hgLASSO** inverse solution (equations 2-11) or the second step in the multistep **hgLASSO** inverse solution (equations 2-6).

$$
\begin{aligned}
p\left(unb\left(\widehat{\mathbf{\Theta}}_{\boldsymbol{u}}(f;i,j)\right)\right) &= N_{1}^{\mathbb{C}}\left(unb\left(\widehat{\mathbf{\Theta}}_{\boldsymbol{u}}(f;i,j)\right)\middle|\mathbf{\Theta}_{\boldsymbol{u}}(f;i,j),\widehat{\mathbf{\Sigma}}_{\mathbf{\Theta}}(f;i,j)/\sqrt{T}\right) \\
\widehat{\mathbf{\Sigma}}_{\mathbf{\Theta}}(f;i,j) &= \widehat{\mathbf{\Theta}}_{\boldsymbol{u}}(f;i,i)\widehat{\mathbf{\Theta}}_{\boldsymbol{u}}(f;j,j) + \widehat{\mathbf{\Theta}}_{\boldsymbol{u}}(f;i,j)
\end{aligned}
\qquad \text{(II-30)}
$$

With the fixed value of the regularization parameter $\alpha_{\iota} = \sqrt{T\,log(\mathrm{q})}$ and $T \gg \mathrm{q}$, whose z-statistic (equation 2-16) possesses a Rayleigh distribution with variance $1/\sqrt{2}$. We zero all values of $\widehat{\mathbf{\Theta}}_{\boldsymbol{u}}(f;j,j)$ with $\boldsymbol{Z}(f;i,j)$ lower than a threshold to ensure a familywise error of type I. It should be noted that this debiasing and thresholding yields every iteration of the EM maximization in the **onestep hgLASSO** inverse solution (equations 2-11) or the second step in the multistep **hgLASSO** inverse solution (equations 2-6) by substituting $\widehat{\mathbf{\Sigma}}_{\boldsymbol{u}}(f)$ in $\widehat{\mathbf{\Psi}}_{\boldsymbol{u}}(f)$, a statistically guaranteed thresholded connectivity matrix.

$$
\begin{aligned}
p\left(\boldsymbol{Z}(f;i,j)\right) &= 2e^{-\boldsymbol{Z}(f;i,j)^{2}}\boldsymbol{Z}(f;i,j) \\
\boldsymbol{Z}(f;i,j) &= \sqrt{\left|unb\left(\widehat{\mathbf{\Theta}}_{\boldsymbol{u}}(f;i,j)\right)\sqrt{T}/\widehat{\mathbf{\Sigma}}_{\mathbf{\Theta}}(f;i,j)\right|}
\end{aligned}
\qquad \text{(II-31)}
$$

### 2.3.2 Scalable and stable MAP1 inverse solution with the hgLASSO algorithm

The solution for the *Hermitian Graphical LASSO* (**hgLASSO**) that we present here is a transformation of its prior (equation 2-3) by means of the extension to the complex variable of the scaled Gaussian mixture procedure (equation 2-17) (Andrews 1974); see Lemma 2 in *V. Complex-valued Andrews and Mallows Lemma: Local Quadratic Approximation (LQA) of the hermitian graphical LASSO (hgLASSO) prior (Lemma 2)*.

$$p\left(\boldsymbol{\Theta}_{\boldsymbol{u}}(f)|\boldsymbol{\Gamma}(f)\right)=\prod_{i,j=1}^{\mathrm{q}} N_{1}^{T}\left(|\boldsymbol{\Theta}_{\boldsymbol{u}}(f;i,j)|\middle|0,\boldsymbol{\Gamma}^{2}(f;i,j)\right)$$
$$p\left(\boldsymbol{\Gamma}(f)\right)=\prod_{i,j}^{\mathrm{q}} Ga^{T}\left(\boldsymbol{\Gamma}^{2}(f;i,j)|1,\alpha_{\iota}^{2}\mathbf{A}_{\boldsymbol{u}}^{2}(i,j)/2\right) \quad \text{(II-32)}$$

The scaled Gaussian mixture (equations 2-17) is a type of representation of the hgLASSO prior (equations 2-3) by a Gaussian distribution of $\boldsymbol{\Theta}_{\boldsymbol{u}}(f)$ conditioned to other hyperparameters (variances $\boldsymbol{\Gamma}(f)$) with a gamma distribution. As a result, we obtain a *Local Quadratic Approximation* (LQA) of the target function used in (equations 2-6 or 2-11) $\mathcal{L}\left(\boldsymbol{\Theta}(f),\boldsymbol{\Gamma}(f)\right)$ in terms of the weighted hermitian graphical Ridge (hgRidge) due to the modified LASSO prior (equations 2-18). The estimation derived from the LQA poses a unique solution, given by Lemma 3 in *VI. Concavity of the first-type maximum a posteriori with hgLASSO LQA prior (Lemma 3)*.

$$\left(\widehat{\boldsymbol{\Theta}}_{\boldsymbol{u}}(f),\widehat{\boldsymbol{\Gamma}}(f)\right)=argmin_{\boldsymbol{\Theta}(f),\boldsymbol{\Gamma}(f)}\left\{\mathcal{L}\left(\boldsymbol{\Theta}_{\boldsymbol{u}}(f),\boldsymbol{\Gamma}(f)\right)\right\}$$
$$\mathcal{L}\left(\boldsymbol{\Theta}_{\boldsymbol{u}}(f),\boldsymbol{\Gamma}(f)\right)=-T\,log|\boldsymbol{\Theta}_{\boldsymbol{u}}(f)|\ +T\,tr\left(\widehat{\boldsymbol{\Sigma}}_{\boldsymbol{u}}(f)\boldsymbol{\Theta}_{\boldsymbol{u}}(f)\right)+\frac{T}{2}\|\boldsymbol{\Theta}_{\boldsymbol{u}}(f)\oslash\boldsymbol{\Gamma}(f)\|_{2}^{2} \quad \text{(II-33)}$$
$$+\frac{1}{2}\sum_{i,j=1}^{\mathrm{q}} log\,\boldsymbol{\Gamma}^{2}(f;i,j)+\frac{T\alpha^{2}}{2}\|\boldsymbol{\Gamma}(f)\odot\mathbf{A}_{\boldsymbol{u}}\|_{2}^{2}$$

We can easily estimate the entries of the variance matrix $\boldsymbol{\Gamma}(f)$ (equation 2-19) iteratively upon any estimated precision-matrix an inner cycle iteration (say the $l$-th) $\widehat{\boldsymbol{\Theta}}_{\boldsymbol{u}}^{(l)}(\nu)$.

$$\widehat{\boldsymbol{\Gamma}}^{(l+1)}(f)=argmin_{\boldsymbol{\Gamma}(f)}\left\{\mathcal{L}\left(\widehat{\boldsymbol{\Theta}}_{\boldsymbol{u}}^{(l)}(f),\boldsymbol{\Gamma}(f)\right)\right\}$$
$$\widehat{\boldsymbol{\Gamma}}^{(l+1)}(f)=\left(\left(-\mathbf{1}_{\mathrm{q}}+\left(\mathbf{1}_{\mathrm{q}}+4\alpha_{\iota}^{2}\mathbf{A}_{\boldsymbol{u}}^{.2}\odot\left|\widehat{\boldsymbol{\Theta}}_{\boldsymbol{u}}^{(l)}(f)\right|^{.2}\right)^{.\frac{1}{2}}\right)\oslash\left(2\alpha_{\iota}\mathbf{A}_{\boldsymbol{u}}^{.2}\right)\right)^{.\frac{1}{2}} \quad \text{(II-34)}$$

Unfortunately, there is no explicit solution for $\widehat{\boldsymbol{\Theta}}_{\boldsymbol{u}}^{(l)}(f)$ through (equations 2-18), something we solve by redefining it into standard precision-matrix $\widetilde{\boldsymbol{\Theta}}_{\boldsymbol{u}}(\nu)$ (equations 2-20), which are weighted locally by the estimated variance matrix $\boldsymbol{\Gamma}(f)$ from the previous iteration. This precision matrix also requires transforming the Wishart distribution associated with the GGS of (equation 2-2) and the HIGGS (equation 2-10) in terms of the standard sampled covariance matrix $\widetilde{\widehat{\boldsymbol{\Sigma}}}_{\boldsymbol{u}}(f)$ due to Lemma 4 in *VII. Standardization of the first-type likelihood with hgLASSO LQA prior (Lemma 4)*.

$$\widetilde{\boldsymbol{\Theta}}_{\boldsymbol{u}}(f)=\boldsymbol{\Theta}_{\boldsymbol{u}}(f)\oslash\boldsymbol{\Gamma}(f)$$
$$\widetilde{\widehat{\boldsymbol{\Sigma}}}_{\boldsymbol{u}}(f)=\left(\widehat{\boldsymbol{\Sigma}}_{\boldsymbol{u}}^{-1}(f)\oslash\boldsymbol{\Gamma}(f)\right)^{-1} \quad \text{(II-35)}$$

This standardization poses the typical hermitian graphical Ridge (hgRidge) problem (van Wieringen 2019), without weights (equation 2-21).

$$p\left(\widetilde{\widehat{\boldsymbol{\Sigma}}}_{\boldsymbol{u}}(f)\middle|\widetilde{\boldsymbol{\Theta}}_{\boldsymbol{u}}(f)\right)=W_{\mathrm{q}}^{\mathbb{C}}\left(\widetilde{\widehat{\boldsymbol{\Sigma}}}_{\boldsymbol{u}}(f)\middle|T^{-1}\widetilde{\boldsymbol{\Theta}}_{\boldsymbol{u}}^{-1}(\nu),T\right)$$
$$\widetilde{\boldsymbol{\Theta}}_{\boldsymbol{u}}(f)\sim exp\left(\left\|\widetilde{\boldsymbol{\Theta}}_{\boldsymbol{u}}(f)\right\|_{2}^{2}\middle|T/2\right) \quad \text{(II-36)}$$

The solution to the hgRidge with standard empirical covariance $\widetilde{\widehat{\boldsymbol{\Sigma}}}_{\boldsymbol{u}}(f)$ (equation 2-20) allows obtaining the standard precision-matrix $\widetilde{\widehat{\boldsymbol{\Theta}}}_{\boldsymbol{u}}^{(l)}(f)$ (equation 2-21) and thus retrieving the actual precision-matrix $\boldsymbol{\Theta}_{\boldsymbol{u}}(f)$ (equation 2-18). See Lemma 5 in *VIII. Hermitian graphical Ridge (hgRidge) and hgLASSO LQA estimator (Lemma 5)*. This solution is in terms of the matrix square root operation (equation 2-22) that bounds the computational cost of this procedure, which requires a single cycle for the mutual estimation of the quantities $\boldsymbol{\Theta}_{\boldsymbol{u}}(f)$ and $\boldsymbol{\Gamma}(f)$.

$$\widetilde{\widehat{\boldsymbol{\Theta}}}_{\boldsymbol{u}}^{(l)}(f)=-\frac{1}{2}\widetilde{\widehat{\boldsymbol{\Sigma}}}_{\boldsymbol{u}}^{(l)}(f)+\frac{1}{2}\sqrt{\left(\widetilde{\widehat{\boldsymbol{\Sigma}}}_{\boldsymbol{u}}^{(l)}(f)\right)^{2}+4\mathbf{I}_{\mathrm{q}}}$$
$$\widetilde{\widehat{\boldsymbol{\Sigma}}}_{\boldsymbol{u}}^{(l)}(f)=\left(\widehat{\boldsymbol{\Sigma}}_{\boldsymbol{u}}^{-1}(f)\oslash\widehat{\boldsymbol{\Gamma}}^{(l)}(f)\right)^{-1} \quad \text{(II-37)}$$
$$\widehat{\boldsymbol{\Theta}}_{\boldsymbol{u}}^{(l)}(f)=\widetilde{\widehat{\boldsymbol{\Theta}}}_{\boldsymbol{u}}^{(l)}(f)\odot\widehat{\boldsymbol{\Gamma}}^{(l)}(f)$$

## 2.4 Validating performance for HIGGS inverse solutions

### 2.4.1 Simulations

We created simulations (Fig. 4) based on second-order stochastic stationary dynamics (equation 2-2) following from an underlying *Gaussian Graphical Spectral* (GGS) model. This hermitian precision matrix (Fig. 4 a) defining this GGS produces brain oscillations that resemble the so-called EEG Xi-Alpha model in the spectrum (Fig. 4 b) (Pascual-marqui, Valdes-sosa, and Alvarez-amador 1988). In other words, this autoregressive dynamic $\boldsymbol{\iota}(t)$ is defined from $\boldsymbol{\iota}(t,f)$ (equations 2-4), produced by a hermitian precision tensor (connectivity) $\boldsymbol{\Theta}_{\boldsymbol{u}}(f)$ at all frequencies which also causes the oscillations in the alpha peak (Fig. 4 c). This hermitian precision tensor originates from spectral factors $\mathbf{K}_{\boldsymbol{u}}(f)$ (Fig. 4 d) or the Hilbert transform of multiply lagged connectivity $\mathbf{K}_{\boldsymbol{u}}(\tau)$ producing broad band dynamics (Fig. 4 e).

This connectivity $\boldsymbol{\Theta}_{\boldsymbol{u}}(f)$ (Fig. 4 a) is employed as a baseline to compare the results and is more in correspondence with the actual target for state-of-the-art methods. The simulation pipeline that was repeated 100 times to produce the same amount of possible ground truth and observations is as follows.

1. Definition of random hermitian precision-matrix $\boldsymbol{\Theta}_{aa}(f_0)$ (Fig. 4a) (ground truth functional connectivity) given for a referential frequency component $f_0$ meant to be the center of the alpha peak (Fig. 4b) ($f_0 = 10\mathrm{Hz}$) in the cortical network subspace. The elements of this matrix must lie in the unitary circle of the complex plane for the reliability of binary classification measures used in this validation.
2. Construction of a synthetic spectral factorization for $\boldsymbol{\Theta}_{aa}(f_0)$ (Fig. 4c) or spectral directed model used to design a precision tensor at all frequencies ($f \in \mathbb{F}$). This factorization was performed assuming that the reference spectral factors at ($f_0 = 10\mathrm{Hz}$) are derived from a hermitian eigendecomposition $\boldsymbol{\Theta}_{aa}(f_0) = \mathbf{UDU}^{\dagger}$ (equation 2-38), where $\mathbf{K}_{\boldsymbol{u}}(f_0)$ is the directed transfer function at $f_0$ extracted from the $\mathbf{UDU}^{\dagger}$ elements.

$$\begin{gathered}\boldsymbol{\Theta}_{aa}(f_0) = \left(\mathbf{I} - \mathbf{K}_{\boldsymbol{u}}(f_0)\right)^{\dagger}\left(\mathbf{I} - \mathbf{K}_{\boldsymbol{u}}(f_0)\right) \\ \mathbf{K}_{\boldsymbol{u}}(f_0) = \mathbf{I} - \mathbf{U}^{\dagger}\sqrt{\mathbf{D}}\end{gathered} \quad \text{(II-38)}$$

3. Considering discrete-time lags in $\mathbf{K}_{\boldsymbol{u}}(\tau)$ as in equation (equation 2-2) encoded in a matrix $\mathbf{T}_{aa}$, where $\mathbf{T}_{aa}(i,j)$ is the lag for directed interactions directed $\{\dots, \boldsymbol{\iota}(t;i) \leftarrow \boldsymbol{\iota}(t;j), \dots\}$, we represent $\mathbf{K}_{\boldsymbol{u}}(\tau)$ (equation 2-39) as proportional to a constant connectivity matrix $\mathbf{K}_{aa}$ and the Dirac delta matrix function of the lags $\boldsymbol{\Delta}(\tau - \mathbf{T}_{aa})$.

$$\begin{gathered}\mathbf{K}_{\boldsymbol{u}}(\tau) \propto \mathbf{K}_{aa}\boldsymbol{\Delta}(\tau - \mathbf{T}_{aa}) \\ \mathbf{T}_{aa} = phase\left(\mathbf{K}_{\boldsymbol{u}}(f_0)\right)/f_0\end{gathered} \quad \text{(II-39)}$$

These lags are associated with the spectral $\mathbf{K}_{\boldsymbol{u}}(f_0)$ at $f_0$ given the phase relation (equation 2-40), and the directed transfer function tensor $\mathbf{K}_{\boldsymbol{u}}(f)$ at all frequencies ($f \in \mathbb{F}$) (Fig. 4d) as in (equation 2-3) is then obtained from the definition for $\mathbf{K}_{\boldsymbol{u}}(\tau)$ (equation 2-39) modulated by a Gaussian spectral form with center in the alpha peak ($f_0 = 10\mathrm{Hz}$).

$$\begin{gathered}\mathbf{K}_{\boldsymbol{u}}(f_0) = \mathbf{K}_{aa}exp(\mathbf{j}\mathbf{T}_{aa}f_0) \\ \mathbf{K}_{\boldsymbol{u}}(f) = \mathbf{K}_{aa}exp(\mathbf{j}\mathbf{T}_{aa}f)exp((f - f_0)^2|\theta_{aa})\end{gathered} \quad \text{(II-40)}$$

4. The precision tensor $\boldsymbol{\Theta}_{\boldsymbol{u}}(f)$ (Fig. 4c) is then recomposed from the spectral factors (equation 2-41) assuming directed alpha transfer function tensor $\mathbf{K}_{\boldsymbol{u}}(f)$ (equation 2-40) and Xi innovations precision tensor $\boldsymbol{\Theta}_{\zeta\zeta}(f)$ as in (equation 2-4). The Xi factor was construed by following steps similar to alpha but modulated by a right-sided Gaussian spectral form with central frequency ($f_0 = 0\mathrm{Hz}$). The precision matrix $\boldsymbol{\Theta}_{xx}$ was the inverse of the surface Laplacian in the whole cortical space, and the lags for the precision tensor construction were set to zero, corresponding to the properties of this process.

$$\begin{gathered}\boldsymbol{\Theta}_{\boldsymbol{u}}(f) = \left(\mathbf{I} - \mathbf{K}_{\boldsymbol{u}}(f)\right)^{\dagger}\boldsymbol{\Theta}_{\zeta\zeta}(f)\left(\mathbf{I} - \mathbf{K}_{\boldsymbol{u}}(f)\right) \\ \boldsymbol{\Theta}_{\zeta\zeta}(f) = \boldsymbol{\Theta}_{xx}exp(f^2|\theta_{xx})\end{gathered} \quad \text{(II-41)}$$

5. The precision tensor slicewise inverted $\boldsymbol{\Sigma}_{\boldsymbol{u}}(f) = \boldsymbol{\Theta}_{\boldsymbol{u}}^{-1}(f)$ (equation 2-41) (this is the ground truth covariance or cross-spectrum $\boldsymbol{\Sigma}_{\boldsymbol{u}}(f)$) is used to obtain complex-valued vectors of the Fourier transform (equation 2-42). Equivalent in this case to the Hilbert transform $\boldsymbol{\iota}(t,f)$ ($t \in \mathbb{T}$) by means of a hermitian Gaussian random generator with sample number $T = 600$ at every frequency ($f \in \mathbb{F}$). Brain oscillation time series $\boldsymbol{\iota}(t)$ (Fig. 4e) are obtained by means of the inverse Fourier transform ($\mathcal{F}^{-1}$) across frequencies ($f \in \mathbb{F}$) of these samples $\boldsymbol{\iota}(t,f)$ to obtain the same number of time instances ($t \in \mathbb{T}$).

$$\begin{gathered}p(\boldsymbol{\iota}(t,f)|\boldsymbol{\Theta}_{\boldsymbol{u}}(f)) = N^{\mathbb{C}}\left(\boldsymbol{\iota}(t,f)\middle|\mathbf{0}, \boldsymbol{\Theta}_{\boldsymbol{u}}^{-1}(f)\right) \\ \boldsymbol{\iota}(t) = \mathcal{F}^{-1}\left(\boldsymbol{\iota}(t,f)\right)\end{gathered} \quad \text{(II-42)}$$

6. The Xi-Alpha process $\boldsymbol{\iota}(t,f)$ in the cortical network subspace (equation 2-42) was projected to obtain observations (Fig. 4g) at the sensor space $\boldsymbol{v}(t)$ (equation 2-43), time-domain equivalent of (equation 2-9) with corresponding forward-operator $\mathbf{L}_{\boldsymbol{v\iota}}$ (Fig. 4f), adding the Xi process and a white noise process in the whole cortical space. A white noise process $\boldsymbol{\xi}(t)$ in the sensor space $\mathbb{E}$ was also added to the projected source space process. The composition of the confounding processes (Xi process defined by $\boldsymbol{\Theta}_{xx}$ and source and white sensor noises) with the alpha process defined by $\boldsymbol{\Theta}_{aa}(f_0)$ was adjusted to keep spectral energy in the alpha band (8-12 Hz) of 10% of the alpha process energy.

$$\boldsymbol{v}(t) = \mathbf{L}_{\boldsymbol{v\iota}}\boldsymbol{\iota}(t) + \boldsymbol{\xi}(t) \quad \text{(II-43)}$$

Two different simulations based on the *Electromagnetic Forward Model* (EFM) with EFM forward-operator $\mathbf{L}_{\boldsymbol{v\iota}}$ (equation 2-43) were defined: For an idealized “planar EFM” head (Fig. 4f1) and realistic “human EFM” head (Fig. 4f2) to produce the Xi-Alpha observations (Fig. 4g). The planar EFM (Fig. 4f1) was computed based on the bidimensional geometry of two concentric circles defining a planar cortex and scalp and a layout of equidistant planar scalp sensors (30 sensors). The human EFM (Fig. 4f2) was computed for the cortical surface extracted from a healthy subject T1 image, with the *Boundary Element Method* (BEM) implemented in SPM and a layout of the extended 10-20 system (30 sensors). A cortical network was defined on a subspace of 22 points for both the planar and human EFMs. For the planar cortex, these were equidistantly located, and for the human cortex, they were randomly located across different areas of the left (L) and right (R) hemispheres: Occipital (LO and RO), Temporal (LT and RT), Parietal (LP and RP) and Frontal (LF and RF).

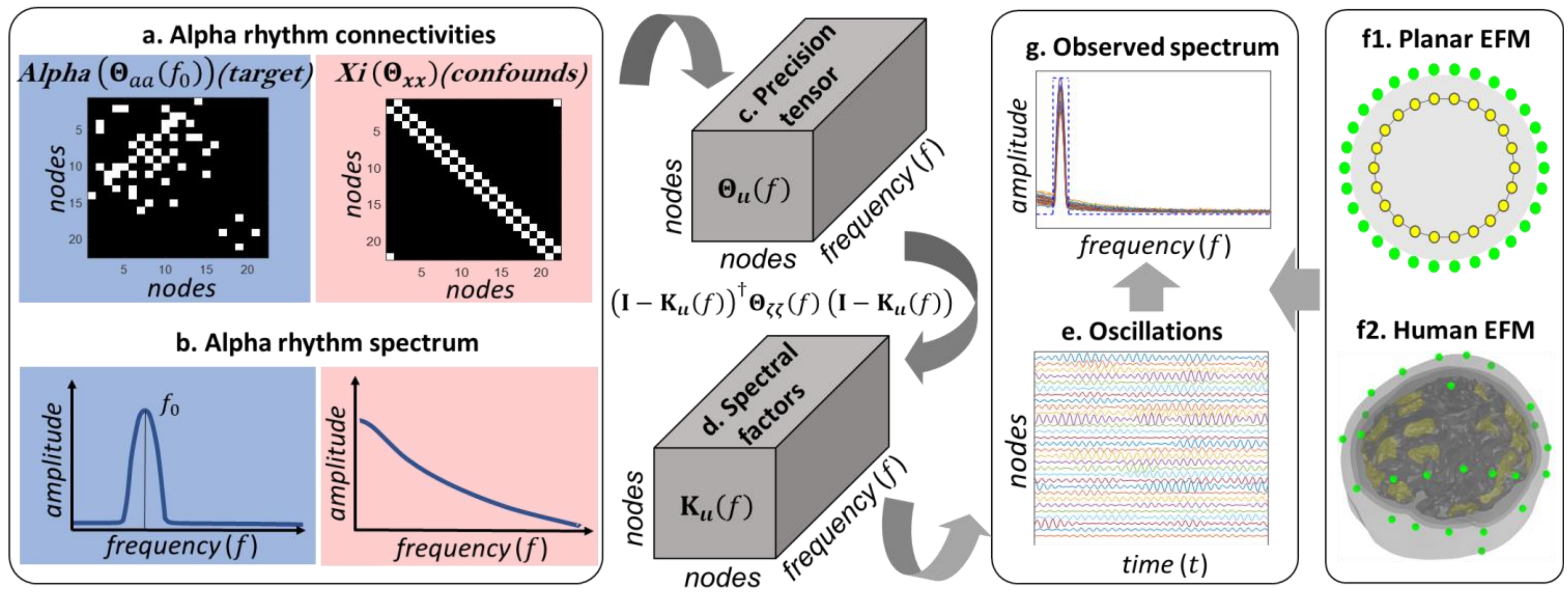

Fig. 4| Specification of a linear autoregressive dynamics in the frequency domain based on the Xi and Alpha model a. Alpha connectivity is the ground true to be mixed later with that of the Xi process. The starting point are these Alpha and Xi precision-matrices at 10Hz for an oscillatory network process represented in a binary map that corresponds to edges in such network graph-elements. b. This is extended for all frequencies via an amplitude-phase transformation to recreate c. the precision-tensor employing d. directed transfer function tensor for the Alpha process by means of its composition in spectral factors with the Xi innovations The factors are derived from the central slice (10Hz) by means of the eigen-decomposition. The process cross-spectrum tensor is obtained by slice-inverting the precision tensor producing e. brain-oscillations by means of a hermitian random Gaussian generator. f. Two types of forward models, f1. planar and f2. human project these oscillations to the sensor space producing g. the Xi Alpha observations.

**2.4.2 Experimental confirmation**

For the confirmation of HIGGS connectivity, we compared EEG against a more direct technique: electrocorticography (ECoG) (Fig. 5) leveraging the unique experimental setup that offers the advantage of large brain coverage ECoG on a healthy macaque (Nagasaka, Shimoda, and Fujii 2011). This comparison scenario is particularly interpretable in terms of the effect of volume conduction heterogeneities since the EEG is hidden under several tissue layers (Fig. 5 e2), and the ECoG is instead hidden under only one (Fig. 5 e1). See the macaque preparation for the surgical implantation of this ECoG in (Fig. 5 a) that, which is shown in the postsurgical X-ray (Fig. 5 b). The macaque sensor layouts for the simultaneous ECoG/EEG recordings consisted of 128 ECoG sensors placed upon the cortical surface in the left hemisphere and a low-density array of 20 EEG scalp sensors. The relative distribution of the macaque cortical surface segmented from the MRI is shown in (Fig. 5 c).

The observations for EEG $\boldsymbol{v}^{EEG}(t)$ and ECoG $\boldsymbol{v}^{ECoG}(t)$ were recorded simultaneously in the resting state. During the experimental session, the monkey was awake, blindfolded, and constrained to a sitting position. Both EEG and ECoG were synchronized to the trigger signal and downsampled to 1000 Hz, keeping 2 minutes of recordings. The artifact removal procedure included linear detrending with the L1TF package, average DC subtraction, and 50 Hz notch filtering. The spectral analysis (Fig. 5 d) of both ECoG (Fig. 5 d1) and EEG (Fig. 5 d2) signals reveals a larger power spectral density within the band (8-14 Hz) associated with the macaque alpha oscillations. The forward operators were obtained from a head conductivity model (Fig. 5 e) for the ECoG $\mathbf{L}_{v\iota}^{ECoG}$ (Fig. 5 e1) and EEG $\mathbf{L}_{v\iota}^{EEG}$ (Fig. 5 e2) through FEM computations in SimBio using the macaque individual T1 MRI segmentation. The head model included five conductivity compartments: cortex, ECoG silicon layer, inner skull, outer skull, and scalp.

We work at a subspace of cortical sources detected from ECoG, and in a similar fashion to the simulation study (Fig. 5 f), extracting the connectivity (Fig. 5 g) with all methods from both ECoG (Fig. 5 g1) and EEG (Fig. 5 g2) observations. We screen out the significant cortical network nodes (Fig. 5 f) from the ECoG by means of implementation for spectral analysis of the *Structured Sparse Bayesian Learning* (SSBL) (Paz-Linares et al. 2017). This implementation uses assumptions similar to those of HIGGS but with a diagonal covariance structure. We offer technical details in (Gonzalez-Moreira et al. 2020).

The detection (Fig. 5 f) corresponding to the alpha band oscillations (Fig. 5 d1) yielded a large-scale network of distributed nodes, which strongly correlates with the findings for these ECoG cortical patterns in the left hemisphere (Q. Wang et al. 2019). The more extended Occipital (O) oscillations were accompanied by secondary cortical oscillations extended over the Temporal (T), Parietal (P), and Frontal (F) areas. This large-scale analysis differs from previously reported studies, which were limited to P <-> F interactions at ECoG sensor space as in (Papadopoulou, Friston, and Marinazzo 2019). We consider eight regions of interest (ROI) extracted manually from the left hemisphere cortical segmentation. These are Inferior Occipital (IO), Superior Occipital (SO), Posterior Temporal (PT), Anterior Temporal (AT), Inferior Parietal (IP), Superior Parietal (SP), Inferior Frontal (IF), and Superior Frontal (SF).

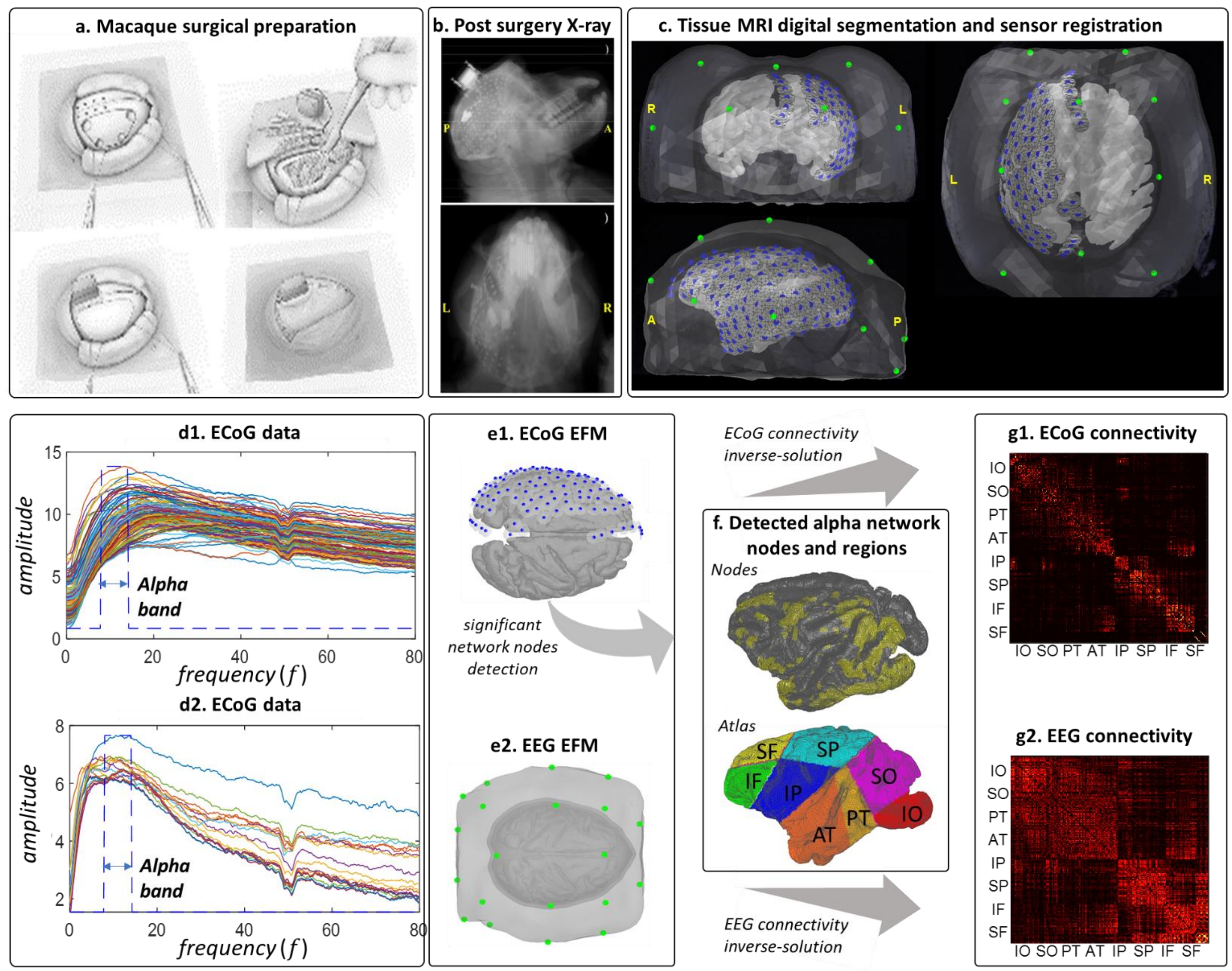

Fig. 5| Confirmation of HIGGS connectivity based in EEG that is recorded simultaneously with ECoG implanted in the macaque. a. Surgical preparation for implantation onto the macaque cortex macaque of a high-density ECoG array embedded in a silicon layer. b. A post-surgical X-ray image showing this implantation. c. Digital preparation based on the macaque MRI segmentation of the cortex, inner skull, outer skull and scalp, and conductance model for this segmentation based on registration with the MRI of the high-density ECoG implantation (128 blue sensors) and low-density EEG layout (20 green sensors). d. Power spectral density of the simultaneous recordings for ECoG (d1) and EEG (d2) highlighting the alpha band within 8-14Hz employed later as data for the identification of network connectivity. e. Electromagnetic Forward Model (EFM) employed to compute the connectivity inverse-solutions from the ECoG (e1), with Lead Field dependent on two conductance layers (cortex and silicon), and EEG (e2), with Lead Field dependent on four conductance layers (cortex, inner skull, outer skull and scalp). ECoG recordings and their Lead Fields provide a more fine-grained reference for confirming connectivity estimators and measures of distortions for the EEG f. Significant cortical subspace based on the ECoG alpha band data to facilitate the comparison in terms of connectivity due to low-density of the EEG. The detected subspace is revealing a large-scale alpha cortical network that extended over the Inferior Occipital (IO), Superior Occipital (SO), Posterior Temporal (PT), Anterior Temporal (AT), Inferior Parietal (IP), Superior Parietal (SP), Inferior Frontal (IF) and Superior Frontal (SF) areas. g) For the subspace covered by this network the HIGGS connectivity inverse-solutions were computed from the ECoG (g1) and EEG (g2).

# 3 Results

### 3.1 Unbiased functional connectivity via hgLASSO inverse solution of the GGS model

As an essential step (Fig. 6), we verify the properties of the MAP1 estimator for the precision matrix $\boldsymbol{\Theta}_{\boldsymbol{u}}(f)$ representing functional connectivity as described by a Gaussian Graphical Spectral (GGS) model of oscillatory brain networks at a generic frequency $f$. See Materials and Methods section 2.1 for the interpretation of such a functional network model and MAP1 inverse solution $\widehat{\boldsymbol{\Theta}}_{\boldsymbol{u}}(f)$ for the functional connectivity. We remind the reader that the algorithm revindicated for this MAP1 inverse solution places the *hermitian*

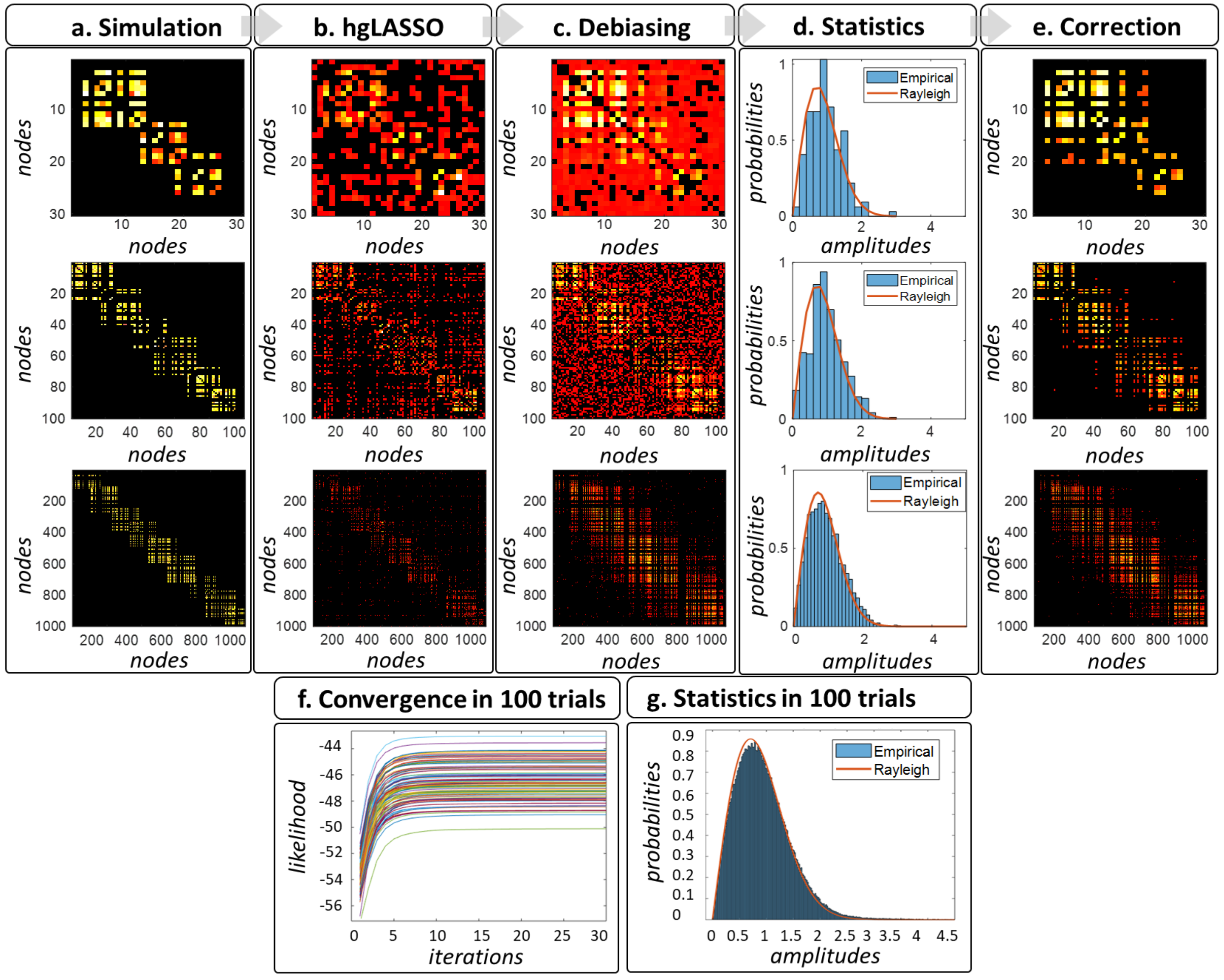

Fig. 6| Experiment to evaluate unbiasedness conditions of the Hermitian Graphical LASSO (hgLASSO) algorithm in different scales at the top: low (first row), high (second row) and ultrahigh (third row). The hgLASSO algorithm performs the MAP1 for the presicion-matrix as described in Materials and Methods sections 2.1 and 2.3. From left to right in hot colormap a. the typical simulated precision-matrix, b. the hgLASSO estimation, c. the debiased estimator, with d. the empirical distribution, compared of the theoretical Rayleigh distribution, and e. the corrections due to thresholds of this distribution. At the bottom for the ultra-high dimensionality the robust convergence pattern in 30 iterations of the hgLASSO likelihood in 100 trials (f) and the seemingly perfect Rayleigh statistics in these 100 trials (g).

*graphical LASSO* (**hgLASSO**) a priori model upon the hermitian graph elements in $\boldsymbol{\Theta}_{\boldsymbol{u}}(f)$, which explains brain-oscillations $\boldsymbol{\iota}(t,f)$ in the GGS model. In Materials and Methods section 2.3, we describe the implementation of the **hgLASSO** algorithm that we demonstrate here for the first time fulfilling stable and scalable unbiasedness conditions. Here, amplitudes $|\boldsymbol{\Theta}_{\boldsymbol{u}}(f)|$ of the hermitian graph elements are employed as the proxy for functional connectivity and to illustrate the algorithmic properties. The **hgLASSO** a priori model is incorporated into both multistep MAP1 and **onestep** MAP2 inverse solutions, as described in Materials and Methods section 2.2 for the *Hidden GGS* (HIGGS) model explaining oscillations $\boldsymbol{v}(t,f)$.

We illustrate the properties of our algorithm with an example at different dimensionalities, 30, 100, and 1000, yielding 900, 10,000, and 1000,000 functional connectivities to be estimated, respectively. Our validation strategy (Fig. 6) is to compare the empirical distribution of the empirical distribution of the unbiased statistic for the estimated precision matrix $\widehat{\boldsymbol{\Theta}}_{\boldsymbol{u}}(f)$ against the Rayleigh distribution for the theoretical $\boldsymbol{\Theta}_{\boldsymbol{u}}(f)$. A Rayleigh distribution is derived from the complex-valued extension of (Janková and van de Geer 2018) for the GGS model described in Materials and Methods section 2.3.

First, we employ a blockwise chained (blocks were mutually dependent) simulated precision-matrix structure $\boldsymbol{\Theta}_{\boldsymbol{u}}(f)$ as in (Tugnait 2019), illustrated here with hot colormaps of a typical trial (Fig. 6 a). These matrices were integrated into a hermitian Gaussian generator generating the sampled covariance matrix from sampled instances (a sample number of 100 times the matrix dimensions). Second, $\widehat{\mathbf{\Theta}}_{\boldsymbol{u}}(f)$, a precision matrix, is estimated with our **hgLASSO** algorithm (Fig. 6 b), and third, from this estimate, the unbiased precision matrix $unb\left(\widehat{\mathbf{\Theta}}_{\boldsymbol{u}}(f)\right)$ (Fig. 6 c) is computed as proposed by the theory. Fourth, we obtain the histogram of a z-statistic $\boldsymbol{Z}(f)$ (Fig. 6 d) for this unbiased estimate at the null hypothesis matrix entries (zero precision-matrix entries $\mathbf{\Theta}_{\boldsymbol{u}}(f;i,j)=0$), and this histogram accurately resembles the theoretical Rayleigh distribution. Computing the z-statistic was performed by scaling the unbiased precision matrix using the theoretical variances $\widehat{\mathbf{\Sigma}}_{\mathbf{\Theta}}(f;i,j)$.
Fifth, we remove the values under the Rayleigh threshold (Rayleigh corrected precision-matrix) obtained from the 0.05 p value of the theoretical Rayleigh distribution. With the Rayleigh threshold, from the originally dense unbiased precision matrix (Fig. 6 c), we obtain an improved corrected result (Fig. 6 e) in comparison to the sparse biased hgLASSO precision matrix (Fig. 6 b). The **hgLASSO** (base of all estimated maps), as we illustrate in the ultrahigh dimensionality, a robust convergence pattern (Fig. 6 f) for multiple repetitions of the experiment (100 trials). The z-statistic histograms for these 100 trials reflect the high coincidence with the theoretical Rayleigh distribution (Fig. 6 g).

### 3.2 Erroneous multistep and exact onestep functional connectivity for the HIGGS model in simulations

The proof of concept for the HIGGS model (Fig. 7) employs random precision-matrices instances $\mathbf{\Theta}_{\boldsymbol{u}}(f)$ similar to the previous results in section 3.1 but yielding binary functional connectivity as defined with the amplitude of hermitan graph-elements $|\mathbf{\Theta}_{\boldsymbol{u}}(f)|$. Our binary "ground-truth" functional connectivity $|\mathbf{\Theta}_{\boldsymbol{u}}(f)|$ (Fig. 7 left-center) is a much more plausible target of estimation errors via classification scores (Fig. 7 right).
Simulations of the HIGGS model producing EEG observations $\boldsymbol{v}(t,f)$ are due to brain oscillations $\boldsymbol{\iota}(t,f)$ steaming from an alpha rhythm peaking at 10 Herts, which follows from an interpretation of the GGS model in terms of second-order stochastic stationary dynamics. See the description of these simulations in Materials and Methods section 2.4. The brain oscillations $\boldsymbol{\iota}(t,f)$ are projected through forward operators $\mathbf{L}_{v\iota}$ that follow two types of EFM geometrical designs: 1) Planar EFM (Fig. 7 left-top), recreating an ideal circumstance where concentric circles recreate the cortical network, head and EEG sensor geometries, and EEG fields lead through homogeneous conductance. 2) Human EFM (Fig. 7 left-bottom), recreating an average human EEG circumstance. Then, the GGS model precision matrix $\mathbf{\Theta}_{\boldsymbol{u}}(f)$ (Fig. 7 left-center), which describes brain oscillations $\boldsymbol{\iota}(t,f)$ at the alpha peak, is the target of the estimation $\widehat{\mathbf{\Theta}}_{\boldsymbol{u}}(f)$ with all types of inverse solutions exposed here (Fig. 7 a1-g1 and Fig. 7 a2-g2). See these inverse solutions in Materials and Methods section 2.2.
First, three **onestep** MAP2 inverse solutions are implemented as successive approximations due to MAP1 inverse solutions within an Expectation-Maximization (EM) loop for 1) brain oscillations $\hat{\boldsymbol{\iota}}(f,t)$ in the EM expectation stage and 2) functional connectivity $\widehat{\mathbf{\Theta}}_{\boldsymbol{u}}(f)$ in the EM maximization stage. The functional connectivity MAP1 (maximization stage) determines a GGS model precision matrix $\widehat{\mathbf{\Theta}}_{\boldsymbol{u}}(f)$ from the expected covariance matrix $\widehat{\mathbf{\Psi}}_{\boldsymbol{u}}(f)$ via the unbiased *Hermitian Graphical LASSO* (**hgLASSO**) algorithm (Fig. 7 a1 and Fig. 7 a2). See this algorithm in Materials and Methods sections 2.1 and 2.3. Here, we also introduce violations to the **hgLASSO** unbiasedness conditions in the functional connectivity MAP1 via the *hermitian graphical Ridge* (**hgRidge**) (Fig. 7 b1 and Fig. 7 b2) and *hermitian graphical Naïve* (**hgNaïve**) (Fig. 7 c1 and Fig. 7 c2).
Second, the **multistep** MAP1 inverse solutions are implemented for 1) brain-oscillations $\hat{\boldsymbol{\iota}}(f,t)$ via **ELORETA** (Fig. 7d1 and Fig. 7d2) (Pascual-Marqui et al. 2006) and **LCMV** (Fig. 7 e1 and Fig. 7 e2) (Van Veen et al. 1997) in the first step and 2) functional connectivity $\widehat{\mathbf{\Theta}}_{\boldsymbol{u}}(f)$ in the second step and from the sampled covariance-matrix $\widehat{\mathbf{\Sigma}}_{\boldsymbol{u}}(f)$ via the unbiased *Hermitian Graphical LASSO* (**hgLASSO**) algorithm. Here, we also introduce a violation to the **hgLASSO** hermitian graph elements in the functional connectivity MAP1 via the approximation of *Graphical LASSO* (**gLASSO**) (J. Friedman, Hastie, and Tibshirani 2008; G. L. Colclough et al. 2015). This violation is incorporated in the second step for either **multistep** MAP1 inverse solutions with the first step via **ELORETA** (Fig. 7f1 and Fig. 7f2) or **LCMV** (Fig. 7g1 and Fig. 7g2). The sampled covariance matrix $\widehat{\mathbf{\Sigma}}_{\boldsymbol{u}}(f)$ is determined from the first-step brain oscillations $\hat{\boldsymbol{\iota}}(f,t)$ defined as a filtered process (G. L. Colclough et al. 2015) and not its Hilbert transform (Wodeyar 2019). To achieve a fair comparison, we employed the optimal regularization parameter for the initial inverse solutions **eLORETA** and **LCMV** selected by generalized cross-validation of the sampled covariance matrix for the data $\widehat{\mathbf{\Sigma}}_{vv}(f)$.
The accuracy of the functional connectivity estimates was assessed by the classification scores of a binary ground true $|\mathbf{\Theta}_{\boldsymbol{u}}(f)|$ (Fig. 7 left-center) by measures of the Receiver Operator Characteristic (ROC) (Fig. 7 right). These measures are given for inverse solutions $\left|\widehat{\mathbf{\Theta}}_{\boldsymbol{u}}(f)\right|$ due to the planar (Fig. 7right-top) and human (Fig. 7right-bottom) EFMs. The measures used in the radar graphs were the global AUC (area under the curve) and partial SENS (sensitivity), SPEC (specificity), PREC (precision), and RECALL (F1 or Fisher scores) measured at the optimal operating point of the ROC curve.

### 3.3 Confirmation of the performance of the functional connectivity for the HIGGS model in macaque EEG/ECoG experiments

We leverage an experimental setup (Fig. 8) offering for macaque 1) large-scale coverage resting-state ECoG observations $\boldsymbol{v}^{ECoG}(t,f)$ onto the left cortical hemisphere and 2) simultaneous EEG observations $\boldsymbol{v}^{EEG}(t,f)$ (Nagasaka, Shimoda, and Fujii 2011). The ECoG observations $\boldsymbol{v}^{ECoG}(t,f)$ are from a high-density subdural sensor array that was placed surgically onto the macaque cortex and provides (Fig. 8 left-top). See the experimental preparation and inverse solutions from ECoG $\widehat{\mathbf{\Theta}}_{\boldsymbol{u}}^{ECoG}(f)$ and EEG $\widehat{\mathbf{\Theta}}_{\boldsymbol{u}}^{EEG}(f)$ in Materials and Methods section 2.4, in this case targeting a functional network underlying the resting-state alpha rhythm.

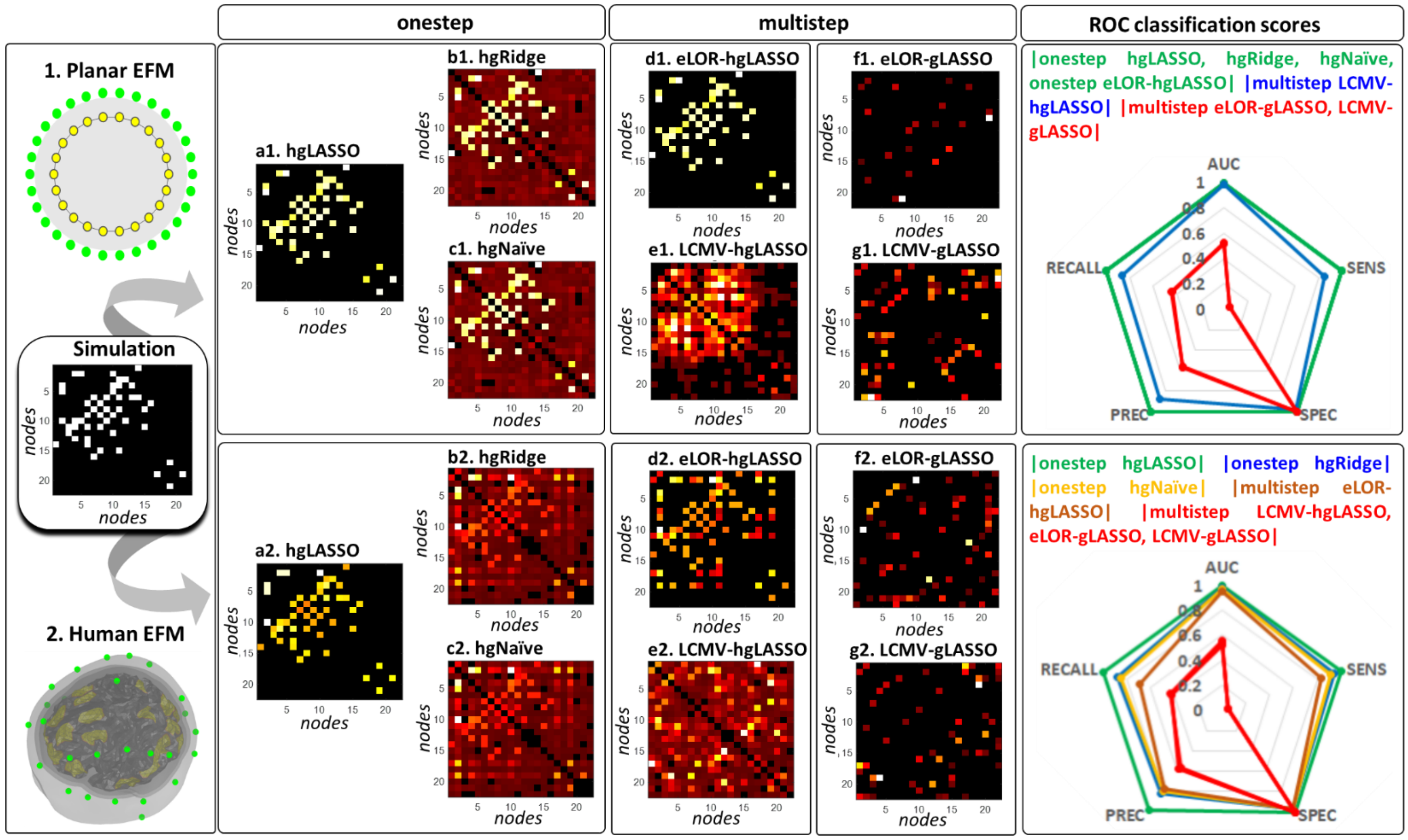

Fig. 7| Simulated experiment of a relatively simple oscillatory network HIGGS model (left-center) producing observations through 1. an idealized "Planar" (left-top) or 2. average "Human" (left-bottom) EFM. Estimation errors regarding a ground-true hermitian precision-matrix with binary amplitudes (left-center) may be judged in the colormaps of the estimation under the Planar (a1-g1) and Human (a2-g2) EFM. These estimation are characteristic of multistep methods (based on first-step ELORETA and LCMV) in an average EEG situation (d2,e2) even employing the hgLASSO unbiased estimator. With same multistep methods the estimation errors of the state-of-the-art gLASSO approximating the hermitian graph-elements in this HIGGS (f1,g1;f2,g2) fall beyond even in ideal circumstances. In 100 trials of a binary connectivity ROC measures show these estimation errors and improvement with our direct solution with hgLASSO (a1,a2) which may not be possible with hgRidge or hgNaïve estimation (b1,c1;b2,c2).

On the one hand, a fine-grained EFM forward-operator $\mathbf{L}_{v\iota}^{ECoG}$ for the ECoG observations provides accurate functional network detection (Fig. 8 left-center), as well as functional connectivity by any **onestep** (Fig. 8a1) or **multistep** (Fig. 8 b1) inverse solutions $\widehat{\boldsymbol{\Theta}}_{\boldsymbol{\iota}}^{ECoG}(f)$. On the other hand, sensor low density and several tissue layers yield a coarse-grained EFM forward operator $\mathbf{L}_{v\iota}^{EEG}$ for the EEG observations (Fig. 8 left-bottom) and therefore as expected incongruent inverse solutions $\widehat{\boldsymbol{\Theta}}_{\boldsymbol{\iota}}^{EEG}(f)$ (Fig. 8 a2 and Fig. 8 b2) regarding an ECoG reference $\widehat{\boldsymbol{\Theta}}_{\boldsymbol{\iota}}^{ECoG}(f)$ (Fig. 8 a1 and Fig. 8 b1) as proposed in (Q. Wang et al. 2019).

We employ the *Kullback–Leibler Divergence* (KLD) (Kullback 1968) as the proxy to measure this incongruence. The KLD distance measures is the relative entropy between pairs of multivariate probability distributions. In this case, the distributions are *Gaussian Graphical Spectral* (GGS) models $p\left(\boldsymbol{\iota}(t,f)|\widehat{\boldsymbol{\Theta}}_{\boldsymbol{\iota}}^{EEG}(f)\right)$ and $p\left(\boldsymbol{\iota}(t,f)|\widehat{\boldsymbol{\Theta}}_{\boldsymbol{\iota}}^{ECoG}(f)\right)$ expressed as a function of the estimated precision matrices $\widehat{\boldsymbol{\Theta}}_{\boldsymbol{\iota}}^{EEG}(f)$ and $\widehat{\boldsymbol{\Theta}}_{\boldsymbol{\iota}}^{ECoG}(f)$ summarizing the GGS properties. A multivariate effect of distortions locally (for each graph element) for EEG $\widehat{\boldsymbol{\Theta}}_{\boldsymbol{\iota}}^{EEG}(f)$ and regarding ECoG $\widehat{\boldsymbol{\Theta}}_{\boldsymbol{\iota}}^{ECoG}(f)$ may be measured by means of the KLD kernel for **onestep** (Fig. 8 a3) and **multistep** (Fig. 8 b3) inverse solutions. For a quantitative analysis of the performance (Fig. 8 right), we report distances based on the KLD metric and others (Riemannian and Log-Euclid).

# 4 Discussion

## 4.1 Theory

The state-of-the-art connectivity inverse solution MEG-ROI-nets *Graphical LASSO* (**gLASSO**) (G. L. Colclough et al. 2015), which is employed here as a baseline here for comparison, was completely limited to only real-valued operations. The flaw is twofold due to instability in large scale and the lack of MEG-ROI-nets **gLASSO** algorithms capable of dealing with models that dwell in the complex-variable space of the Fourier or Hilbert transform under the GGS assumptions (K. J. Friston et al. 2012; Bien and Tibshirani 2011; C.-J. Hsieh et al. 2011; C. Hsieh 2014; Tugnait 2019; Schreier and Scharf 2010; Andersen et al. 1995; Bollhöfer et al. 2019; C.-J. Hsieh et al. 2013, 2012). It was therefore our task to provide with a solution to estimation errors produced by such algorithms as discussed in *Introduction section 1.3*.

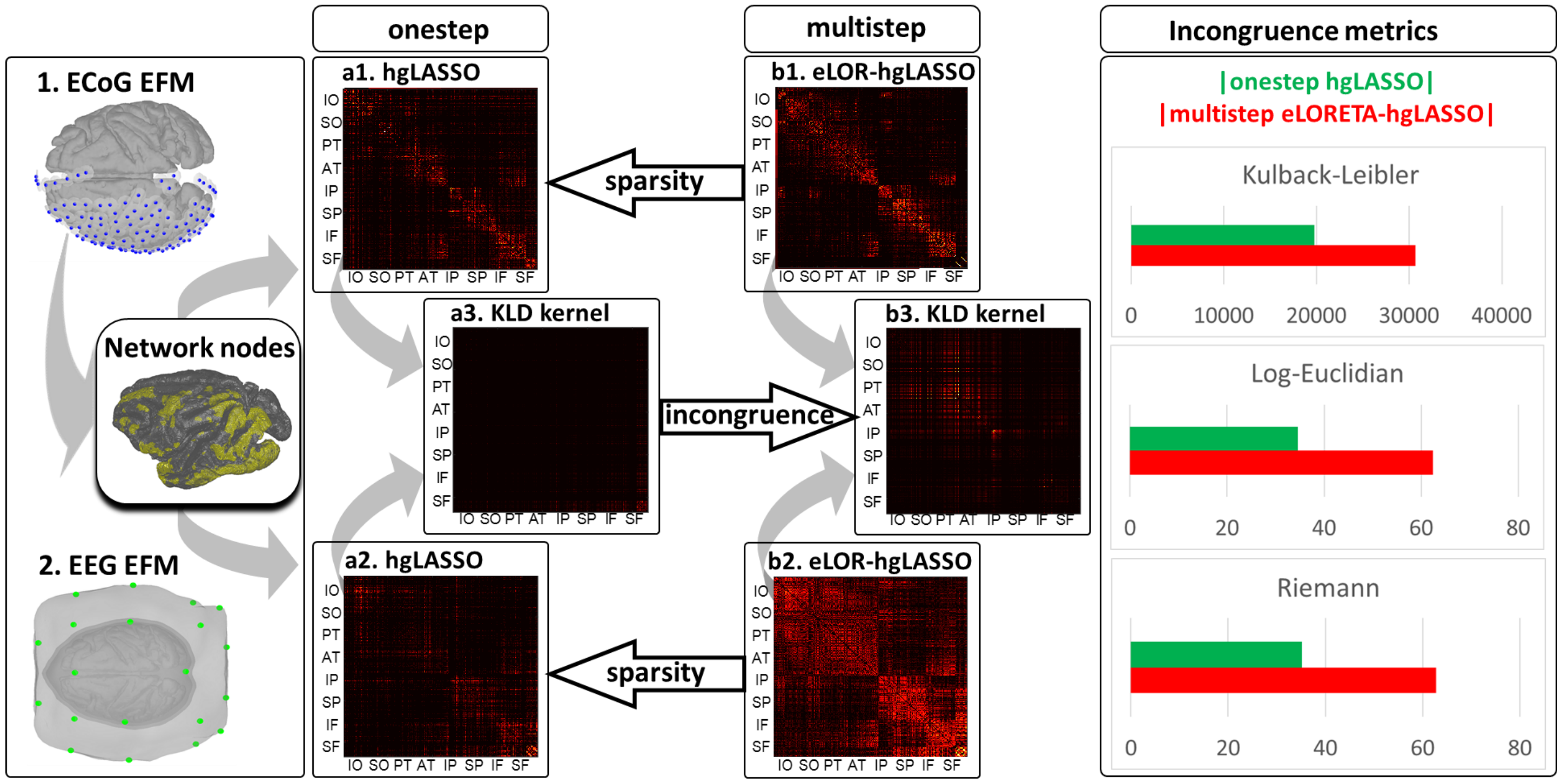

Fig. 8| Experimental confirmation of performance in macaque simultaneous EEG/ECoG recordings (left). A large-scale alpha oscillatory network (left-center) scrrened from ECoG with certain method is the target of HIGGS inverse-solutions. Incongrunce base on comparing these solutions under the ECOG (left-top) and EEG (left-bottom) EFM. Employing as reference the fine-grained ECOG solutions (a1,b1), incongrunce of the EEG solutions are measured with Kulback-Leibler, LogEuclidean and Remaniean metrics (right) confirming the performance in Fig. 2. This incongruence due to leakage and localization errors is represented with the Kulback-Leibler Divergence (KLD) kernel "EEG/ECoG + EEG\ECoG – 2I" (a3,b3), for EEG and ECoG precision-matrices.

A **gLASSO** inverse solution based upon a real-valued sampled covariance $\widehat{\boldsymbol{\Sigma}}_{\boldsymbol{u}}(f)$ for $\boldsymbol{\iota}(t,f)$ defined as the filtered process for $\boldsymbol{\iota}(t)$ or sometimes the Hilbert envelope $-log|\boldsymbol{\iota}(t,f)|$ is claimed to encode in $\widehat{\boldsymbol{\Theta}}_{\boldsymbol{u}}(f)$ all connectivity information in state-of-the-art neuroimaging practice but only does it partially. Here, we employed the MEG-ROI-nets **gLASSO** (G. L. Colclough et al. 2015) upon the real-valued sampled covariance matrix $\widehat{\boldsymbol{\Sigma}}_{\boldsymbol{u}}(f)$ for the filtered process $\boldsymbol{\iota}(t,f)$ which does not uphold the *Gaussian Graphical Spectral* (GGS) model assumptions but is a less gross approximation than Hilbert envelope $-log|\boldsymbol{\iota}(t,f)|$ (Matthew J. Brookes, Hale, et al. 2011; Matthew J. Brookes, Woolrich, et al. 2011b; Giles L Colclough et al. 2015; Vidaurre, Abeysuriya, et al. 2018; Vidaurre, Hunt, et al. 2018; Tewarie et al. 2019; Hillebrand et al. 2012; Matthew J. Brookes et al. 2007; Larson-Prior et al. 2013; Hipp et al. 2012).

For oscillatory network phenomena estimation of connectivity $\widehat{\boldsymbol{\Theta}}_{\boldsymbol{u}}(f)$ required both amplitude and phase information that was determined with the *Hermitian Graphical LASSO* (**hgLASSO**) (Nolte et al. 2020). We remind the reader a **hgLASSO** inverse solution introduced in *Materials and Methods section 2.1 and section 2.3* is only resource to find estimates $\widehat{\boldsymbol{\Theta}}_{\boldsymbol{u}}(f)$ for the complex-valued precision matrix of such GGS model. At every frequency, the complex-valued sampled covariance matrix $\widehat{\boldsymbol{\Sigma}}_{\boldsymbol{u}}(f)$ for $\boldsymbol{\iota}(t,f)$ defined as the Hilbert transform for $\boldsymbol{\iota}(t)$ conforms to the so-called cross-spectrum. Only this quantity, or equivalently its matrix inverse or precision-matrix $\widehat{\boldsymbol{\Theta}}_{\boldsymbol{u}}(f)$, fully encodes the dynamical properties of oscillatory components $\boldsymbol{\iota}(t,f)$ in activity $\boldsymbol{\iota}(t)$.

Our **hgLASSO** algorithm showed consistency in simulations to the theoretical Rayleigh tendency of amplitudes for the hermitian graph-elements in complex-valued *Gaussian Graphical Spectral* (GGS) models (Fig. 6). Such Rayleigh tendency in hermitian graph-elements (Andersen et al. 1995; Tugnait 2019; Schreier and Scharf 2010) was a natural extension of the Chi-Square tendency in the real-valued case (J. Friedman, Hastie, and Tibshirani 2007; Janková and van de Geer 2018). The complexity of the proposed algorithm was bounded by the hermitian matrix eigen decomposition and we shall note no **gLASSO** algorithm was available which fulfills statistical guarantees, performs stably in the large dimensionality, and deals with a complex-valued cross-spectrum. Our **hgLASSO** is therefore an optimal procedure amongst state-of-the-art algorithms for this type of graphical model either in real or complex-variable (W. Liu and Luo 2015; C. Hsieh 2014; Zhang and Zou 2014; N. Friedman and Koller 2003; Bien and Tibshirani 2011; Qiao, Guo, and James 2019; Janková and van de Geer 2018; Andersen et al. 1995; J. Friedman, Hastie, and Tibshirani 2008; Tugnait 2019; K. J. Friston et al. 2012; G. L. Colclough et al. 2015).

Estimation of the *Hidden GGS* (HIGGS) connectivity $\widehat{\boldsymbol{\Theta}}_{\boldsymbol{u}}(f)$ in simulations (Fig. 7) and macaque real data (Fig. 8) as in typical real life situations assumed prescreened network nodes by some method determining the statistically significant cortical regions or nodes for a given resting state or task activity transient $\boldsymbol{\iota}(t,f)$ (Matthew J. Brookes, Woolrich, et al. 2011a; M. J. Brookes, Woolrich, and Barnes 2012; Smith et al. 2013; G. L. Colclough et al. 2015). Implementing **onestep** and **multistep** connectivity estimates $\widehat{\boldsymbol{\Theta}}_{\boldsymbol{u}}(f)$ in the subspace of prescreened network nodes ruled out other sources of distortions and allowed us to evaluate in isolation the estimation errors exposed in *Introduction section 1.2*. These estimation errors are caused in theory by misspecification of the HIGGS model in inverse solution targeting connectivity estimates $\widehat{\boldsymbol{\Theta}}_{\boldsymbol{u}}(f)$. Other sources of distortion were the localization errors and leakage in activity estimates $\hat{\boldsymbol{\iota}}(t,f)$ that would be produced by **multistep** (in the first step) or **onestep** (successive approximations) upon the whole cortical space (G. L. Colclough et al. 2015; Palva et al. 2018; M. J. Brookes, Woolrich, and Barnes 2012; Wens et al. 2015).
Inverse solutions for HIGGS connectivity $\widehat{\boldsymbol{\Theta}}_{\boldsymbol{u}}(f)$ were implemented via the **hgLASSO**, for both **onestep**, in successive approximations in an iterative fashion, and **multistep**, in a first step, estimating the cross-spectrum $\widehat{\boldsymbol{\Sigma}}_{\boldsymbol{u}}(f)$ as described in *Materials and Methods section 2.2*. Our **onestep** connectivity inverse solution was an extension to the *Expectation Maximization* (EM) algorithm as with the identification of the covariance matrix $\widehat{\boldsymbol{\Sigma}}_{\boldsymbol{u}}(f)$ in a *Covariance Component Model* (CCM) (C. Liu and Rubin 1994; Galka, Yamashita, and Ozaki 2004; K. Friston et al. 2006, 2007) or identification of the autoregressive-coefficients $\widehat{\mathbf{K}}_{\boldsymbol{u}}(f)$ in a *State Space Model* (SSM) (Jorge J. Riera et al. 2004; Faes, Stramaglia, and Marinazzo, n.d.; Galka et al. 2004). Implementation of the **hgLASSO** algorithm defined the flavor of the HIGGS model **onestep** identification connectivity defined as the complex-valued precision matrix $\widehat{\boldsymbol{\Theta}}_{\boldsymbol{u}}(f)$. HIGGS obtaining an unbiased, scalable and stable inverse solution for a locally optima GGS within EM iterations (McLachlan and Krishnan 2007) was not considered by previous CCM or SSM identification approaches.

**4.2 Validation**

We spoused our validation of estimation errors in HIGGS inverse solutions in *electroencephalogram* (EEG) in simulations (Fig. 7) and macaque real data (Fig. 8), since this is the "worst case scenario" compared to *magnetoencephalogram* (MEG) or *electrocorticogram* (ECoG). EEG is exquisitely sensitive to conductivity heterogeneity of head tissue layers, which produces quite large distortions of the electric lead fields $\mathbf{L}_{v\iota}$ (Jörge Javier Riera and Fuentes 1998; Baillet, Mosher, and Leahy 2001; Grech et al. 2008). Additionally, the high conductivity of the scalp leads to blurring of the EEG potential, which aggravates the estimation errors of any activity estimates $\hat{\boldsymbol{\iota}}(t,f)$ or its cross-spectrum $\widehat{\boldsymbol{\Sigma}}_{\boldsymbol{u}}(f)$, that are transferred to connectivity estimates $\widehat{\boldsymbol{\Theta}}_{\boldsymbol{u}}(f)$ (Van De Steen, Valdes-Sosa, and Marinazzo 2016). Thus, EEG results for connectivity $\widehat{\boldsymbol{\Theta}}_{\boldsymbol{u}}(f)$ presented here can be considered a lower bound for those of MEG or ECoG.
The methods under evaluation fell into three different modalities: I) the **onestep** successive approximations, employing our **hgLASSO** unbiased estimator for $\widehat{\boldsymbol{\Theta}}_{\boldsymbol{u}}(f)$ (Fig. 7 a1, a1), and employing estimators hgRidge (Fig. 7 b1, b2) and hgNaïve (Fig. 7 c1, c2) which violated this unbiasedness condition in $\widehat{\boldsymbol{\Theta}}_{\boldsymbol{u}}(f)$. II) the **multistep** methods with **eLORETA** (Pascual-Marqui et al. 2006) (Fig. 7 d1, d2) and **LCMV** (Van Veen et al. 1997) (Fig. 7 e1, e2), both employing the same **hgLASSO** unbiased estimator for $\widehat{\boldsymbol{\Theta}}_{\boldsymbol{u}}(f)$. It has been argued (Wodeyar 2019) that these types of **multistep** estimators reduce connectivity estimation errors. III) the same **multistep** methods with **eLORETA** (Fig. 7 f1, f2) and **LCMV** (Fig. 7 g1, g2) but employing MEG-ROI-nets **gLASSO** estimator (J. Friedman, Hastie, and Tibshirani 2008) for the real-valued $\widehat{\boldsymbol{\Theta}}_{\boldsymbol{u}}(f)$ implemented in the MEG-ROI-nets package (G. L. Colclough et al. 2015) which does not uphold the GGS assumptions.
Even with the relatively simple ground truth some methods produced large estimation errors in these simulations (Fig. 7) indistinctly, employing I) an idealized or planar (Fig. 7 left top) and II) a human (Fig. 7 left bottom) *Electromagnetic Forward Model* (EFM). In simulations (Fig. 7 right) partial classification measures RECALL, PREC, and SENS quantify and confirm the essential flaw with **multistep** methods employing II) **hgLASSO** or III) **gLASSO**. The perfect **multistep** estimation in the planar EFM (Fig. 7 a1), which benefited from the renowned statistical goodness of eLORETA and our unbiased **hgLASSO** algorithm, was undone in the human EFM (Fig. 7 a2). Another **multistep** solution based on LCMV (Fig. 7 e1, e2), with excellent performance anywhere else, does not reach expectations even in the planar EFM ideal circumstances (Fig. 7 e1). Only the **onestep hgLASSO** sparse inverse solution showed no major estimation errors (Fig. 7 a1, a2) and outscores any other non-sparse **onestep** (Fig. 7 b1, b2, c1, c2) or sparse **multistep** (Fig. 7 d1, d2, e1, e2) solutions in both planar (right-top) and human (right-bottom) models.
Connectivity leakage measured as 20% in PREC or RECALL (shown as colormap false positives) is a systematic property of these **multistep** methods appearing even in a relatively simple human network model. The situation is even worse when using the **gLASSO** approximation for these methods, which corrects leakage in either simulation model (Fig. 7 f1, f2, g1, g2) as claimed before (Wodeyar 2019; G. L. Colclough et al. 2015) but produces extreme estimation errors, yielding random classification according to all classification measures. **gLASSO** fails under the specific assumption of a GGS model, which bases connectivity on hermitian graph elements. Previous simulation studies validating this correction method did not corroborate them under the GGS assumption and were limited to even simpler network models (up to 5 nodes) (G. L. Colclough et al. 2015).
The qualitatively sparse pattern (Fig. 7) obtained in simulations via the **onestep** (Fig. 7 a1, a2) and **multistep** eLORETA (Fig. 7 d1, d2), with both methods employing the unbiased **hgLASSO** inverse solution, resembled the sparsity simulated in the ground truth (Fig. 7 left center) and minimized connectivity estimation errors. Sparsity renders factual biological circumstances such as efficiency in large-scale network phenomena (Bullmore and Sporns 2009), as we illustrated with the macaque ECoG (Fig. 8) with the sparse alpha network (Fig. 8 left center) and sparse connectivities (Fig. 8 a1, b1) obtained from any (**onestep** or **multistep**) method employing the unbiased **hgLASSO** estimator.
Remarkably, the dimensionality of the macaque ECoG alpha network (Fig. 8 left-center) ruled out the possibility of using current implementations of the **gLASSO** algorithm implemented in (J. Friedman, Hastie, and Tibshirani 2008; C. Hsieh 2014; C.-J. Hsieh et al. 2013), or by MEG-ROI-nets (G. L. Colclough et al. 2015). Confirmation in macaque shows the connectivity determined for this type of network from the ECoG (Fig. 8 left-top) and EEG (Fig. 8 left-bottom) by means of the **onestep hgLASSO** (Fig. 8 a1, a2) and **multistep eLORETA hgLASSO** (Fig. 8 b1, b2).
The macaque experiment (Fig. 8) confirms the benefit in our **onestep hgLASSO** inverse solution, targeting a severely ill-posed inverse problem in recovering connectivities for a large-scale alpha network (Fig. 8 left-center) based only on EEG 19-sensor observations (Fig. 8 left-bottom). Indeed, for both the **onestep** and **multistep** methods, there are striking similarities in the ECoG (Fig. 8 a1, b2) and EEG (Fig. 8 a2, b2) solutions. Our **onestep** hgLASSO solutions (Fig. 8 a1, a2) maximize the sparsity. With the Kullback–Leibler divergence (KLD) kernel (Fig. 8 a3, b3), we illustrate contraposition between these sparsity and distortion levels, measured as multivariate relative entropy for GGS models. This measure represents divergence (incongruence) from the identity matrix for the product between the EEG precision matrix and ECoG covariance that would express a multivariate effect of distortions locally (for each graph element). The norm for this matrix, the KLD metric and others (Reimanian and Log-Euclid) shows a 1/3 congruence improvement.

## 5 Acknowledgments

This work was supported in part by the National Nature and Science Foundation of China (NSFC) under founding No. 61871105 and the CNS program of UESTC under founding No. Y0301902610100201.

## 6 Supplementary Information

### I. Notation for variables and mathematical operations

| | | |
|---|---|---|
| (I-1) | $x$, or $\boldsymbol{x}$, or $\mathbf{X}$, or $\mathbb{X}$ | Respectively denote a scalar $x$ (italic lowercase), or vector $\boldsymbol{x}$ (bold italic lowercase), or matrix $\mathbf{X}$ (bold capital), or set $\mathbb{X}$ (double struck capital). |
| (I-2) | $\boldsymbol{x}(i)$, or $\mathbf{X}(i,j)$ | For a vector $\boldsymbol{x}$ or matrix $\mathbf{X}$ (as defined in I-1) the $i$-th, or $i,j$-th element. |
| (I-3) | $x(t)$, or $\boldsymbol{x}(t)$, or $\mathbf{X}(t)$, and $x(t,f)$, or $\boldsymbol{x}(t,f)$, or $\mathbf{X}(t,f)$ | A scalar $x$ function, or vector $\boldsymbol{x}$ function, or matrix $\mathbf{X}$ function (as defined in I-1) with argument $(t)$, and when these are a function of a second argument $(f)$. |
| (I-4) | $\boldsymbol{x}(i,t)$, or $\mathbf{X}(i,j,t)$, and $\boldsymbol{x}(i,t,f)$, or $\mathbf{X}(i,j,t,f)$ | The $i$-th, or $i,j$-th element (as defined in I-2) for (as defined in I-3) a vector $\boldsymbol{x}(t)$ function, or matrix function $\mathbf{X}(t)$ with argument $(t)$, and a vector function $\boldsymbol{x}(t,f)$, or matrix function $\mathbf{X}(t,f)$. |
| (I-5) | $\lvert\mathbf{X}\rvert$ | Determinant of (as defined in I-1) the matrix $\mathbf{X}$. |
| (I-6) | $tr(\mathbf{X})$ | Trace of (as defined in I-1) the matrix $\mathbf{X}$. |
| (I-7) | $\mathbf{X}^{-1}$ | Inverse of (as defined in I-1) the matrix $\mathbf{X}$. |
| (I-8) | $\boldsymbol{x}^{\top}$, or $\mathbf{X}^{\top}$ | Transpose of (as defined in I-1) the vector $\boldsymbol{x}$, or matrix $\mathbf{X}$. |
| (I-9) | $\boldsymbol{x}^{\dagger}$, or $\mathbf{X}^{\dagger}$ | Conjugate transpose of (as defined in I-1) the complex-valued vector $\boldsymbol{x}$, or matrix $\mathbf{X}$. |
| (I-10) | $p(x)$, or $p(\boldsymbol{x})$, or $p(\mathbf{X})$ | Types of probability density functions for (as defined in I-1) a random scalar $x$, or random vector $\boldsymbol{x}$, or random matrix $\mathbf{X}$. |
| (I-11) | $p(x,y)$, or $p(\boldsymbol{x},\boldsymbol{y})$, or $p(\mathbf{X},\mathbf{Y})$, and $p(x,\boldsymbol{y})$, or $p(x,\mathbf{Y})$, or $p(\boldsymbol{x},\mathbf{Y})$ | Types of joint probability density functions for random scalars $x,y$, or random vectors $\boldsymbol{x},\boldsymbol{y}$, or random matrices $\mathbf{X},\mathbf{Y}$ (as defined in I-1 and I-10) as well as their possible combinations. |
| (I-12) | $p(x\vert y)$, or $p(\boldsymbol{x}\vert\boldsymbol{y})$, or $p(\boldsymbol{x}\vert\mathbf{Y})$, and $p(\mathbf{X}\vert y)$, or $p(\mathbf{X}\vert\mathbf{Y})$ | Types of conditional probability density functions for random scalars $x,y$, or random vectors $\boldsymbol{x},\boldsymbol{y}$, or random matrices $\mathbf{X},\mathbf{Y}$ (as defined in I-1, I-10 and I-11) as well as their possible combinations. |
| (I-13) | $N(\boldsymbol{x}\vert\boldsymbol{\mu}_{\boldsymbol{x}},\boldsymbol{\Sigma}_{\boldsymbol{x}})$, or $N(\boldsymbol{x}\vert\boldsymbol{\mu}_{\boldsymbol{x}},\boldsymbol{\Theta}_{\boldsymbol{x}}^{-1})$ | Real-valued Gaussian distribution for a vector $\boldsymbol{x}$ conditioned to (as defined in I-12) the mean $\boldsymbol{\mu}$ and real valued symmetric ensemble covariance-matrix $\boldsymbol{\Sigma}_{\boldsymbol{x}}$ or precision-matrix $\boldsymbol{\Theta}_{\boldsymbol{x}}$ (as defined in I-12). |
| (I-14) | $N^{\mathbb{C}}(\boldsymbol{x}\vert\boldsymbol{\mu}_{\boldsymbol{x}},\boldsymbol{\Sigma}_{\boldsymbol{x}})$, or $N^{\mathbb{C}}(\boldsymbol{x}\vert\boldsymbol{\mu}_{\boldsymbol{x}},\boldsymbol{\Theta}_{\boldsymbol{x}}^{-1})$ | Hermitian (complex-valued circularly symmetric) Gaussian distribution for a vector $\boldsymbol{x}$ with conditioned to (as defined in I-12) mean $\boldsymbol{\mu}_{\boldsymbol{x}}$ and complex valued hermitian ensemble covariance-matrix $\boldsymbol{\Sigma}_{\boldsymbol{x}}$ or precision-matrix $\boldsymbol{\Theta}_{\boldsymbol{x}}$. |
| (I-15) | $exp(x\vert\alpha)$ | Exponential distribution for the scalar $x$ conditioned to (as defined in I-12) the scalar parameter of shape $\alpha$. |
| (I-16) | $Ga(x\vert\rho,\alpha)$ | Gamma distribution for the scalar $x$ conditioned (as defined in I-12) to the scalar parameters of shape $\rho$ and rate $\alpha$. |
| (I-17) | $\hat{x}$, or $\hat{\boldsymbol{x}}$, or $\hat{\mathbf{X}}$, and, $\check{x}$, or $\check{\boldsymbol{x}}$, or $\check{\mathbf{X}}$ | Estimator (hat) for (as defined in I-1) a scalar $x$, or vector $\boldsymbol{x}$, or matrix $\mathbf{X}$, also when these estimators denote a class of auxiliarly quantities (inverted hat). |
| (I-18) | $\hat{x}^{(k)}$, or $\hat{\boldsymbol{x}}^{(k)}$, or $\hat{\mathbf{X}}^{(k)}$, and $\check{x}^{(k)}$, or $\check{\boldsymbol{x}}^{(k)}$, $\check{\mathbf{X}}^{(k)}$ | The estimators (as defined in 4-14) given in the $k$-th iteration within a loop. |

| (I-19) | $\sum_{t\in\mathbb{T}}$ , $\sum_{t=1}^{T}$ | Sum operator along index $t$, valid for $i, j, t, f$ in matrices or vectors (as defined in 4.4) that can be defined within a countable set $\mathbb{T}$ (as defined in 4-1) or from 1 up to a maximum $T$. |
|---|---|---|
| (I-20) | $\prod_{t\in\mathbb{T}}$ , $\prod_{t=1}^{T}$ | Product operator along index $t$, valid for $i, j, t, f$ in matrices or vectors (as defined in 4.4) that can be defined within a countable set $\mathbb{T}$ (as defined in 4-1) or from 1 up to a maximum $T$. |
| (I-21) | $\|\boldsymbol{x}\|_n, n = 1,2$, *or* $\|\mathbf{X}\|_n, n = 1,2$ | L1 or L2 norms of the vector $\boldsymbol{x}$ or matrix $\mathbf{X}$. |
| (I-22) | $\mathbf{I}$, $\mathbf{1}$, $\mathbf{0}$ | Respectively Identity, Ones and Ceros matrices. |
| (I-23) | $\odot, \oslash$ | Elementwise (Hadamard) product a division operators for matrices or vectors. |
| (I-24) | $argmin_{\mathbf{X}}\{f(\mathbf{X})\}$ or $argmax_{\mathbf{X}}\{f(\mathbf{X})\}$ | Extreme values of the scalar function $f$, correspondingly minimum or maximum, in the matrix argument $\mathbf{X}$ , valid for a scalar $x$ or a vector $\boldsymbol{x}$. |
| (I-25) | $zeros_{\mathbf{X}}\{f(\mathbf{X})\}$ | Zeros of the scalar function $f$ in the matrix argument $\mathbf{X}$, valid for a scalar $x$ or a vector $\boldsymbol{x}$. |

## II. Nomenclature for theoretical quantities

| (II-1) | $\mathbb{E}$, *with* $\mathrm{E} = \|\mathbb{E}\|$ *and* $e \in \mathbb{E}$ | Physical space of MEG/EEG/ECoG observations $\mathbb{E}$, with size $\mathrm{E}$, sensors in this space are defined as $e \in \mathbb{E}$. |
|---|---|---|
| (II-2) | $\mathbb{G}$, *with* $\mathrm{G} = \|\mathbb{G}\|$ *and* $g \in \mathbb{G}$ | Physical space of local neural currents or cortical generators of the MEG/EEG/ECoG $\mathbb{G}$, with size $\mathrm{G}$, cortical generators in this space are defined as $g \in \mathbb{G}$. |
| (II-3) | $\mathbb{T}$, *with* $\mathrm{T} = \|\mathbb{T}\|$ *and* $t \in \mathbb{T}$ | Domain of time for processes, local neural currents or MEG/EEG/ECoG observations and their perturbations, $\mathbb{T}$ with size $\mathrm{T}$, instances in this space are defined as $t \in \mathbb{T}$. |
| (II-4) | $\mathbb{F}$, *with* $\mathrm{F} = \|\mathbb{F}\|$ *and* $f \in \mathbb{F}$ | Domain of frequency for spectral processes, spectral local neural currents or spectral MEG/EEG/ECoG observations and their perturbations, $\mathbb{F}$ with size $\mathrm{F}$, instances in this space are defined as $f \in \mathbb{F}$. |
| (II-5) | $\boldsymbol{v}(t)$, *and* $\boldsymbol{v}(t,f)$ | MEG/EEG/ECoG observations $\boldsymbol{v}(t)$, in time $\mathbb{T}$ (as defined in II-3), and its Hilbert transform, MEG/EEG/ECoG spectral observations or MEG/EEG/ECoG oscillations $\boldsymbol{v}(t,f)$, in time $\mathbb{T}$ (as defined in II-3) and frequency $\mathbb{F}$ (as defined in II-4). Either vectors are defined in the space $\mathbb{E}$ (as defined in 5-1). |
| (II-6) | $\boldsymbol{\xi}(t)$, *and* $\boldsymbol{\xi}(t,f)$ | For the MEG/EEG/ECoG observations $\boldsymbol{v}(t)$ (as defined in II-5) an instrumental noise process $\boldsymbol{\xi}(t)$, and its Hilbert transform or spectral instrumental noise $\boldsymbol{\xi}(t,f)$. |
| (II-7) | $\boldsymbol{\iota}(t)$, *and* $\boldsymbol{\iota}(t,f)$ | Local neural currents or cortical activity $\boldsymbol{\iota}(t)$, in time $\mathbb{T}$ (as defined in II-3), and its Hilbert transform, cortical spectral activity or cortical oscillations $\boldsymbol{\iota}(t,f)$, in time $\mathbb{T}$ (as defined in II-3) and frequency $\mathbb{F}$ (as defined in II-4). Either vectors are defined in the space $\mathbb{E}$ (as defined in II-1). |
| (II-8) | $\boldsymbol{\zeta}(t)$, *and* $\boldsymbol{\zeta}(t,f)$ | For the cortical activity $\boldsymbol{\iota}(t)$ (as defined in II-7) a biological noise process $\boldsymbol{\zeta}(t)$, and its Hilbert transform or spectral biological noise $\boldsymbol{\zeta}(t,f)$. |
| (II-9) | $\mathbf{L}$, *with* $\boldsymbol{v}(t) = \mathbf{L}\boldsymbol{\iota}(t) + \boldsymbol{\xi}(t)$, *or* $\boldsymbol{v}(t,f) = \mathbf{L}\boldsymbol{\iota}(t,f) + \boldsymbol{\xi}(t,f)$ | Lead Field of the Electromagnetic Forward Model for MEG/EEG/ECoG $\mathbf{L}$, an $\mathrm{E} \times \mathrm{G}$-size matrix in the space product $\mathbb{E} \times \mathbb{G}$ (as defined in II-1 and II-2). With the observations $\boldsymbol{v}(t)$ or spectral observations $\boldsymbol{v}(t,f)$ (as defined in II-5) produced from cortical activity $\boldsymbol{\iota}(t)$ or spectral cortical activity $\boldsymbol{\iota}(t,f)$ (as defined in II-7). |
| (II-10) | $\mathbf{T}_{\xi v}$, *with* $\hat{\boldsymbol{\xi}}(t,f) = \mathbf{T}_{\xi v}\boldsymbol{v}(t,f)$ | Inverse operator $\mathbf{T}_{\xi v}$ estimating (hat) the spectral instrumental noise $\hat{\boldsymbol{\xi}}(t,f)$ (as defined in 5-6) from MEG/EEG/ECoG spectral observations $\boldsymbol{v}(t,f)$ (as defined in 5-5), $\mathrm{E} \times \mathrm{E}$-size matrix in the space product $\mathbb{E} \times \mathbb{E}$ (as defined in 5-1). |
| (II-11) | $\mathbf{T}_{\iota v}$, *with* $\hat{\boldsymbol{\iota}}(t,f) = \mathbf{T}_{\iota v}\boldsymbol{v}(t,f)$ | Inverse operator $\mathbf{T}_{\iota v}$ estimating (hat) the spectral cortical activity or cortical oscillations $\hat{\boldsymbol{\iota}}(t,f)$ (as defined in 5-7) from MEG/EEG/ECoG spectral observations $\boldsymbol{v}(t,f)$ (as defined in 5-5), $\mathrm{G} \times \mathrm{E}$-size matrix in the space product $\mathbb{G} \times \mathbb{E}$ (as defined in 5-1 and 5-2). |
| (II-12) | $\boldsymbol{\Sigma}_{v}(f)$, *and* $\boldsymbol{\Sigma}_{\xi}(f)$, *and* $\boldsymbol{\Sigma}_{\iota}(f)$, *and* $\boldsymbol{\Sigma}_{\zeta}(f)$ | Ensemble covariance-matrix $\boldsymbol{\Sigma}$ for a process at a given frequency $f \in \mathbb{F}$ and for all time instances $t \in \mathbb{T}$. Respectively for the MEG/EEG/ECoG spectral observations $\boldsymbol{v}(t,f)$ (as defined in 5-5), the spectral instrumental noise $\boldsymbol{\xi}(t,f)$ (as defined in 5-6), the spectral |
| | | activity $\boldsymbol{\iota}(t,f)$ (as defined in 5-7), and the spectral biological noise $\boldsymbol{\zeta}(t,f)$ (as defined in 5-8). |
| (II-13) | $\widehat{\mathbf{S}}_{v}(f)$, *and* $\widehat{\boldsymbol{\Sigma}}_{\xi}(f)$ T, *and* $\widehat{\boldsymbol{\Sigma}}_{\boldsymbol{\iota}}(f)$, *and* $\widehat{\boldsymbol{\Sigma}}_{\zeta}(f)$ | Sampled covariance-matrix $\widehat{\boldsymbol{\Sigma}}$ for a process at a given frequency $f \in \mathbb{F}$ and for all time instances $t \in \mathbb{T}$. Respectively for the MEG/EEG/ECoG spectral observations $\boldsymbol{v}(t,f)$ (as defined in 5-5), the spectral instrumental noise $\boldsymbol{\xi}(t,f)$ (as defined in 5-6), the spectral activity $\boldsymbol{\iota}(t,f)$ (as defined in 5-7), and the spectral biological noise $\boldsymbol{\zeta}(t,f)$ (as defined in 5-8). |
| (II-14) | $\overline{\boldsymbol{\Sigma}}_{\xi}(f)$, *and* $\overline{\boldsymbol{\Sigma}}_{\boldsymbol{\iota}}(f)$ | First type maximum a posteriori estimator (bar $\overline{\boldsymbol{\Sigma}}$) of a covariance-matrix $\boldsymbol{\Sigma}$ (as defined in 5-12), and which if dependent of the second type maximum a posteriori estimators. Respectively, for the spectral instrumental noise $\boldsymbol{\xi}(t,f)$ (as defined in 5-6), spectral activity $\boldsymbol{\iota}(t,f)$ (as defined in 5-7). |
| (II-15) | $\overline{\boldsymbol{\Sigma}}_{\xi}(f)$, *and* $\overline{\boldsymbol{\Sigma}}_{\boldsymbol{\iota}}(f)$ | First type maximum a posteriori estimator (bar $\overline{\boldsymbol{\Sigma}}$) of a covariance-matrix $\boldsymbol{\Sigma}$ (as defined in 5-12), and which if dependent of the second type maximum a posteriori estimators. Respectively, for the spectral instrumental noise $\boldsymbol{\xi}(t,f)$ (as defined in 5-6), spectral activity $\boldsymbol{\iota}(t,f)$ (as defined in 5-7). |
| (II-16) | $\check{\mathbf{Z}}_{\xi\xi}(f)$, *and* $\check{\boldsymbol{\Psi}}_{\boldsymbol{\iota\iota}}(f)$ | Auxiliarly estimator (inverted hat $\check{\boldsymbol{\Psi}}$) Second type maximum a posteriori effective estimator of the covariance-matrices (as defined in 5-12) for the spectral instrumental noise $\boldsymbol{\xi}(t,f)$ (as defined in 5-6), spectral activity $\boldsymbol{\iota}(t,f)$ (as defined in 5-7). This estimation depends on the sampled estimator (as defined in 5-13) and a posteriori estimator (as defined in 5-14). |
| (II-17) | $\boldsymbol{\Theta}_{vv}(f)$, *and* $\boldsymbol{\Theta}_{\xi\xi}(f)$, *and* $\boldsymbol{\Theta}_{\boldsymbol{\iota\iota}}(f)$, *and* $\boldsymbol{\Theta}_{\zeta\zeta}(f)$ | Ensemble precision-matrices $\boldsymbol{\Theta}$ (covariance-matrix inverse $\boldsymbol{\Sigma}^{-\mathbf{1}}$, as defined in 5-12) in the time domain $t$ for the spectral observations $\boldsymbol{v}(t,f)$ (as defined in 5-5), spectral instrumental noise $\boldsymbol{\xi}(t,f)$ (as defined in 5-6), spectral activity $\boldsymbol{\iota}(t,f)$ (as defined in 5-7), and spectral biological noise $\boldsymbol{\zeta}(t,f)$ (as defined in 5-8). |
| (II-18) | $\widehat{\boldsymbol{\Theta}}_{\xi\xi}(f)$, *with* $\widehat{\boldsymbol{\Theta}}_{\xi\xi}(f) = \mathbf{W}_{\xi\xi}\left(\widehat{\boldsymbol{\Sigma}}_{\xi\xi}(f)\right)$ *or* $\widehat{\boldsymbol{\Theta}}_{\xi\xi}(f) = \mathbf{W}_{\xi\xi}\left(\check{\boldsymbol{\Psi}}_{\xi\xi}(f)\right)$, *and* $\widehat{\boldsymbol{\Theta}}_{\boldsymbol{\iota\iota}}(f)$, *with* $\widehat{\boldsymbol{\Theta}}_{\boldsymbol{\iota\iota}}(f) = \mathbf{W}_{\boldsymbol{\iota\iota}}\left(\widehat{\boldsymbol{\Sigma}}_{\boldsymbol{\iota\iota}}(f)\right)$ *or* $\widehat{\boldsymbol{\Theta}}_{\boldsymbol{\iota\iota}}(f) = \mathbf{W}_{\boldsymbol{\iota\iota}}\left(\check{\boldsymbol{\Psi}}_{\boldsymbol{\iota\iota}}(f)\right)$ | Estimator (hat $\widehat{\boldsymbol{\Theta}}$) of the precision-matrices (as defined in 5-16) for the spectral instrumental noise $\boldsymbol{\xi}(t,f)$ (as defined in 5-6), spectral activity $\boldsymbol{\iota}(t,f)$ (as defined in 5-7). This estimation is via an inverse operator via an inverse-operator $\mathbf{W}$ from the sampled covariance-matrix estimators (hat $\widehat{\boldsymbol{\Sigma}}$) (as defined in 5-13) or the efective covariance-matrix estimators (inverted hat $\check{\boldsymbol{\Psi}}$) (as defined in 5-15). |
| (II-19) | $\boldsymbol{\sigma}^{\mathbf{2}}(f) = diag\big(\boldsymbol{\Sigma}(f)\big)$, *or* $\boldsymbol{\theta}^{2}(f) = diag\big(\boldsymbol{\Theta}(f)\big)$ | Variances $\boldsymbol{\sigma}^{\mathbf{2}}$ from the covariance-matrices $\boldsymbol{\Sigma}$ (as defined 5-12) or precisions from the precision-matrices $\boldsymbol{\Theta}$ (as defined in 5-16). The same notation is extensible to the estimators (as defined in 5-14, 5-15 and 5-17). Special cases are the variances (or precisions) for the spectral instrumental noise $\boldsymbol{\sigma}_{\xi}^{2}$ (or $\boldsymbol{\theta}_{\xi}^{2}$) and biological noise $\boldsymbol{\sigma}_{\zeta}^{2}$ (or $\boldsymbol{\theta}_{\zeta}^{2}$). |
| (II-20) | $\Pi\big(\mathbf{A}\odot\boldsymbol{\Theta}(f)\big)$ | Penalization function or exponent of the precision-matrix $\boldsymbol{\Theta}$ (as defined in 5-16) Gibbs prior parametrized in the regularization parameters or mask matrix $\mathbf{A}$. As a penalization function $\Pi$ we may use the hermitian graphical LASSO (hgLASSO) $\Pi = \|\cdot\|_{1}$, hermitian graphical Ridge (hgRidge) $\|\cdot\|_{2}^{2}$, or hermitian graphical Naïve $\Pi = 0$. |
| (II-21) | $\boldsymbol{\Omega}(f)$, *and* $\widehat{\boldsymbol{\Omega}}(f)$ | Set of hyperparameters $\boldsymbol{\Omega}$ and its estimator (hat $\widehat{\boldsymbol{\Omega}}$) in a model. This set may be conformed by the covariance-matrices $\boldsymbol{\Sigma}$ (as defined in 5-12) or precision-matrices $\boldsymbol{\Theta}$ (as defined in 5-16). |
| (II-22) | $Q\left(\boldsymbol{\Omega}(f), \widehat{\boldsymbol{\Omega}}(f)\right)$ | Expected $-log$ second-type likelihood of the data $\boldsymbol{v}(t,f)$ (as defined in 5-5), obtained after the expectation operation upon parameters $\boldsymbol{\iota}(t,f)$ (as defined in 5-7) of the data $\boldsymbol{v}(t,f)$ and parameters $\boldsymbol{\iota}(t,f)$ $-log$ joint conditional probability density function over the hyperparameters $\boldsymbol{\Omega}(f)$ (as defined in 5-20). the expectarion is based on the posterior distribution for the parameters $\boldsymbol{\iota}(t,f)$ parametrized in the estimated hyperparameters $\widehat{\boldsymbol{\Omega}}(f)$. |
| (II-23) | $\mathcal{L}\big(\boldsymbol{\Omega}(f)\big)$ | Additive combination of $Q\left(\boldsymbol{\Omega}(f), \widehat{\boldsymbol{\Omega}}(f)\right)$ (as defined in 5-21) and the $-log$ prior for the hyperparameters $\boldsymbol{\Omega}(f)$ (as defined in 5-20). |

# III. Bimodal Wishart form of the HIGGS expected second-type likelihood (Lemma 1)

***Lemma 1:*** *For a three-level Bayesian hierarchy due to the Hidden Gaussian Graphical Spectral (HIGGS) model (equation III-1) conditionally dependent to the precision-matrices (hyperparameters)* $\boldsymbol{\Omega} = \{\boldsymbol{\Theta}_{\boldsymbol{\iota}}, \boldsymbol{\Theta}_{\xi\xi}\}$

$$
\begin{gathered}
p\left(\boldsymbol{v}(t) \middle| \boldsymbol{\iota}(t), \boldsymbol{\Theta}_{\xi\xi}\right) = N_{\mathrm{p}}^{\mathbb{C}}\left(\boldsymbol{v}(t) \middle| \mathbf{L}_{\boldsymbol{v}\boldsymbol{\iota}}\boldsymbol{\iota}(t), \boldsymbol{\Theta}_{\xi\xi}^{-1}\right) \\
p\left(\boldsymbol{\iota}(t) \middle| \boldsymbol{\Theta}_{\boldsymbol{\iota}}\right) = N_{\mathrm{p}}^{\mathbb{C}}\left(\boldsymbol{\iota}(t) \middle| \mathbf{0}, \boldsymbol{\Theta}_{\boldsymbol{\iota}}^{-1}\right)
\end{gathered}
\tag{III-1}
$$

*a local approximation to the intractable second-type likelihood* $p\left(\widehat{\boldsymbol{\Sigma}}_{\boldsymbol{vv}} \middle| \boldsymbol{\Omega}\right)$*, which explains the observed cross-spectrum, sampled covariance-matrix* $\widehat{\boldsymbol{\Sigma}}_{\boldsymbol{vv}}$ *(equation III-2) in terms of these hyperparameters* $\boldsymbol{\Omega}$*, at every iteration* $k$*-th of the Expectation Maximization (EM) algorithm*

$$
\widehat{\boldsymbol{\Sigma}}_{\boldsymbol{vv}} = \frac{1}{T}\sum_{t\in\mathbb{T}} \boldsymbol{v}(t)\boldsymbol{v}(t)^{\dagger}
\tag{III-2}
$$

*denominated expected second-type likelihood* $p^{(k)}\left(\widehat{\boldsymbol{\Sigma}}_{\boldsymbol{vv}} \middle| \boldsymbol{\Omega}\right)$*, admits a decomposition into two locally independent Wishart factors, a bimodal Wishart form (equation III-3) in terms of the effective sampled covariance-matrices* $\breve{\boldsymbol{\Psi}}_{\boldsymbol{\iota}}^{(k)}$ *and* $\breve{\boldsymbol{\Psi}}_{\xi\xi}^{(k)}$

$$
\begin{gathered}
p^{(k)}\left(\widehat{\boldsymbol{\Sigma}}_{\boldsymbol{vv}} \middle| \boldsymbol{\Omega}\right) = p^{(k)}\left(\widehat{\boldsymbol{\Sigma}}_{\boldsymbol{vv}} \middle| \boldsymbol{\Theta}_{\boldsymbol{\iota}}\right) p^{(k)}\left(\widehat{\boldsymbol{\Sigma}}_{\boldsymbol{vv}} \middle| \boldsymbol{\Theta}_{\xi\xi}\right) \\
p^{(k)}\left(\widehat{\boldsymbol{\Sigma}}_{\boldsymbol{vv}} \middle| \boldsymbol{\Theta}_{\boldsymbol{\iota}}\right) = W_{\mathrm{q}}^{\mathbb{C}}\left(\breve{\boldsymbol{\Psi}}_{\boldsymbol{\iota}}^{(k)} \middle| T^{-1}\boldsymbol{\Theta}_{\boldsymbol{\iota}}^{-1}, T\right) \\
p^{(k)}\left(\widehat{\boldsymbol{\Sigma}}_{\boldsymbol{vv}} \middle| \boldsymbol{\Theta}_{\xi\xi}\right) = W_{\mathrm{p}}^{\mathbb{C}}\left(\breve{\boldsymbol{\Psi}}_{\xi\xi}^{(k)} \middle| T^{-1}\boldsymbol{\Theta}_{\xi\xi}^{-1}, T\right)
\end{gathered}
\tag{III-3}
$$

*due to the Gibbs exponential form expressing* $p^{(k)}\left(\widehat{\boldsymbol{\Sigma}}_{\boldsymbol{vv}} \middle| \boldsymbol{\Omega}\right)$ *as function of the expected* $-log$ *second-type likelihood from the EM integral operation* $Q\left(\boldsymbol{\Omega}, \widehat{\boldsymbol{\Omega}}^{(k)}\right)$ *(equation III-4) which may be expressed explicitly as an additive functions of* $\boldsymbol{\Theta}_{\boldsymbol{\iota}}$ *and* $\boldsymbol{\Theta}_{\xi\xi}$ *(equation III-5)*

$$
\begin{gathered}
p^{(k)}\left(\widehat{\boldsymbol{\Sigma}}_{\boldsymbol{vv}} \middle| \boldsymbol{\Omega}\right) = exp\left(Q\left(\boldsymbol{\Omega}, \widehat{\boldsymbol{\Omega}}^{(k)}\right) \middle| 1\right) \\
Q\left(\boldsymbol{\Omega}, \widehat{\boldsymbol{\Omega}}^{(k)}\right) = -\sum_{t\in\mathbb{T}} E\left(log\left(p\left(\boldsymbol{v}(t), \boldsymbol{\iota}(t) \middle| \boldsymbol{\Omega}\right)\right) \middle| \boldsymbol{v}(t), \widehat{\boldsymbol{\Omega}}^{(k)}\right) \\
E\left(log\left(p\left(\boldsymbol{v}(t), \boldsymbol{\iota}(t) \middle| \boldsymbol{\Omega}\right)\right) \middle| \boldsymbol{v}(t), \widehat{\boldsymbol{\Omega}}^{(k)}\right) = \int_{\mathbb{C}^{\mathrm{q}}} log\left(p\left(\boldsymbol{v}(t), \boldsymbol{\iota}(t) \middle| \boldsymbol{\Omega}\right)\right) p\left(\boldsymbol{\iota}(t) \middle| \boldsymbol{v}(t), \widehat{\boldsymbol{\Omega}}^{(k)}\right) d\boldsymbol{\iota}(t)
\end{gathered}
\tag{III-4}
$$

*where the explicit* $Q\left(\boldsymbol{\Omega}, \widehat{\boldsymbol{\Omega}}^{(k)}\right)$ *(equation III-5) is*

$$
Q\left(\boldsymbol{\Omega}, \widehat{\boldsymbol{\Omega}}^{(k)}\right) = -T\, log\left|\boldsymbol{\Theta}_{\boldsymbol{\iota}}\right| + T\, tr\left(\breve{\boldsymbol{\Psi}}_{\boldsymbol{\iota}}^{(k)}\boldsymbol{\Theta}_{\boldsymbol{\iota}}\right) - T\, log\left|\boldsymbol{\Theta}_{\xi\xi}\right| + T\, tr\left(\breve{\boldsymbol{\Psi}}_{\xi\xi}^{(k)}\boldsymbol{\Theta}_{\xi\xi}\right)
\tag{III-5}
$$

***Proof of Lemma 1***

The general strategy for the identification of the HIGGS model (equation III-1) is based on the EM algorithm: a second-type maximum a posteriori estimation, performed iteratively for $\boldsymbol{\iota}(t)$ (parameters) and $\boldsymbol{\Omega}$ (hyperparameters) (Dempster, Laird, and Rubin 1977; C. Liu and Rubin 1994; McLachlan and Krishnan 2007; Wills, Ninness, and Gibson 2009). The EM expectation obtains a second-type likelihood (equation III-3) based on the posterior distribution of the parameters $p\left(\boldsymbol{\iota}(t) \middle| \boldsymbol{v}(t), \widehat{\boldsymbol{\Omega}}^{(k)}\right)$ (equation III-4) expressed upon fixed values of $\widehat{\boldsymbol{\Omega}}^{(k)}$ (hyperparameters).

This posterior distribution is proportional to the factor of the first-type likelihood $N_{\mathrm{p}}^{\mathbb{C}}\left(\boldsymbol{v}(t) \middle| \mathbf{L}_{\boldsymbol{v}\boldsymbol{\iota}}\boldsymbol{\iota}(t), {\widehat{\boldsymbol{\Theta}}_{\xi\xi}^{(k)}}^{-1}\right)$ and the prior $N_{\mathrm{p}}^{\mathbb{C}}\left(\boldsymbol{\iota}(t) \middle| \mathbf{0}, {\widehat{\boldsymbol{\Theta}}_{\boldsymbol{\iota}}^{(k)}}^{-1}\right)$ (equation III-1) due to the Bayes theorem por conditional probabilities.

$$
p\left(\boldsymbol{\iota}(t) \middle| \boldsymbol{v}(t), \boldsymbol{\Omega}^{(k)}\right) = N_{\mathrm{p}}^{\mathbb{C}}\left(\boldsymbol{v}(t) \middle| \mathbf{L}_{\boldsymbol{v}\boldsymbol{\iota}}\boldsymbol{\iota}(t), {\widehat{\boldsymbol{\Theta}}_{\xi\xi}^{(k)}}^{-1}\right) N_{\mathrm{p}}^{\mathbb{C}}\left(\boldsymbol{\iota}(t) \middle| \mathbf{0}, {\widehat{\boldsymbol{\Theta}}_{\boldsymbol{\iota}}^{(k)}}^{-1}\right) \Big/ p\left(\boldsymbol{v}(t)\right)
\tag{III-6}
$$

Analyzing the exponent (equation III-7) from the Bayesian expression (equation III-6) is enough to find the structure of the posterior distribution $p\left(\boldsymbol{\iota}(t) \middle| \boldsymbol{v}(t), \boldsymbol{\Omega}^{(k)}\right)$.

$$
-\left(\boldsymbol{v}(t) - \mathbf{L}_{\boldsymbol{v}\boldsymbol{\iota}}\boldsymbol{\iota}(t)\right)^{\dagger} \widehat{\boldsymbol{\Theta}}_{\xi\xi}^{(k)} \left(\boldsymbol{v}(t) - \mathbf{L}_{\boldsymbol{v}\boldsymbol{\iota}}\boldsymbol{\iota}(t)\right) - \boldsymbol{\iota}^{\dagger}(t)\widehat{\boldsymbol{\Theta}}_{\boldsymbol{\iota}}^{(k)}\boldsymbol{\iota}(t)
\tag{III-7}
$$

Reorganizing terms in this exponent (equation III-7) we obtain the following expression (equation III-8), where $\mathbf{L}_{\boldsymbol{\iota}\boldsymbol{v}}$ is the transpose for $\mathbf{L}_{\boldsymbol{v}\boldsymbol{\iota}}$

$$
-\boldsymbol{\iota}^{\dagger}(t)\left(\mathbf{L}_{\boldsymbol{\iota}\boldsymbol{v}}\widehat{\boldsymbol{\Theta}}_{\xi\xi}^{(k)}\mathbf{L}_{\boldsymbol{v}\boldsymbol{\iota}} + \widehat{\boldsymbol{\Theta}}_{\boldsymbol{\iota}}^{(k)}\right)\boldsymbol{\iota}(t) + \boldsymbol{v}^{\dagger}(t)\widehat{\boldsymbol{\Theta}}_{\xi\xi}^{(k)}\mathbf{L}_{\boldsymbol{v}\boldsymbol{\iota}}\boldsymbol{\iota}(t) + \boldsymbol{\iota}^{\dagger}(t)\mathbf{L}_{\boldsymbol{\iota}\boldsymbol{v}}\widehat{\boldsymbol{\Theta}}_{\xi\xi}^{(k)}\boldsymbol{v}(t)
\tag{III-8}
$$

This expression (equation III-8) can then be expressed in terms of an auxiliary quantity in the EM iterations denominated source posterior precision-matrix $\breve{\boldsymbol{\Theta}}_{\boldsymbol{\iota}}^{(k)}$ (equation III-9).

$$
-\boldsymbol{\iota}^{\dagger}(t)\breve{\boldsymbol{\Theta}}_{\boldsymbol{\iota}}^{(k)}\boldsymbol{\iota}(t) + \boldsymbol{v}^{\dagger}(t)\widehat{\boldsymbol{\Theta}}_{\xi\xi}^{(k)}\mathbf{L}_{\boldsymbol{v}\boldsymbol{\iota}}\boldsymbol{\iota}(t) + \boldsymbol{\iota}^{\dagger}(t)\mathbf{L}_{\boldsymbol{\iota}\boldsymbol{v}}\widehat{\boldsymbol{\Theta}}_{\xi\xi}^{(k)}\boldsymbol{v}(t)
\tag{III-9}
$$

Where this source posterior precision-matrix $\breve{\boldsymbol{\Theta}}_{\boldsymbol{\iota}}^{(k)}$ in (equation III-9), or source posterior covariance-matrix given by its inverse $\breve{\boldsymbol{\Sigma}}_{\boldsymbol{\iota}}^{(k)} = {\breve{\boldsymbol{\Theta}}_{\boldsymbol{\iota}}^{(k)}}^{-1}$, is given by (equation III-10).

$$
\begin{gathered}
\breve{\boldsymbol{\Theta}}_{\boldsymbol{\iota}}^{(k)} = \mathbf{L}_{\boldsymbol{\iota}\boldsymbol{v}}\widehat{\boldsymbol{\Theta}}_{\xi\xi}^{(k)}\mathbf{L}_{\boldsymbol{v}\boldsymbol{\iota}} + \widehat{\boldsymbol{\Theta}}_{\boldsymbol{\iota}}^{(k)} \\
\breve{\boldsymbol{\Sigma}}_{\boldsymbol{\iota}}^{(k)} \leftarrow \left(\mathbf{L}_{\boldsymbol{\iota}\boldsymbol{v}}\widehat{\boldsymbol{\Theta}}_{\xi\xi}^{(k)}\mathbf{L}_{\boldsymbol{v}\boldsymbol{\iota}} + \widehat{\boldsymbol{\Theta}}_{\boldsymbol{\iota}}^{(k)}\right)^{-1}
\end{gathered}
\tag{III-10}
$$

Introducing in the second and third term of the exponent (equation III-9) the product $\breve{\boldsymbol{\Theta}}_{\boldsymbol{\iota\iota}}^{(k)}\breve{\boldsymbol{\Sigma}}_{\boldsymbol{\iota\iota}}^{(k)} = \mathbf{I}_{\mathrm{q}}$, of the source posterior precision-matrix and posterior covariance-matrix as given in (equation III-10) we obtain

$$-\boldsymbol{\iota}^{\dagger}(t)\breve{\boldsymbol{\Theta}}_{\boldsymbol{\iota\iota}}^{(k)}\boldsymbol{\iota}(t) + \boldsymbol{v}^{\dagger}(t)\widehat{\boldsymbol{\Theta}}_{\xi\xi}^{(k)}\mathbf{L}_{\boldsymbol{v\iota}}\breve{\boldsymbol{\Sigma}}_{\boldsymbol{\iota\iota}}^{(k)}\breve{\boldsymbol{\Theta}}_{\boldsymbol{\iota\iota}}^{(k)}\boldsymbol{\iota}(t) + \boldsymbol{\iota}^{\dagger}(t)\breve{\boldsymbol{\Theta}}_{\boldsymbol{\iota\iota}}^{(k)}\breve{\boldsymbol{\Sigma}}_{\boldsymbol{\iota\iota}}^{(k)}\mathbf{L}_{\boldsymbol{\iota v}}\widehat{\boldsymbol{\Theta}}_{\xi\xi}^{(k)}\boldsymbol{v}(t) \qquad \text{(III-11)}$$

The second and third terms in the exponent (equation III-11) can be then expressed in terms of the estimates of the parameters in the EM iterations $\hat{\boldsymbol{\iota}}^{(k)}(t)$ and the source posterior precision-matrix $\breve{\boldsymbol{\Theta}}_{\boldsymbol{\iota\iota}}^{(k)}$ (equation III-12).

$$-\boldsymbol{\iota}^{\dagger}(t)\breve{\boldsymbol{\Theta}}_{\boldsymbol{\iota\iota}}^{(k)}\boldsymbol{\iota}(t) + {\hat{\boldsymbol{\iota}}^{(k)}}^{\dagger}(t)\breve{\boldsymbol{\Theta}}_{\boldsymbol{\iota\iota}}^{(k)}\boldsymbol{\iota}(t) + \boldsymbol{\iota}^{\dagger}(t)\breve{\boldsymbol{\Theta}}_{\boldsymbol{\iota\iota}}^{(k)}\hat{\boldsymbol{\iota}}^{(k)}(t) \qquad \text{(III-12)}$$

This parameter estimates $\hat{\boldsymbol{\iota}}^{(k)}(t)$ in (equation III-12) may be expressed locally linearly at EM iterations from the data $\boldsymbol{v}(t)$ in terms of another auxiliary quantity denominated source inverse-operator $\breve{\mathbf{T}}_{\boldsymbol{\iota v}}^{(k)}$ for the lead field $\mathbf{L}_{\boldsymbol{v\iota}}$.

$$\hat{\boldsymbol{\iota}}^{(k)}(t) = \breve{\mathbf{T}}_{\boldsymbol{\iota v}}^{(k)}\boldsymbol{v}(t) \qquad \text{(III-13)}$$

Where the inverse-operator $\breve{\mathbf{T}}_{\boldsymbol{\iota v}}^{(k)}$ in (equation III-13) is given in terms of the lead field $\mathbf{L}_{\boldsymbol{v\iota}}$, the source posterior covariance-matrix $\breve{\boldsymbol{\Sigma}}_{\boldsymbol{\iota\iota}}^{(k)}$, and the residuals precision-matrix $\widehat{\boldsymbol{\Theta}}_{\xi\xi}^{(k)}$ as follows (equation III-14)

$$\breve{\mathbf{T}}_{\boldsymbol{\iota v}}^{(k)} = \breve{\boldsymbol{\Sigma}}_{\boldsymbol{\iota\iota}}^{(k)}\mathbf{L}_{\boldsymbol{\iota v}}\widehat{\boldsymbol{\Theta}}_{\xi\xi}^{(k)} \qquad \text{(III-14)}$$

Completing the exponent in (equation III-12) with the term ${\hat{\boldsymbol{\iota}}^{(k)}}^{\dagger}(t)\breve{\boldsymbol{\Theta}}_{\boldsymbol{\iota\iota}}^{(k)}\hat{\boldsymbol{\iota}}^{(k)}(t)$ it is straightforward that the parameters posterior distribution can be defined as a circularly symmetric complex-valued Gaussian $N^{\mathbb{C}}$ (equation III-15) parametrized with the posterior covariance-matrix $\breve{\boldsymbol{\Sigma}}_{\boldsymbol{\iota\iota}}^{(k)}$ from (equation III-10) and posterior mean $\hat{\boldsymbol{\iota}}^{(k)}(t)$ from (equation III-13).

$$p\left(\boldsymbol{\iota}(t)\middle|\boldsymbol{v}(t), \widehat{\boldsymbol{\Omega}}^{(k)}\right) = N_{\mathrm{q}}^{\mathbb{C}}\left(\boldsymbol{\iota}(t)\middle|\hat{\boldsymbol{\iota}}^{(k)}(t), \breve{\boldsymbol{\Sigma}}_{\boldsymbol{\iota\iota}}^{(k)}\right) \qquad \text{(III-15)}$$

The EM local approximation in HIGGS (equation III-1) for the intractable second-type likelihood $p^{(k)}\left(\widehat{\boldsymbol{\Sigma}}_{\boldsymbol{vv}}\middle|\boldsymbol{\Omega}\right)$ (equation III-3) is then expressed by means of the integral $Q\left(\boldsymbol{\Omega}, \widehat{\boldsymbol{\Omega}}^{(k)}\right)$ (equation III-4). This integral is interpreted as an expected $-log$ second-type likelihood that yields a local approximation to the second-type likelihood as a Gibb's exponential form that must be combined with priors $p(\boldsymbol{\Omega})$ for the second-type maximum a posteriori of the hyperparameters $\boldsymbol{\Omega}$. Expressing $Q\left(\boldsymbol{\Omega}, \widehat{\boldsymbol{\Omega}}^{(k)}\right)$ is much simpler by means of the additive terms $Q_t\left(\boldsymbol{\Omega}, \widehat{\boldsymbol{\Omega}}^{(k)}\right) = E\left(log\left(p(\boldsymbol{v}(t), \boldsymbol{\iota}(t)|\boldsymbol{\Omega})\right)\middle|\boldsymbol{v}(t), \widehat{\boldsymbol{\Omega}}^{(k)}\right)$ (equation III-16), which are dependent on the data and parameters joint distribution $p(\boldsymbol{v}(t), \boldsymbol{\iota}(t)|\boldsymbol{\Omega})$ and the parameters posterior distribution $p\left(\boldsymbol{\iota}(t)\middle|\boldsymbol{v}(t), \widehat{\boldsymbol{\Omega}}^{(k)}\right)$.

$$\begin{gathered} Q\left(\boldsymbol{\Omega}, \widehat{\boldsymbol{\Omega}}^{(k)}\right) = \textstyle\sum_{t\in\mathbb{T}} Q_t\left(\boldsymbol{\Omega}, \widehat{\boldsymbol{\Omega}}^{(k)}\right) \\ Q_t\left(\boldsymbol{\Omega}, \widehat{\boldsymbol{\Omega}}^{(k)}\right) = -E\left(log\left(p(\boldsymbol{v}(t), \boldsymbol{\iota}(t)|\boldsymbol{\Omega})\right)\middle|\boldsymbol{v}(t), \widehat{\boldsymbol{\Omega}}^{(k)}\right) \end{gathered} \qquad \text{(III-16)}$$

This Bayesian formalism equivalent to the missing data problem ($\boldsymbol{\iota}(t)$), where the pair $\{\boldsymbol{v}(t), \boldsymbol{\iota}(t)\}$ is denominated complete data and the joint distribution $p(\boldsymbol{v}(t), \boldsymbol{\iota}(t)|\boldsymbol{\Omega})$ is denominated complete data likelihood. The complete data likelihood is factorized due to the Bayes theorem (equation III-17) into the observed data likelihood $p(\boldsymbol{v}(t)|\boldsymbol{\iota}(t), \boldsymbol{\Omega})$ and the missing data (parameters) likelihood $p(\boldsymbol{\iota}(t)|\boldsymbol{\Omega})$ that are according to the HIGGS model (equation III-1).

$$p(\boldsymbol{v}(t), \boldsymbol{\iota}(t)|\boldsymbol{\Omega}) = p(\boldsymbol{v}(t)|\boldsymbol{\iota}(t), \boldsymbol{\Omega})p(\boldsymbol{v}(t)|\boldsymbol{\iota}(t), \boldsymbol{\Omega}) = N_{\mathrm{p}}^{\mathbb{C}}\left(\boldsymbol{v}(t)\middle|\mathbf{L}_{\boldsymbol{v\iota}}\boldsymbol{\iota}(t), \boldsymbol{\Theta}_{\xi\xi}^{-1}\right)N_{\mathrm{p}}^{\mathbb{C}}\left(\boldsymbol{\iota}(t)|\mathbf{0}, \boldsymbol{\Theta}_{\boldsymbol{\iota\iota}}^{-1}\right) \qquad \text{(III-17)}$$

The additive terms $Q_t\left(\boldsymbol{\Omega}, \widehat{\boldsymbol{\Omega}}^{(k)}\right)$ (equation III-16) due to the missing data posterior distribution $p\left(\boldsymbol{\iota}(t)\middle|\boldsymbol{v}(t), \widehat{\boldsymbol{\Omega}}^{(k)}\right)$ (equation III-15) and the complete data likelihood expression $p(\boldsymbol{v}(t), \boldsymbol{\iota}(t)|\boldsymbol{\Omega})$ (equation III-17) and can be then expressed as follows (equation III-18).

$$\begin{gathered} Q_t\left(\boldsymbol{\Omega}, \widehat{\boldsymbol{\Omega}}^{(k)}\right) = -E\left(log\left(p(\boldsymbol{v}(t), \boldsymbol{\iota}(t)|\boldsymbol{\Omega})\right)\middle|\boldsymbol{v}(t), \widehat{\boldsymbol{\Omega}}^{(k)}\right) \\ E\left(log\left(p(\boldsymbol{v}(t), \boldsymbol{\iota}(t)|\boldsymbol{\Omega})\right)\middle|\boldsymbol{v}(t), \widehat{\boldsymbol{\Omega}}^{(k)}\right) = \int_{\mathbb{C}^{\mathrm{q}}} log\left(N_{\mathrm{p}}^{\mathbb{C}}\left(\boldsymbol{v}(t)\middle|\mathbf{L}_{\boldsymbol{v\iota}}\boldsymbol{\iota}(t), \boldsymbol{\Theta}_{\xi\xi}^{-1}\right)N_{\mathrm{p}}^{\mathbb{C}}\left(\boldsymbol{\iota}(t)|\mathbf{0}, \boldsymbol{\Theta}_{\boldsymbol{\iota\iota}}^{-1}\right)\right)N_{\mathrm{q}}^{\mathbb{C}}\left(\boldsymbol{\iota}(t)\middle|\hat{\boldsymbol{\iota}}^{(k)}(t), \breve{\boldsymbol{\Sigma}}_{\boldsymbol{\iota\iota}}^{(k)}\right)d\boldsymbol{\iota}(t) \end{gathered} \qquad \text{(III-18)}$$

Where the logarithm (equation III-18) of the factors $log\left(N_{\mathrm{p}}^{\mathbb{C}}\left(\boldsymbol{v}(t)\middle|\mathbf{L}_{\boldsymbol{v\iota}}\boldsymbol{\iota}(t), \boldsymbol{\Theta}_{\xi\xi}^{-1}\right)N_{\mathrm{p}}^{\mathbb{C}}\left(\boldsymbol{\iota}(t)|\mathbf{0}, \boldsymbol{\Theta}_{\boldsymbol{\iota\iota}}^{-1}\right)\right)$ (equation III-17) can be expressed additively as (equation III-19).

$$\begin{gathered} log\left(N_{\mathrm{p}}^{\mathbb{C}}\left(\boldsymbol{v}(t)\middle|\mathbf{L}_{\boldsymbol{v\iota}}\boldsymbol{\iota}(t), \boldsymbol{\Theta}_{\xi\xi}^{-1}\right)N_{\mathrm{p}}^{\mathbb{C}}\left(\boldsymbol{\iota}(t)|\mathbf{0}, \boldsymbol{\Theta}_{\boldsymbol{\iota\iota}}^{-1}\right)\right) = \\ log\left|\boldsymbol{\Theta}_{\xi\xi}\right| - \left(\boldsymbol{v}(t) - \mathbf{L}_{\boldsymbol{v\iota}}\boldsymbol{\iota}(t)\right)^{\dagger}\boldsymbol{\Theta}_{\xi\xi}\left(\boldsymbol{v}(t) - \mathbf{L}_{\boldsymbol{v\iota}}\boldsymbol{\iota}(t)\right) + log\left|\boldsymbol{\Theta}_{\boldsymbol{\iota\iota}}\right| - \boldsymbol{\iota}^{\dagger}(t)\boldsymbol{\Theta}_{\boldsymbol{\iota\iota}}\boldsymbol{\iota}(t) \end{gathered} \qquad \text{(III-19)}$$

Employing the trace properties to transform (equation III-19) we obtain (equation III-20) where an expansion of the second term in (III-19) is included from the second term to the fifth in (equation III-20).

$$\begin{aligned} log\left(N_{\mathrm{p}}^{\mathbb{C}}\left(\boldsymbol{v}(t)\middle|\mathbf{L}_{\boldsymbol{v\iota}}\boldsymbol{\iota}(t), \boldsymbol{\Theta}_{\xi\xi}^{-1}\right)N_{\mathrm{p}}^{\mathbb{C}}\left(\boldsymbol{\iota}(t)|\mathbf{0}, \boldsymbol{\Theta}_{\boldsymbol{\iota\iota}}^{-1}\right)\right) &= log\left|\boldsymbol{\Theta}_{\xi\xi}\right| - tr\left(\boldsymbol{v}(t)\boldsymbol{v}^{\dagger}(t)\boldsymbol{\Theta}_{\xi\xi}\right) + tr\left(\boldsymbol{v}(t)\boldsymbol{\iota}^{\dagger}(t)\mathbf{L}_{\boldsymbol{\iota v}}\boldsymbol{\Theta}_{\xi\xi}\right) \\ &+ tr\left(\mathbf{L}_{\boldsymbol{v\iota}}\boldsymbol{\iota}(t)\boldsymbol{v}(t)\boldsymbol{\Theta}_{\xi\xi}\right) - tr\left(\mathbf{L}_{\boldsymbol{v\iota}}\boldsymbol{\iota}(t)\boldsymbol{\iota}^{\dagger}(t)\mathbf{L}_{\boldsymbol{\iota v}}\boldsymbol{\Theta}_{\xi\xi}\right) + log\left|\boldsymbol{\Theta}_{\boldsymbol{\iota\iota}}\right| - tr\left(\boldsymbol{\iota}(t)\boldsymbol{\iota}^{\dagger}(t)\boldsymbol{\Theta}_{\boldsymbol{\iota\iota}}\right) \end{aligned} \qquad \text{(III-20)}$$

From the expected $-log$ second-type likelihood (equation III-18) and the logarithm (equation III-20) we can then express $Q_t\left(\boldsymbol{\Omega}, \widehat{\boldsymbol{\Omega}}^{(k)}\right)$ as (equation III-21).

$$\begin{aligned} Q_t\left(\boldsymbol{\Omega}, \widehat{\boldsymbol{\Omega}}^{(k)}\right) = &-log\left|\boldsymbol{\Theta}_{\xi\xi}\right| + tr\left(\boldsymbol{v}(t)\boldsymbol{v}^{\dagger}(t)\boldsymbol{\Theta}_{\xi\xi}\right) - tr\left(\boldsymbol{v}(t)E\left(\boldsymbol{\iota}^{\dagger}(t)\middle|\boldsymbol{v}(t), \widehat{\boldsymbol{\Omega}}^{(k)}\right)\mathbf{L}_{\boldsymbol{\iota v}}\boldsymbol{\Theta}_{\xi\xi}\right) - tr\left(\mathbf{L}_{\boldsymbol{v\iota}}E\left(\boldsymbol{\iota}(t)\middle|\boldsymbol{v}(t), \widehat{\boldsymbol{\Omega}}^{(k)}\right)\boldsymbol{v}(t)\boldsymbol{\Theta}_{\xi\xi}\right) + \\ &tr\left(\mathbf{L}_{\boldsymbol{v\iota}}E\left(\boldsymbol{\iota}(t)\boldsymbol{\iota}^{\dagger}(t)\middle|\boldsymbol{v}(t), \widehat{\boldsymbol{\Omega}}^{(k)}\right)\mathbf{L}_{\boldsymbol{\iota v}}\boldsymbol{\Theta}_{\xi\xi}\right) - log\left|\boldsymbol{\Theta}_{\boldsymbol{\iota\iota}}\right| + tr\left(E\left(\boldsymbol{\iota}(t)\boldsymbol{\iota}^{\dagger}(t)\middle|\boldsymbol{v}(t), \widehat{\boldsymbol{\Omega}}^{(k)}\right)\boldsymbol{\Theta}_{\boldsymbol{\iota\iota}}\right) \end{aligned} \qquad \text{(III-21)}$$

Where it can be easily checked that for the expectation terms in (equation III-21) that following expressions hold (equation III-22).

$$
\begin{gathered}
E\left(\boldsymbol{\iota}^{\dagger}(t)\middle|\boldsymbol{v}(t),\widehat{\boldsymbol{\Omega}}^{(k)}\right) = \hat{\boldsymbol{\iota}}^{(k)^{\dagger}}(t) \\
E\left(\boldsymbol{\iota}(t)\middle|\boldsymbol{v}(t),\widehat{\boldsymbol{\Omega}}^{(k)}\right) = \hat{\boldsymbol{\iota}}^{(k)}(t) \\
E\left(\boldsymbol{\iota}(t)\boldsymbol{\iota}^{\dagger}(t)\middle|\boldsymbol{v}(t),\widehat{\boldsymbol{\Omega}}^{(k)}\right) = \breve{\boldsymbol{\Sigma}}_{\boldsymbol{\iota}}^{(k)} + \hat{\boldsymbol{\iota}}^{(k)}(t)\hat{\boldsymbol{\iota}}^{(k)^{\dagger}}(t)
\end{gathered}
\tag{III-22}
$$

Then, plugging the expectation terms (equation III-22) into $Q_t\left(\boldsymbol{\Omega},\widehat{\boldsymbol{\Omega}}^{(k)}\right)$ (equation III-21) we obtain (equation III-23).

$$
\begin{gathered}
Q_t\left(\boldsymbol{\Omega},\widehat{\boldsymbol{\Omega}}^{(k)}\right) = -log\left|\boldsymbol{\Theta}_{\xi\xi}\right| + tr\left(\boldsymbol{v}(t)\boldsymbol{v}^{\dagger}(t)\boldsymbol{\Theta}_{\xi\xi}\right) - tr\left(\boldsymbol{v}(t)\hat{\boldsymbol{\iota}}^{(k)^{\dagger}}(t)\mathbf{L}_{\iota v}\boldsymbol{\Theta}_{\xi\xi}\right) - tr\left(\mathbf{L}_{v\iota}\hat{\boldsymbol{\iota}}^{(k)}(t)\boldsymbol{v}(t)\boldsymbol{\Theta}_{\xi\xi}\right) + tr\left(\mathbf{L}_{v\iota}\left(\breve{\boldsymbol{\Sigma}}_{\boldsymbol{\iota}}^{(k)} + \right.\right. \\
\left.\left.\hat{\boldsymbol{\iota}}^{(k)}(t)\hat{\boldsymbol{\iota}}^{(k)^{\dagger}}(t)\right)\mathbf{L}_{\iota v}\boldsymbol{\Theta}_{\xi\xi}\right) - log\left|\boldsymbol{\Theta}_{\boldsymbol{\iota}}\right| + tr\left(\left(\breve{\boldsymbol{\Sigma}}_{\boldsymbol{\iota}}^{(k)} + \hat{\boldsymbol{\iota}}^{(k)}(t)\hat{\boldsymbol{\iota}}^{(k)^{\dagger}}(t)\right)\boldsymbol{\Theta}_{\boldsymbol{\iota}}\right)
\end{gathered}
\tag{III-23}
$$

Merging from the second term to the fifth in (equation III-23) we obtain (equation III-24).

$$
\begin{gathered}
Q_t\left(\boldsymbol{\Omega},\widehat{\boldsymbol{\Omega}}^{(k)}\right) = -log\left|\boldsymbol{\Theta}_{\xi\xi}\right| + tr\left(\left(\mathbf{L}_{v\iota}\breve{\boldsymbol{\Sigma}}_{\boldsymbol{\iota}}^{(k)}\mathbf{L}_{\iota v} + \boldsymbol{v}(t)\boldsymbol{v}^{\dagger}(t) - \boldsymbol{v}(t)\hat{\boldsymbol{\iota}}^{(k)^{\dagger}}(t)\mathbf{L}_{\iota v} - \mathbf{L}_{v\iota}\hat{\boldsymbol{\iota}}^{(k)}(t)\boldsymbol{v}(t) + \mathbf{L}_{v\iota}\hat{\boldsymbol{\iota}}^{(k)}(t)\hat{\boldsymbol{\iota}}^{(k)^{\dagger}}(t)\mathbf{L}_{\iota v}\right)\boldsymbol{\Theta}_{\xi\xi}\right) \\
-log\left|\boldsymbol{\Theta}_{\boldsymbol{\iota}}\right| + tr\left(\left(\breve{\boldsymbol{\Sigma}}_{\boldsymbol{\iota}}^{(k)} + \hat{\boldsymbol{\iota}}^{(k)}(t)\hat{\boldsymbol{\iota}}^{(k)^{\dagger}}(t)\right)\boldsymbol{\Theta}_{\boldsymbol{\iota}}\right)
\end{gathered}
\tag{III-24}
$$

Then (equation III-24) may be expressed in terms of residual posterior covariance-matrix $\breve{\boldsymbol{\Sigma}}_{\xi\xi}^{(k)} = \mathbf{L}_{v\iota}\breve{\boldsymbol{\Sigma}}_{\boldsymbol{\iota}}^{(k)}\mathbf{L}_{\iota v}$ (a projection by the Lead Field $\mathbf{L}_{v\iota}$ of the posterior covariance-matrix for source cortical oscillations $\breve{\boldsymbol{\Sigma}}_{\boldsymbol{\iota}}^{(k)}$) and the residuals estimates $\hat{\boldsymbol{\xi}}^{(k)}(t) = \boldsymbol{v}(t) - \mathbf{L}_{v\iota}\hat{\boldsymbol{\iota}}^{(k)}(t)$ (equation III-25) compact and locally linear relations at EM iterations.

$$
\begin{gathered}
Q_t\left(\boldsymbol{\Omega},\widehat{\boldsymbol{\Omega}}^{(k)}\right) = -log\left|\boldsymbol{\Theta}_{\xi\xi}\right| + tr\left(\left(\breve{\boldsymbol{\Sigma}}_{\xi\xi}^{(k)} + \hat{\boldsymbol{\xi}}^{(k)}(t)\hat{\boldsymbol{\xi}}^{(k)^{\dagger}}(t)\right)\boldsymbol{\Theta}_{\xi\xi}\right) \\
-log\left|\boldsymbol{\Theta}_{\boldsymbol{\iota}}\right| + tr\left(\left(\breve{\boldsymbol{\Sigma}}_{\boldsymbol{\iota}}^{(k)} + \hat{\boldsymbol{\iota}}^{(k)}(t)\hat{\boldsymbol{\iota}}^{(k)^{\dagger}}(t)\right)\boldsymbol{\Theta}_{\boldsymbol{\iota}}\right)
\end{gathered}
\tag{III-25}
$$

Then $Q\left(\boldsymbol{\Omega},\widehat{\boldsymbol{\Omega}}^{(k)}\right)$ (equation III-16) can be expressed from the additive terms $Q_t\left(\boldsymbol{\Omega},\widehat{\boldsymbol{\Omega}}^{(k)}\right)$ (equation III-25) to obtain the expected $-log$ second-type likelihood in terms of the sampled covariances for the residuals $\widehat{\boldsymbol{\Sigma}}_{\xi\xi}^{(k)}$ and source cortical oscillations $\widehat{\boldsymbol{\Sigma}}_{\boldsymbol{\iota}}^{(k)}$, given for $T$ samples or time instances ($t \in \mathbb{T}$) (equation III-26).

$$
Q\left(\boldsymbol{\Omega},\widehat{\boldsymbol{\Omega}}^{(k)}\right) = -Tlog\left|\boldsymbol{\Theta}_{\xi\xi}\right| + Ttr\left(\left(\breve{\boldsymbol{\Sigma}}_{\xi\xi}^{(k)} + \widehat{\boldsymbol{\Sigma}}_{\xi\xi}^{(k)}\right)\boldsymbol{\Theta}_{\xi\xi}\right) - Tlog\left|\boldsymbol{\Theta}_{\boldsymbol{\iota}}\right| + Ttr\left(\left(\breve{\boldsymbol{\Sigma}}_{\boldsymbol{\iota}}^{(k)} + \widehat{\boldsymbol{\Sigma}}_{\boldsymbol{\iota}}^{(k)}\right)\boldsymbol{\Theta}_{\boldsymbol{\iota}}\right)
\tag{III-26}
$$

Where in (equation III-26) the sampled covariance-matrices for the residuals $\widehat{\boldsymbol{\Sigma}}_{\xi\xi}^{(k)}$ and source cortical oscillations $\widehat{\boldsymbol{\Sigma}}_{\boldsymbol{\iota}}^{(k)}$ are defined as (equation III-27).

$$
\begin{gathered}
\widehat{\boldsymbol{\Sigma}}_{\xi\xi}^{(k)} = \frac{1}{T}\sum_{t\in\mathbb{T}}\hat{\boldsymbol{\xi}}^{(k)}(t)\hat{\boldsymbol{\xi}}^{(k)}(t)^{\dagger} \\
\widehat{\boldsymbol{\Sigma}}_{\boldsymbol{\iota}}^{(k)} = \frac{1}{T}\sum_{t\in\mathbb{T}}\hat{\boldsymbol{\iota}}^{(k)}(t)\hat{\boldsymbol{\iota}}^{(k)^{\dagger}}(t)
\end{gathered}
\tag{III-27}
$$

These sampled covariances (equation III-27) may be expressed locally linearly at EM iterations from the data sampled covariance-matrix $\widehat{\boldsymbol{\Sigma}}_{vv}$ in terms of the inverse-operators for the residuals $\breve{\mathbf{T}}_{\xi v}^{(k)}$ and source cortical oscillations $\breve{\mathbf{T}}_{\iota v}^{(k)}$ (equation III-28).

$$
\begin{gathered}
\widehat{\boldsymbol{\Sigma}}_{\xi\xi}^{(k)} = \breve{\mathbf{T}}_{\xi v}^{(k)}\widehat{\boldsymbol{\Sigma}}_{vv}\breve{\mathbf{T}}_{\xi v}^{(k)^{\dagger}} \\
\widehat{\boldsymbol{\Sigma}}_{\boldsymbol{\iota}}^{(k)} = \breve{\mathbf{T}}_{\iota v}^{(k)}\widehat{\boldsymbol{\Sigma}}_{vv}\breve{\mathbf{T}}_{\iota v}^{(k)^{\dagger}}
\end{gathered}
\tag{III-28}
$$

This compact expression (equation III-28) is due to the relations of the residuals $\hat{\boldsymbol{\xi}}^{(k)}(t)$ and source cortical oscillations $\hat{\boldsymbol{\iota}}^{(k)}(t)$ with the data $\boldsymbol{v}(t)$ (equation III-29).

$$
\begin{gathered}
\hat{\boldsymbol{\xi}}^{(k)}(t) = \breve{\mathbf{T}}_{\xi v}^{(k)}\boldsymbol{v}(t) \\
\hat{\boldsymbol{\iota}}^{(k)}(t) = \breve{\mathbf{T}}_{\iota v}^{(k)}\boldsymbol{v}(t)
\end{gathered}
\tag{III-29}
$$

The inverse-operator $\breve{\mathbf{T}}_{\xi v}^{(k)}$ and relations for the residuals in (equation III-28) and (equation III-29) are given the residuals definition (equation III-30).

$$
\hat{\boldsymbol{\xi}}^{(k)}(t) = \boldsymbol{v}(t) - \mathbf{L}_{v\iota}\hat{\boldsymbol{\iota}}^{(k)}(t)
\tag{III-30}
$$

This definition (equation III-30) may be reformulated due to the relation $\hat{\boldsymbol{\iota}}^{(k)}(t) = \breve{\mathbf{T}}_{\iota v}^{(k)}\boldsymbol{v}(t)$ (equation III-13) that yields the following expression for $\breve{\mathbf{T}}_{\xi v}^{(k)}$ (equation III-31).

$$
\breve{\mathbf{T}}_{\xi v}^{(k)} = \mathbf{I}_{\mathrm{p}} - \mathbf{L}_{v\iota}\breve{\mathbf{T}}_{\iota v}^{(k)}
\tag{III-31}
$$

The expected $-log$ second-type likelihood $Q\left(\boldsymbol{\Omega},\widehat{\boldsymbol{\Omega}}^{(k)}\right)$ (equation III-26) may be represented in bimodal compact form (equation III-32) through auxiliary quantities denominated effective covariance-matrices for residuals $\breve{\boldsymbol{\Psi}}_{\xi\xi}^{(k)}$ and source cortical oscillations $\breve{\boldsymbol{\Psi}}_{\boldsymbol{\iota}}^{(k)}$.

$$
Q\left(\boldsymbol{\Omega},\widehat{\boldsymbol{\Omega}}^{(k)}\right) = -Tlog\left|\boldsymbol{\Theta}_{\xi\xi}\right| + Ttr\left(\breve{\boldsymbol{\Psi}}_{\xi\xi}^{(k)}\boldsymbol{\Theta}_{\xi\xi}\right) - Tlog\left|\boldsymbol{\Theta}_{\boldsymbol{\iota}}\right| + Ttr\left(\breve{\boldsymbol{\Psi}}_{\boldsymbol{\iota}}^{(k)}\boldsymbol{\Theta}_{\boldsymbol{\iota}}\right)
\tag{III-32}
$$

Where the effective covariance-matrix for source cortical oscillations $\breve{\boldsymbol{\Psi}}_{\boldsymbol{u}}^{(k)}$ (equation III-33) is given by the additive combination of the posterior $\breve{\boldsymbol{\Sigma}}_{\boldsymbol{u}}^{(k)}$ and sampled $\widehat{\boldsymbol{\Sigma}}_{\boldsymbol{u}}^{(k)}$ covariance-matrices for source activity. The effective covariance-matrix for residuals $\breve{\boldsymbol{\Psi}}_{\xi\xi}^{(k)}$ (equation III-34) is given by the additive combination of the posterior $\breve{\boldsymbol{\Sigma}}_{\xi\xi}^{(k)}$ and sampled $\widehat{\boldsymbol{\Sigma}}_{\xi\xi}^{(k)}$ covariance-matrices for the residuals.

$$\breve{\boldsymbol{\Psi}}_{\boldsymbol{u}}^{(k)} = \breve{\boldsymbol{\Sigma}}_{\boldsymbol{u}}^{(k)} + \widehat{\boldsymbol{\Sigma}}_{\boldsymbol{u}}^{(k)} \tag{III-33}$$

$$\breve{\boldsymbol{\Psi}}_{\xi\xi}^{(k)} = \breve{\boldsymbol{\Sigma}}_{\xi\xi}^{(k)} + \widehat{\boldsymbol{\Sigma}}_{\xi\xi}^{(k)} \tag{III-34}$$

The expression $p^{(k)}\left(\widehat{\boldsymbol{\Sigma}}_{\boldsymbol{vv}}\middle|\boldsymbol{\Omega}\right) = exp\left(Q\left(\boldsymbol{\Omega}, \boldsymbol{\Omega}^{(k)}\right)\middle|1\right)$ (equation III-35) based on $Q\left(\boldsymbol{\Omega}, \widehat{\boldsymbol{\Omega}}^{(k)}\right)$ (equation III-32) yields the factor, where the effective covariance-matrix for source cortical oscillations $\breve{\boldsymbol{\Psi}}_{\boldsymbol{u}}^{(k)}$ (equation III-33) and residuals $\breve{\boldsymbol{\Psi}}_{\xi\xi}^{(k)}$ (equation III-34) are a function of $\widehat{\boldsymbol{\Sigma}}_{\boldsymbol{vv}}$ through (equation III-28) the sampled covariance-matrices for source cortical oscillations $\widehat{\boldsymbol{\Sigma}}_{\boldsymbol{u}}^{(k)}$ and residuals $\widehat{\boldsymbol{\Sigma}}_{\xi\xi}^{(k)}$.

$$\begin{gathered}
p^{(k)}\left(\widehat{\boldsymbol{\Sigma}}_{\boldsymbol{vv}}\middle|\boldsymbol{\Omega}\right) = exp\left(-Tlog\left|\boldsymbol{\Theta}_{\xi\xi}\right| + Ttr\left(\breve{\boldsymbol{\Psi}}_{\xi\xi}^{(k)}\boldsymbol{\Theta}_{\xi\xi}\right)\middle|1\right) exp\left(-Tlog\left|\boldsymbol{\Theta}_{\boldsymbol{u}}\right| + Ttr\left(\breve{\boldsymbol{\Psi}}_{\boldsymbol{u}}^{(k)}\boldsymbol{\Theta}_{\boldsymbol{u}}\right)\middle|1\right) \\
p^{(k)}\left(\widehat{\boldsymbol{\Sigma}}_{\boldsymbol{vv}}\middle|\boldsymbol{\Omega}\right) = p^{(k)}\left(\widehat{\boldsymbol{\Sigma}}_{\boldsymbol{vv}}\middle|\boldsymbol{\Theta}_{\boldsymbol{u}}\right) p^{(k)}\left(\widehat{\boldsymbol{\Sigma}}_{\boldsymbol{vv}}\middle|\boldsymbol{\Theta}_{\xi\xi}\right) \\
p^{(k)}\left(\widehat{\boldsymbol{\Sigma}}_{\boldsymbol{vv}}\middle|\boldsymbol{\Theta}_{\boldsymbol{u}}\right) = W_{\mathrm{q}}^{\mathbb{C}}\left(\breve{\boldsymbol{\Psi}}_{\boldsymbol{u}}^{(k)}\middle|T^{-1}\boldsymbol{\Theta}_{\boldsymbol{u}}^{-1}, T\right) \\
p^{(k)}\left(\widehat{\boldsymbol{\Sigma}}_{\boldsymbol{vv}}\middle|\boldsymbol{\Theta}_{\xi\xi}\right) = W_{\mathrm{p}}^{\mathbb{C}}\left(\breve{\boldsymbol{\Psi}}_{\xi\xi}^{(k)}\middle|T^{-1}\boldsymbol{\Theta}_{\xi\xi}^{-1}, T\right)
\end{gathered} \tag{III-35}$$

# IV. HIGGS second-type maximum a posteriori and priors (Corollary to Lemma 1)

The second-type maximum a posteriori distribution $p^{(k)}\left(\boldsymbol{\Omega}\middle|\widehat{\boldsymbol{\Sigma}}_{\boldsymbol{vv}}\right)$, locally approximated at EM iterations, of the Hidden Gaussian Graphical (HIGGS) model hyperparameters, Lemma 1 (equation III-1), is thus expressed analytically by the Bayes theorem (equation IV-1). This Bayes theorem combines second-type likelihood iterated approximation $p^{(k)}\left(\widehat{\boldsymbol{\Sigma}}_{\boldsymbol{vv}}\middle|\boldsymbol{\Omega}\right)$, as a function of the observed cross-spectrum $\widehat{\boldsymbol{\Sigma}}_{\boldsymbol{vv}}$, and hyperparameters prior $p(\boldsymbol{\Omega})$.

$$p^{(k)}\left(\boldsymbol{\Omega}\middle|\widehat{\boldsymbol{\Sigma}}_{\boldsymbol{vv}}\right) = \frac{p^{(k)}\left(\widehat{\boldsymbol{\Sigma}}_{\boldsymbol{vv}}\middle|\boldsymbol{\Omega}\right)p(\boldsymbol{\Omega})}{p\left(\widehat{\boldsymbol{\Sigma}}_{\boldsymbol{vv}}\right)} \tag{IV-1}$$

As expressed by the expected $-log$ second-type likelihood from the EM integral operation $Q\left(\boldsymbol{\Omega}, \widehat{\boldsymbol{\Omega}}^{(k)}\right)$ the expected second-type likelihood above $p^{(k)}\left(\widehat{\boldsymbol{\Sigma}}_{\boldsymbol{vv}}\middle|\boldsymbol{\Omega}\right)$ is given by the Gibbs form (equation IV-2).

$$p^{(k)}\left(\widehat{\boldsymbol{\Sigma}}_{\boldsymbol{vv}}\middle|\boldsymbol{\Omega}\right) = exp\left(Q\left(\boldsymbol{\Omega}, \boldsymbol{\Omega}^{(k)}\right)\middle|1\right) \tag{IV-2}$$

The second-type maximum a posteriori estimator for hyperparameters $\widehat{\boldsymbol{\Omega}}^{(k+1)}$ may then be obtained maximizing the posterior distribution $p^{(k)}\left(\boldsymbol{\Omega}\middle|\widehat{\boldsymbol{\Sigma}}_{\boldsymbol{vv}}\right)$ that is given by the proportion (equation IV-3) of $p^{(k)}\left(\widehat{\boldsymbol{\Sigma}}_{\boldsymbol{vv}}\middle|\boldsymbol{\Omega}\right)$ as a function of $Q\left(\boldsymbol{\Omega}, \widehat{\boldsymbol{\Omega}}^{(k)}\right)$ and $p(\boldsymbol{\Omega})$.

$$\begin{gathered}
\widehat{\boldsymbol{\Omega}}^{(k+1)} = argmax_{\boldsymbol{\Omega}}\left(p^{(k)}\left(\boldsymbol{\Omega}\middle|\widehat{\boldsymbol{\Sigma}}_{\boldsymbol{vv}}\right)\right) \\
p^{(k)}\left(\boldsymbol{\Omega}\middle|\widehat{\boldsymbol{\Sigma}}_{\boldsymbol{vv}}\right) \propto exp\left(Q\left(\boldsymbol{\Omega}, \widehat{\boldsymbol{\Omega}}^{(k)}\right)\middle|1\right)p(\boldsymbol{\Omega})
\end{gathered} \tag{IV-3}$$

Or equivalently, this second-type maximum a posteriori estimator for hyperparameters $\widehat{\boldsymbol{\Omega}}^{(k+1)}$ (equation IV-3), may then be obtained minimizing the target function $\mathcal{L}^{(k)}(\boldsymbol{\Omega})$ (equation IV-4).

$$\begin{gathered}
\widehat{\boldsymbol{\Omega}}^{(k+1)} = argmin_{\boldsymbol{\Omega}}\left(\mathcal{L}^{(k)}(\boldsymbol{\Omega})\right) \\
\mathcal{L}^{(k)}\left(\boldsymbol{\Omega}(f)\right) = Q\left(\boldsymbol{\Omega}, \widehat{\boldsymbol{\Omega}}^{(k)}\right) - log\, p(\boldsymbol{\Omega})
\end{gathered} \tag{IV-4}$$

Because of the bimodal $-log$ Wishart form of $Q\left(\boldsymbol{\Omega}, \widehat{\boldsymbol{\Omega}}^{(k)}\right)$, a consequence of Lemma 1 (equation III-3), that implies the separability $\mathcal{L}^{(k)}(\boldsymbol{\Omega}) = \mathcal{L}^{(k)}(\boldsymbol{\Theta}_{\boldsymbol{u}}) + \mathcal{L}^{(k)}\left(\boldsymbol{\Theta}_{\xi\xi}\right)$ for the precision-matrices $\boldsymbol{\Omega} = \left\{\boldsymbol{\Theta}_{\boldsymbol{u}}, \boldsymbol{\Theta}_{\xi\xi}\right\}$ in (equation IV-4) and their a priori independency $p(\boldsymbol{\Omega}) = p(\boldsymbol{\Theta}_{\boldsymbol{u}})p\left(\boldsymbol{\Theta}_{\xi\xi}\right)$, the second-type maximum a posteriori estimators $\widehat{\boldsymbol{\Omega}}^{(k+1)} = \left\{\widehat{\boldsymbol{\Theta}}_{\boldsymbol{u}}^{(k+1)}, \widehat{\boldsymbol{\Theta}}_{\xi\xi}^{(k+1)}\right\}$ are given by (equation IV-5) for $\widehat{\boldsymbol{\Theta}}_{\boldsymbol{u}}^{(k+1)}$ and (equation IV-6) for $\widehat{\boldsymbol{\Theta}}_{\xi\xi}^{(k+1)}$.

$$\begin{gathered}
\widehat{\boldsymbol{\Theta}}_{\boldsymbol{u}}^{(k+1)} = argmin_{\boldsymbol{\Theta}_{\boldsymbol{u}}}\left(\mathcal{L}^{(k)}(\boldsymbol{\Theta}_{\boldsymbol{u}})\right) \\
\mathcal{L}^{(k)}(\boldsymbol{\Theta}_{\boldsymbol{u}}) = -Tlog\left|\boldsymbol{\Theta}_{\boldsymbol{u}}\right| + Ttr\left(\breve{\boldsymbol{\Psi}}_{\boldsymbol{u}}^{(k)}\boldsymbol{\Theta}_{\boldsymbol{u}}\right) - log\, p(\boldsymbol{\Theta}_{\boldsymbol{u}})
\end{gathered} \tag{IV-5}$$

$$\begin{gathered}
\widehat{\boldsymbol{\Theta}}_{\xi\xi}^{(k+1)} = argmin_{\boldsymbol{\Theta}_{\xi\xi}}\left(\mathcal{L}^{(k)}\left(\boldsymbol{\Theta}_{\xi\xi}\right)\right) \\
\mathcal{L}^{(k)}\left(\boldsymbol{\Theta}_{\xi\xi}\right) = -Tlog\left|\boldsymbol{\Theta}_{\xi\xi}\right| + Ttr\left(\breve{\boldsymbol{\Psi}}_{\xi\xi}^{(k)}\boldsymbol{\Theta}_{\xi\xi}\right) - log\, p\left(\boldsymbol{\Theta}_{\xi\xi}\right)
\end{gathered} \tag{IV-6}$$

The final expression of the target functions above for the sources $\mathcal{L}^{(k)}(\boldsymbol{\Theta}_{\boldsymbol{u}})$ (equation IV-5), and residuals $\mathcal{L}^{(k)}\left(\boldsymbol{\Theta}_{\xi\xi}\right)$, (equation IV-6), defining the solutions of iterated Gaussian Graphical Spectral (GGS) models, is due to the priors upon the precision-matrices $\boldsymbol{\Theta}_{\boldsymbol{u}}$ (equation IV-7) and $\boldsymbol{\Theta}_{\xi\xi}$ (equation IV-8).

$$p(\boldsymbol{\Theta}_{\boldsymbol{u}}) = exp\left(\Pi(\mathbf{A}_{\boldsymbol{u}} \odot \boldsymbol{\Theta}_{\boldsymbol{u}})\middle|\alpha_{\iota}T\right) \tag{IV-7}$$

$$p\left(\theta_{\xi}^{2}\right) = exp\left(\theta_{\xi}^{2}\middle|\alpha_{\xi}\mathrm{p}T\right) \tag{IV-8}$$

Where the prior structure of the GGS precision-matrices is represented wuith the ad-hoc $\mathbf{A}_{\boldsymbol{u}}$ for the sources (equation IV-6), and $\mathbf{A}_{\xi\xi}$ for the residuals (equation IV-7). The scale (regularization) parameters control the level of penalization exerted by these priors, given by $\alpha_\iota$ that impose $\Pi$ the L1 or L2 norm upon the multivariate $\mathbf{\Theta}_{\boldsymbol{u}}$, and $\alpha_\xi$ that may be interpreted as the residuals inferior limit represented by the scalar precison $\theta_\xi^2$ due to a reparameterization of the residuals precision-matrix $\mathbf{\Theta}_{\xi\xi} = \theta_\xi^2 \mathbf{A}_{\xi\xi}$. The estimators for the second-type maximum a posteriori are then expressed as (equation IV-9), for $\widehat{\mathbf{\Theta}}_{\boldsymbol{u}}^{(k+1)}$ based on the source target function (equation IV-5) and prior (equation IV-7), and (equation IV-10), for $\widehat{\mathbf{\Theta}}_{\xi\xi}^{(k+1)}$ based on the residual target function (equation IV-6) and prior (equation IV-8).

$$\widehat{\mathbf{\Theta}}_{\boldsymbol{u}}^{(k+1)} = argmin_{\mathbf{\Theta}_{\boldsymbol{u}}}\left\{-log|\mathbf{\Theta}_{\boldsymbol{u}}| + tr\left(\breve{\mathbf{\Psi}}_{\boldsymbol{u}}^{(k)}\mathbf{\Theta}_{\boldsymbol{u}}\right) + \alpha_\iota \Pi(\mathbf{A}_{\boldsymbol{u}} \odot \mathbf{\Theta}_{\boldsymbol{u}})\right\} \tag{IV-9}$$

$$\hat{\theta}_\xi^{2(k+1)} = argmin_{\theta_\xi^2}\left\{-\mathrm{p}log\left(\theta_\xi^2\right) + \theta_\xi^2 tr\left(\breve{\mathbf{\Psi}}_{\xi\xi}^{(k)}\mathbf{A}_{\xi\xi}\right) + \alpha_\xi \mathrm{p}\theta_\xi^2\right\} \tag{IV-10}$$

Note that the penalty $\Pi$ defines the particular flavor of the second-type maximum a posteriori estimates for the precision-matrix $\widehat{\mathbf{\Theta}}_{\boldsymbol{u}}^{(k+1)}$ of the hermitian graphical model of the sources placing in the target function (equation IV-9): "hgNaive" with $\Pi = 0$ (equation IV-11), "Ridge" with $\Pi = \|\cdot\|_2^2$ (equation IV-12), and "hgLASSO" with $\Pi = \|\cdot\|_1$ (equation IV-13).

$$\widehat{\mathbf{\Theta}}_{\boldsymbol{u}}^{(k+1)} = argmin_{\mathbf{\Theta}_{\boldsymbol{u}}}\left\{-log|\mathbf{\Theta}_{\boldsymbol{u}}| + tr\left(\breve{\mathbf{\Psi}}_{\boldsymbol{u}}^{(k)}\mathbf{\Theta}_{\boldsymbol{u}}\right)\right\} \tag{IV-11}$$

$$\widehat{\mathbf{\Theta}}_{\boldsymbol{u}}^{(k+1)} = argmin_{\mathbf{\Theta}_{\boldsymbol{u}}}\left\{-log|\mathbf{\Theta}_{\boldsymbol{u}}| + tr\left(\breve{\mathbf{\Psi}}_{\boldsymbol{u}}^{(k)}\mathbf{\Theta}_{\boldsymbol{u}}\right) + \alpha_\iota \|\mathbf{A}_{\boldsymbol{u}} \odot \mathbf{\Theta}_{\boldsymbol{u}}\|_2^2\right\} \tag{IV-12}$$

$$\widehat{\mathbf{\Theta}}_{\boldsymbol{u}}^{(k+1)} = argmin_{\mathbf{\Theta}_{\boldsymbol{u}}}\left\{-log|\mathbf{\Theta}_{\boldsymbol{u}}| + tr\left(\breve{\mathbf{\Psi}}_{\boldsymbol{u}}^{(k)}\mathbf{\Theta}_{\boldsymbol{u}}\right) + \alpha_\iota \|\mathbf{A}_{\boldsymbol{u}} \odot \mathbf{\Theta}_{\boldsymbol{u}}\|_1\right\} \tag{IV-13}$$

The second-type maximum a posteriori estimates for the scalar precision $\hat{\theta}_\xi^{2(k+1)}$, a reparameterization $\mathbf{\Theta}_{\xi\xi} = \theta_\xi^2 \mathbf{A}_{\xi\xi}$ for the hermitian graphical model of the residuals, can be obtained in terms of the zero (equation IV-14) due to the direct differentiation of the target function (equation IV-10).

$$\begin{gathered}\hat{\theta}_\xi^{2(k+1)} = 1/\left(tr\left(\breve{\mathbf{\Psi}}_{\xi\xi}^{(k)}(f)\mathbf{A}_{\xi\xi}\right)/\mathrm{p} + \alpha_\xi\right) \\ \hat{\theta}_\xi^{2(k+1)} = zero_{\theta_\xi^2}\left\{-\mathrm{p}\theta_\xi^{-2} + tr\left(\breve{\mathbf{\Psi}}_{\xi\xi}^{(k)}\mathbf{A}_{\xi\xi}\right) + \alpha_\xi \mathrm{p}\right\}\end{gathered} \tag{IV-14}$$

## V. Complex-valued Andrews and Mallows Lemma: Local Quadratic Approximation (LQA) of the hermitian graphical LASSO (hgLASSO) prior (Lemma 2)

***Lemma 2 (corollary of Andrews and Mallows for the hermitian graphical LASSO):*** *The measurable space of the random matrix $\mathbf{\Theta}$ with Gibbs probability density function $p(\mathbf{\Theta}) = exp(\Pi(\mathbf{A} \odot \mathbf{\Theta})|\alpha T)$, with penalization function of the hermitian graphical LASSO model ($\Pi = \|\cdot\|_1$), admits a hierarchical (conditional) representation in the product of measurable spaces for the random variables: 1) a conditional expectation $\mathbf{\Theta}|\mathbf{\Gamma}$ (equation V-1) with Gaussian improper probability density function (a density but not a probability), and 2) $\mathbf{\Gamma}$ (equation V-2) with Gamma probability density function.*

$$p(\mathbf{\Theta}|\mathbf{\Gamma}) \propto \prod_{i,j=1}^{\mathrm{q}} N_1(|\mathbf{\Theta}(i,j)||0, \mathbf{\Gamma}^2(i,j)/T) \tag{V-1}$$

$$p(\mathbf{\Theta}|\mathbf{\Gamma}) = \prod_{i,j}^{\mathrm{q}} Ga(\mathbf{\Gamma}^2(i,j)|1, T\alpha^2\mathbf{A}^2(i,j)/2) \tag{V-2}$$

***Proof of Lemma 2***

The hierarchical representation of the Gibbs prior with hermitian graphical LASSO exponent $\Pi(\mathbf{A} \odot \mathbf{\Theta}) = \|\mathbf{A} \odot \mathbf{\Theta}\|_1$, can be built on corollaries of the Andrews and Mallows Lemma (Andrews 1974), for the extension of real-valued Laplace probability density function to the real-valued or complex-valued matrix case, by considering improper density functions or simply more general measurable spaces. By the Andrews and Mallows Lemma in the real-valued Laplace, also for the real-valued or complex-valued case the integral representation holds (equation V-3).

$$e^{-\alpha|z|} = \int N_1(|z||0, x) Ga(x|1, \alpha^2/2) dx \tag{V-3}$$

The measurable space in which the variable $z|x$ is given by the Gaussian improper probability density function $p(z|x) = N_1(|z||0, x)$, where its variance $x$ has Gamma probability density function $p(x) = Ga(x|1, \alpha^2/2)$ (equation V-3). Therefore, the measure in the product space of $z|x$ and $\tau$ is has density represented as an improper product of Gaussian and Gamma probability density functions (equation V-4). We denominate this the generalization of Andrews and Mallows Lemma for real-valued or complex-valued Laplace probability density function.

$$p(z, x) \propto N_1(|z||0, x) Ga(x|1, \alpha^2/2) \tag{V-4}$$

The hermitian graphical LASSO Gibbs probability density function, that may be expressed for $\mathbf{\Theta}$ as (equation V-5), there is a priori independency of the precision-matrix elements $p(\mathbf{\Theta}) = \prod_{i,j=1}^{q} p\left(\mathbf{\Theta}(i,j)\right)$, and therefore $\mathbf{\Theta}(i,j)$ possesses a Gibbs probability density function as follows (equation V-6).

$$p(\mathbf{\Theta}) = exp(\|\mathbf{A} \odot \mathbf{\Theta}\|_1|\alpha T) \tag{V-5}$$

$$p\big(\mathbf{\Theta}(i,j)\big) = exp(|\mathbf{A}(i,j)\mathbf{\Theta}(i,j)||\alpha T) \tag{V-6}$$

If we apply the generalization of Andrews and Mallows Lemma for real-valued or complex-valued Laplace probability density function (equation V-4) to $exp(|\mathbf{A}(i,j)\mathbf{\Theta}(i,j)||\alpha T)$ in (equation V-6), by substituting $\alpha = \alpha T^{1/2}\mathbf{A}(i,j)$, $z = T^{1/2}\mathbf{\Theta}(i,j)$ and $x = \mathbf{\Gamma}^2(i,j)$ for the measure in the product space of the precision-matrix elements conditional expectation $\mathbf{\Theta}(i,j)|\mathbf{\Gamma}(i,j)$ and variance matrix elements $\mathbf{\Gamma}(i,j)$ we obtain (equation V-7).

$$p\big(\mathbf{\Theta}(i,j),\mathbf{\Gamma}(i,j)\big) \propto N_1^T(|\mathbf{\Theta}(i,j)||0,\mathbf{\Gamma}^2(i,j)/T)Ga^T(\mathbf{\Gamma}^2(i,j)|1,T\alpha^2\mathbf{A}^2(i,j)/2) \tag{V-7}$$

For $exp(\|\mathbf{A} \odot \mathbf{\Theta}\|_1|\alpha T)$ in (equation V-5) the measure in the product space of the precision-matrix conditional expectation $\mathbf{\Theta}|\mathbf{\Gamma}$ and variances matrix $\mathbf{\Gamma}$ we obtain (equation V-8).

$$p(\mathbf{\Theta},\mathbf{\Gamma}) = \prod_{i,j}^{q} N_1^T(|\mathbf{\Theta}(i,j)||0,\mathbf{\Gamma}^2(i,j)/T)Ga^T(\mathbf{\Gamma}^2(i,j)|1,T\,\alpha^2\mathbf{A}^2(i,j)/2)\ \blacksquare \tag{V-8}$$

**Remark:** In Lemma 2 we stablish a statistical equivalence between the Gibbs probability density function with argument in the hermitian graphical LASSO prior and a hierarchical improper representation through a Gaussian probability density function of the conditional expectation $\mathbf{\Theta}|\mathbf{\Gamma}$, and variances (weights) $\mathbf{\Gamma}$ with Gamma probability density function. Remarkably, this representation we are not using a probability density function in the strict mathematical sense but a measure density of a product of measurable spaces for $\mathbf{\Theta}|\mathbf{\Gamma}$ and $\mathbf{\Gamma}$.

# VI. Concavity of the first-type maximum a posteriori with hgLASSO LQA prior (Lemma 3)

***Lemma 3 (concavity of the Local Quadratic Approximation for the hermitian graphical LASSO prior):*** *The target function $\mathcal{L}(\mathbf{\Theta},\mathbf{\Gamma})$ ($-log$ of the improper probability density function for the measurable product space $\mathbf{\Theta}|\mathbf{\Gamma}\times\mathbf{\Gamma}$) may be represented as (equation VI-1) an it is strictly concave on the intercept of the region of positive definiteness of its arguments (equation VI-2) and the region comprehended by the set of inequalities (equation VI-3).*

$$\mathcal{L}(\mathbf{\Theta},\mathbf{\Gamma}) = -T\,log|\mathbf{\Theta}| \;\; + T\,tr(\mathbf{\Psi\Theta}) + \frac{T}{2}\|\mathbf{\Theta} \oslash \mathbf{\Gamma}\|_2^2 + \frac{1}{2}\sum_{i,j=1}^{q} log\,\mathbf{\Gamma}^2(i,j) + \frac{T\alpha^2}{2}\sum_{i,j=1}^{q}\mathbf{\Gamma}^2(i,j)\,\mathbf{A}^2(i,j) \tag{VI-1}$$

$$\{\mathbf{\Theta} \succcurlyeq 0, \mathbf{\Gamma} \succcurlyeq 0\} \tag{VI-2}$$

$$\{3T|\mathbf{\Theta}(i,j)|^2 - \mathbf{\Gamma}^2(i,j) + T\alpha^2\mathbf{A}^2(i,j)\mathbf{\Gamma}^4(i,j) \geq 0\}_{i,j=1}^{q} \tag{VI-3}$$

*Then, $\mathcal{L}(\mathbf{\Theta},\mathbf{\Gamma})$ has a minimum within these regions defined in (equation VI-2) and (equation VI-3) given by the intercept of the system (equation VI-4) and (equation VI-5).*

$$-\mathbf{\Theta}^{-1} + \mathbf{\Psi} + \mathbf{\Theta} \oslash \mathbf{\Gamma}^{\cdot 2} = \mathbf{0}_q \tag{VI-4}$$

$$\{-T|\mathbf{\Theta}(i,j)|^2 + \mathbf{\Gamma}^2(i,j) + T\alpha^2\mathbf{A}^2(i,j)\mathbf{\Gamma}^4(i,j) = 0\}_{i,j=1}^{q} \tag{VI-5}$$

***Proof of Lemma 3***

With the hierarchical representation of the hermitian graphical LASSO precision-matrix $\mathbf{\Theta}$ prior, extension of Andrews and Mallows Lemma (Lemma 2), we attain a modified target function $\mathcal{L}(\mathbf{\Theta},\mathbf{\Gamma})$ of the precision-matrix that builds in the combination of the sampled covariance-matrix $p(\mathbf{\Psi}|\mathbf{\Theta})$ Wishart likelihood dependent on the local quadratic approximation (improper probability density function of $\mathbf{\Theta}|\mathbf{\Gamma}\times\mathbf{\Gamma}$) (equation VI-1). Other terms are related to the normalization constant of the precision-matrix $\mathbf{\Theta}|\mathbf{\Gamma}$ Gaussian prior and the precision-matrix variances (weights) $\mathbf{\Gamma}$ Gamma prior. To build the target function $\mathcal{L}(\mathbf{\Theta},\mathbf{\Gamma})$ we may employ the Bayes theorem for the improper posterior probability density function $p(\mathbf{\Theta},\mathbf{\Gamma}|\mathbf{\Psi})$ based on the Wishart likelihood $p(\mathbf{\Psi}|\mathbf{\Theta})$ and improper probability density function for $p(\mathbf{\Theta}|\mathbf{\Gamma}\times\mathbf{\Gamma}) = p(\mathbf{\Theta}|\mathbf{\Gamma})p(\mathbf{\Gamma})$ (equation VI-6).

$$\begin{gathered} p(\mathbf{\Theta},\mathbf{\Gamma}|\mathbf{\Psi}) \propto p(\mathbf{\Psi}|\mathbf{\Theta})p(\mathbf{\Theta}|\mathbf{\Gamma})p(\mathbf{\Gamma}) \\ p(\mathbf{\Theta},\mathbf{\Gamma}|\mathbf{\Psi}) \propto W_q^{\mathbb{C}}(\mathbf{\Psi}|T^{-1}\mathbf{\Theta}^{-1},T)\prod_{i,j=1}^{q} N_1(|\mathbf{\Theta}(i,j)||0,\mathbf{\Gamma}^2(i,j)/T)Ga(\mathbf{\Gamma}^2(i,j)|1,T\,\alpha^2\mathbf{A}^2(i,j)/2) \end{gathered} \tag{VI-6}$$

We then define the target function $\mathcal{L}(\mathbf{\Theta},\mathbf{\Gamma}) = -\log p(\mathbf{\Theta},\mathbf{\Gamma}|\mathbf{\Psi})$ in (equation VI-6) as function dependent on the precision-matrix $\mathbf{\Theta}$ and precision-matrix variances $\mathbf{\Gamma}$ (equation VI-7),

$$\mathcal{L}(\mathbf{\Theta},\mathbf{\Gamma}) = -T\log|\mathbf{\Theta}| \;\; + T\,tr(\mathbf{\Psi\Theta}) - \sum_{i,j=1}^{q} log\,N_1(|\mathbf{\Theta}(i,j)||0,\mathbf{\Gamma}^2(i,j)/T) - \sum_{i,j=1}^{q} log\,Ga(\mathbf{\Gamma}^2(i,j)|1,T\,\alpha_\iota^2\mathbf{A}^2(i,j)/2) \tag{VI-7}$$

Given the definition of the univariate Gaussian probability density function in the third term of $\mathcal{L}(\mathbf{\Theta},\mathbf{\Gamma})$ (equation VI-7) we may express (equation VI-8), where we omit other normalization terms not dependent on $\mathbf{\Theta}$ or $\mathbf{\Gamma}$.

$$\sum_{i,j=1}^{q} log\,N_1(|\mathbf{\Theta}(i,j)||0,\mathbf{\Gamma}^2(i,j)/T) = -\frac{1}{2}\sum_{i,j=1}^{q} log\,\mathbf{\Gamma}^2(i,j) - \frac{T}{2}\sum_{i,j=1}^{q}|\mathbf{\Theta}(i,j)|^2/\mathbf{\Gamma}^2(i,j) \tag{VI-8}$$

Where above (equation VI-8), we may redefine the term involving $\mathbf{\Theta}$ as an L2 norm scaled by $\mathbf{\Gamma}$ (equation VI-9).

$$\|\mathbf{\Theta} \oslash \mathbf{\Gamma}\|_2^2 = \sum_{ij=1}^{q}|\mathbf{\Theta}(i,j)|^2/\mathbf{\Gamma}^2(i,j) \tag{VI-9}$$

Given the definition of the univariate Gamma probability density function in the fourth term of $\mathcal{L}(\mathbf{\Theta},\mathbf{\Gamma})$ (equation 6-7) we may express (equation VI-10), where we omit other normalization terms not dependent on $\mathbf{\Gamma}$.

$$\sum_{i,j=1}^{q} log\,Ga(\mathbf{\Gamma}^2(i,j)|1,T\alpha_\iota^2\mathbf{A}^2(i,j)/2) = -\frac{T\alpha^2}{2}\sum_{i,j=1}^{q}\mathbf{\Gamma}^2(i,j)\mathbf{A}^2(i,j) \tag{VI-10}$$

Then employing the expressions (equation VI-8) and (equation VI-10) and substituting the L2 norm (equation VI-9) we may express the target function $\mathcal{L}(\mathbf{\Theta},\mathbf{\Gamma})$ (equation VI-7) as a compact form of the precision-matrix $\mathbf{\Theta}$ precision-matrix, $\mathbf{\Gamma}$ and its prior (equation VI-11).

$$\mathcal{L}(\mathbf{\Theta},\mathbf{\Gamma}) = -T\log|\mathbf{\Theta}| \;\; + T\, tr(\mathbf{\Psi}\mathbf{\Theta}) + \frac{T}{2}\|\mathbf{\Theta}\oslash\mathbf{\Gamma}\|_2^2 + \frac{1}{2}\sum_{i,j=1}^{\mathrm{q}} log\,\mathbf{\Gamma}^2(i,j) + \frac{T\alpha^2}{2}\sum_{i,j=1}^{\mathrm{q}}\mathbf{\Gamma}^2(i,j)\,\mathbf{A}^2(i,j) \tag{VI-11}$$

The concavity of the target function $\mathcal{L}(\mathbf{\Theta},\mathbf{\Gamma})$ (equation VI-11) may be analyzed in terms of positive definiteness of the Hessian operator $\mathcal{H}$ computed by a block array of the second order matrix derivatives over the arguments $\mathbf{\Theta}$ and $\mathbf{\Gamma}$ (equation VI-12).

$$\mathcal{H}\big(\mathcal{L}(\mathbf{\Theta},\mathbf{\Gamma})\big) = \begin{bmatrix} \frac{\partial^2\mathcal{L}(\mathbf{\Theta},\mathbf{\Gamma})}{\partial\mathbf{\Theta}\partial\mathbf{\Theta}} & \frac{\partial^2\mathcal{L}(\mathbf{\Theta},\mathbf{\Gamma})}{\partial\mathbf{\Gamma}\partial\mathbf{\Theta}} \\ \frac{\partial^2\mathcal{L}(\mathbf{\Theta},\mathbf{\Gamma})}{\partial\mathbf{\Theta}\partial\mathbf{\Gamma}} & \frac{\partial^2\mathcal{L}(\mathbf{\Theta},\mathbf{\Gamma})}{\partial\mathbf{\Gamma}\partial\mathbf{\Gamma}} \end{bmatrix} \tag{VI-12}$$

The block hessian operator (equation VI-12) is positive definite if and only if its diagonal blocks are positive definite. Given the expression of $\mathcal{L}(\mathbf{\Theta},\mathbf{\Gamma})$ in (equation VI-11) we can deduce the structure of the hessian diagonal blocks from the following expressions (equation VI-13) and (equation VI-14).

$$\frac{\partial^2\mathcal{L}(\mathbf{\Theta},\mathbf{\Gamma})}{\partial\mathbf{\Theta}\partial\mathbf{\Theta}} = \frac{\partial^2}{\partial\mathbf{\Theta}\partial\mathbf{\Theta}}\left(-T\log|\mathbf{\Theta}| \;\; + T\, tr(\mathbf{\Psi}\mathbf{\Theta}) + \frac{T}{2}\|\mathbf{\Theta}\oslash\mathbf{\Gamma}\|_2^2\right) \tag{VI-13}$$

$$\frac{\partial^2\mathcal{L}(\mathbf{\Theta},\mathbf{\Gamma})}{\partial\mathbf{\Gamma}\partial\mathbf{\Gamma}} = \frac{\partial^2}{\partial\mathbf{\Gamma}\partial\mathbf{\Gamma}}\left(\frac{T}{2}\|\mathbf{\Theta}\oslash\mathbf{\Gamma}\|_2^2 + \frac{1}{2}\sum_{i,j=1}^{\mathrm{q}} log\,\mathbf{\Gamma}^2(i,j) + \frac{T\alpha^2}{2}\sum_{i,j=1}^{\mathrm{q}}\mathbf{\Gamma}^2(i,j)\,\mathbf{A}^2(i,j)\right) \tag{VI-14}$$

Effectuating the derivatives, $\partial\mathbf{\Theta}\partial\mathbf{\Theta}$ for (equation VI-13) above we obtain the expression in (equation VI-15) and $\partial\mathbf{\Gamma}\partial\mathbf{\Gamma}$ for (equation VI-14) above we obtain the expression in (equation VI-16).

$$\frac{\partial^2\mathcal{L}(\mathbf{\Theta},\mathbf{\Gamma})}{\partial\mathbf{\Theta}\partial\mathbf{\Theta}} = T\mathbf{\Theta}^{-1}\otimes\mathbf{\Theta}^{-1} \;\; + \frac{T}{2}\, diag\left(vect\big(\mathbf{1}_{\mathrm{q}}\oslash\mathbf{\Gamma}^{\cdot 2}\big)\right) \tag{VI-15}$$

$$\frac{\partial^2\mathcal{L}(\mathbf{\Theta},\mathbf{\Gamma})}{\partial\mathbf{\Gamma}\partial\mathbf{\Gamma}} = 3T diag\big(vect(|\mathbf{\Theta}|^{\cdot 2}\oslash\mathbf{\Gamma}^{\cdot 4})\big) - diag\left(vect\big(\mathbf{1}_{\mathrm{q}}\oslash\mathbf{\Gamma}^{\cdot 2}\big)\right) + T\alpha^2 diag\big(vect(\mathbf{A}^{\cdot 2})\big) \tag{VI-16}$$

For the derivative $\partial\mathbf{\Theta}\partial\mathbf{\Theta}$ in (equation VI-15) within the region of positive definiteness of $\mathbf{\Theta}\succcurlyeq 0$ the first term (Kronecker product obtained from $\log|\mathbf{\Theta}|$) is also positive definite $\mathbf{\Theta}^{-1}\otimes\mathbf{\Theta}^{-1}\succcurlyeq 0$, and since the elements $\mathbf{\Gamma}(i,j) > 0$ are positive making the second term positive definite anywhere, then the whole expression (equation VI-15) is positive definite within the region of positive definiteness of $\mathbf{\Theta}\succcurlyeq 0$. The derivative $\partial\mathbf{\Gamma}\partial\mathbf{\Gamma}$ in (equation VI-16) is positive definite within the region of positive derivative elements $\partial\mathbf{\Gamma}(i,j)\partial\mathbf{\Gamma}(i,j)$ (equation VI-17) yielding the following set of inequalities (equation 6-18).

$$\frac{\partial^2\mathcal{L}(\mathbf{\Theta},\mathbf{\Gamma})}{\partial\mathbf{\Gamma}(i,j)\partial\mathbf{\Gamma}(i,j)} = \frac{\partial^2}{\partial\mathbf{\Gamma}(i,j)\partial\mathbf{\Gamma}(i,j)}\left(\frac{T}{2}\frac{|\mathbf{\Theta}(i,j)|^2}{\mathbf{\Gamma}^2(i,j)} + \frac{1}{2}log\,\mathbf{\Gamma}^2(i,j) + \frac{T\alpha^2}{2}\mathbf{\Gamma}^2(i,j)\mathbf{A}^2(i,j)\right) = 3T\frac{|\mathbf{\Theta}(i,j)|^2}{\mathbf{\Gamma}^4(i,j)} - \frac{1}{\mathbf{\Gamma}^2(i,j)} + T\alpha^2\mathbf{A}^2(i,j) \tag{VI-17}$$

$$\left\{3T\frac{|\mathbf{\Theta}(i,j)|^2}{\mathbf{\Gamma}^4(i,j)} + -\frac{1}{\mathbf{\Gamma}^2(i,j)} + T\alpha^2\mathbf{A}^2(i,j) > 0\right\}_{i,j=1}^{\mathrm{q}} \tag{VI-18}$$

Therefore, solutions for the minimum of $\mathcal{L}(\mathbf{\Theta},\mathbf{\Gamma})$ in (equation VI-11) may be obtained by means of the derivatives $\partial\mathbf{\Theta}$ (equation VI-19) and $\partial\mathbf{\Gamma}$ (equation VI-20) in the region of the product space $\mathbf{\Theta}\times\mathbf{\Gamma}$.

$$\frac{\partial\mathcal{L}(\mathbf{\Theta},\mathbf{\Gamma})}{\partial\mathbf{\Theta}} = \frac{\partial}{\partial\mathbf{\Theta}}\left(-T\log|\mathbf{\Theta}| \;\; + T\, tr(\mathbf{\Psi}\mathbf{\Theta}) + \frac{T}{2}\|\mathbf{\Theta}\oslash\mathbf{\Gamma}\|_2^2\right) \tag{VI-19}$$

$$\frac{\partial\mathcal{L}(\mathbf{\Theta},\mathbf{\Gamma})}{\partial\mathbf{\Gamma}} = \frac{\partial}{\partial\mathbf{\Gamma}}\left(\frac{T}{2}\|\mathbf{\Theta}\oslash\mathbf{\Gamma}\|_2^2 + \frac{1}{2}\sum_{i,j=1}^{\mathrm{q}} log\,\mathbf{\Gamma}^2(i,j) + \frac{T\alpha^2}{2}\sum_{i,j=1}^{\mathrm{q}}\mathbf{\Gamma}^2(i,j)\,\mathbf{A}^2(i,j)\right) \tag{VI-20}$$

Effectuating the derivatives, $\partial\mathbf{\Theta}$ for (equation VI-19) above we obtain the expression in (equation VI-21) and $\partial\mathbf{\Gamma}$ for (equation VI-20) above we obtain the expression in (equation VI-22).

$$\frac{\partial\mathcal{L}(\mathbf{\Theta},\mathbf{\Gamma})}{\partial\mathbf{\Theta}} = -T(\mathbf{\Theta}^{-1})^{\mathcal{T}} + T\mathbf{\Psi}^{\mathcal{T}} + T\mathbf{\Theta}^{\mathcal{T}}\oslash\mathbf{\Gamma}^{\mathbf{2}} \tag{VI-21}$$

$$\frac{\partial\mathcal{L}(\mathbf{\Theta},\mathbf{\Gamma})}{\partial\mathbf{\Gamma}} = -T diag\big(vect(|\mathbf{\Theta}|^{\cdot 2}\oslash\mathbf{\Gamma}^{\cdot 3})\big) + diag\left(vect\big(\mathbf{1}_{\mathrm{q}}\oslash\mathbf{\Gamma}\big)\right) + T\alpha^2 diag\big(vect(\mathbf{\Gamma}\odot\mathbf{A}^{\cdot\mathbf{2}})\big) \tag{VI-22}$$

It can be checked that for the derivative $\partial\mathbf{\Gamma}$ (equation VI-22) the zero also makes positive the hessian $\partial\mathbf{\Gamma}\partial\mathbf{\Gamma}$ (equation 2-16). By substituting this zero given in (equation VI-23), in the hessian (equation 2-16) this yields a positive-definite expression (equation VI-24) anywhere the elements $\mathbf{\Gamma}(i,j) > 0$ are positive.

$$-T diag\big(vect(|\mathbf{\Theta}|^{\cdot 2}\oslash\mathbf{\Gamma}^{\cdot 4})\big) + diag\left(vect\big(\mathbf{1}_{\mathrm{q}}\oslash\mathbf{\Gamma}^{\cdot\mathbf{2}}\big)\right) + T\alpha^2 diag\big(vect(\mathbf{A}^{\cdot 2})\big) = 0 \tag{VI-23}$$

$$\frac{\partial^2\mathcal{L}(\mathbf{\Theta},\mathbf{\Gamma})}{\partial\mathbf{\Gamma}\partial\mathbf{\Gamma}} = 2diag\left(vect\big(\mathbf{1}_{\mathrm{q}}\oslash\mathbf{\Gamma}\big)\right) + 4T\alpha^2 diag\big(vect(\mathbf{A}^{\cdot\mathbf{2}})\big) \succcurlyeq 0 \tag{VI-24}$$

From these derivatives in (equation VI-21) and (equation VI-22) the target function $\mathcal{L}(\mathbf{\Theta},\mathbf{\Gamma})$ and possess a minimum in the point due to the systems in (equation VI-25) and (equation VI-26).

$$-\mathbf{\Theta}^{-1} + \mathbf{\Psi} + \mathbf{\Theta}\oslash\mathbf{\Gamma}^{\mathbf{2}} = \mathbf{0}_{\mathrm{q}} \tag{VI-25}$$

$$\{-T|\mathbf{\Theta}(i,j)|^2 + \mathbf{\Gamma}^2(i,j) + T\alpha^2\mathbf{A}^2(i,j)\mathbf{\Gamma}^4(i,j) = 0\}_{i,j=1}^{\mathrm{q}} \tag{VI-26}$$

This minimum is unique due to the convexity within the region defined in the product space $\mathbf{\Theta}\times\mathbf{\Gamma}$ since the following inequalities are fulfilled (equation VI-27) and (equation VI-28).

$$\{\boldsymbol{\Theta} \succcurlyeq 0, \boldsymbol{\Gamma} \succcurlyeq 0\} \tag{VI-27}$$

$$\left\{3T\frac{|\boldsymbol{\Theta}(i,j)|^2}{\boldsymbol{\Gamma}^4(i,j)} + -\frac{1}{\boldsymbol{\Gamma}^2(i,j)} + T\alpha^2\mathbf{A}^2(i,j) > 0\right\}_{i,j=1}^{\mathrm{q}} \tag{VI-28}$$

For every element $\boldsymbol{\Gamma}(i,j)$ of the fourth order system (equation VI-26) the minimum is given by the discriminant formula, roots of second order polynomials (equation VI-29). In matrix form this solution is expressed as (equation VI-30).

$$\boldsymbol{\Gamma}(i,j) = \left(\left(-1 + (1 + 4T^2\alpha^2\mathbf{A}^2(i,j)|\boldsymbol{\Theta}(i,j)|^2)^{\frac{1}{2}}\right)/\left(2T\alpha^2\mathbf{A}^2(i,j)\right)\right)^{\frac{1}{2}} \tag{VI-29}$$

$$\boldsymbol{\Gamma} = \left(\left(-\mathbf{1}_{\mathrm{q}} + \left(\mathbf{1}_{\mathrm{q}} + 4T^2\alpha^2\mathbf{A}^{\cdot 2} \odot |\boldsymbol{\Theta}|^{\cdot 2}\right)^{\cdot\frac{1}{2}}\right) \oslash \left(2T\alpha^2\mathbf{A}^{\cdot 2}\right)\right)^{\cdot\frac{1}{2}} \tag{VI-30}$$

Given the positive definiteness of $\mathbf{A}$ and $\boldsymbol{\Theta}$, it can also be deduced that the solution $\boldsymbol{\Gamma}$ (equation VI-30) is positive definite by following the steps in proposition 3 of Lemma 4. ■

According to this local quadratic approximation, we can define a new random matrix through the scaling transformation $\widetilde{\boldsymbol{\Theta}} = \boldsymbol{\Theta} \oslash \boldsymbol{\Gamma}$ (standard precision-matrix), so that its prior is a Gibbs probability density function of the squared L2 norm (equation VI-31).

$$p(\widetilde{\boldsymbol{\Theta}}) \propto e^{-\frac{T}{2}\|\widetilde{\boldsymbol{\Theta}}\|_2^2} \tag{VI-31}$$

For the solution based of this standard precision-matrix we also standardize the sampled covariance-matrix due to the transformation $\widetilde{\boldsymbol{\Psi}} = (\boldsymbol{\Psi}^{-1} \oslash \boldsymbol{\Gamma})^{-1}$ that keeps the stochastic properties of the Wishart distribution when conditioned to $\widetilde{\boldsymbol{\Theta}}$.

# VII. Standardization of the first-type likelihood with hgLASSO LQA prior (Lemma 4)

***Lemma 4 (standardization of the Wishart distribution):*** *Let $\boldsymbol{\Psi}$ and $\widetilde{\boldsymbol{\Psi}}$ (referred as standardized sampled covariance matrix) be $(\mathrm{q} \times \mathrm{q})$ hermitian random matrices with complex-valued Wishart probability density functions $W_{\mathrm{q}}^{\mathbb{C}}$ defined in (equation VII-1) and (equation VII-2). The Wishart $W_{\mathrm{q}}^{\mathbb{C}}$ has $T$ degrees of freedom and positive definite hermitic scale matrices, $\boldsymbol{\Sigma}$ for (equation VII-1), and $\widetilde{\boldsymbol{\Sigma}} = (\boldsymbol{\Sigma}^{-1} \oslash \boldsymbol{\Gamma})^{-1}$ (referred as standardized scale matrix) for (equation VII-2), with $\boldsymbol{\Gamma}$ being a $(q \times q)$ positive definite matrix of positive weights.*

$$p(\boldsymbol{\Psi}) = W_{\mathrm{q}}^{\mathbb{C}}(\boldsymbol{\Psi}|\,\boldsymbol{\Sigma}, T) \propto |\boldsymbol{\Psi}|^{(T-\mathrm{q})}|\boldsymbol{\Sigma}|^{-T}e^{-tr(\boldsymbol{\Sigma}^{-1}\boldsymbol{\Psi})} \tag{VII-1}$$

$$p(\widetilde{\boldsymbol{\Psi}}) = W_{\mathrm{q}}^{\mathbb{C}}(\widetilde{\boldsymbol{\Psi}}|\,\widetilde{\boldsymbol{\Sigma}}, T) \propto |\widetilde{\boldsymbol{\Psi}}|^{(T-\mathrm{q})}|\widetilde{\boldsymbol{\Sigma}}|^{-T}e^{-tr(\widetilde{\boldsymbol{\Sigma}}^{-1}\widetilde{\boldsymbol{\Sigma}})} \tag{VII-2}$$

*Then for $\boldsymbol{\Phi}$ (referred de-standardization of $\widetilde{\boldsymbol{\Psi}}$) defined by the relationship to the standard sampled covariance-matrix $\widetilde{\boldsymbol{\Psi}} = (\boldsymbol{\Phi}^{-1} \oslash \boldsymbol{\Gamma})^{-1}$ (or $\boldsymbol{\Phi} = \left(\widetilde{\boldsymbol{\Psi}}^{-1} \odot \boldsymbol{\Gamma}\right)^{-1}$), it can be verified:*

*a) All entries $\boldsymbol{\Phi}^{-1}(i,j)$ of the de-standardization inverse $\boldsymbol{\Phi}^{-1}$ keep the Wishart probability density function independency property: they are stochastically independent among them and from the set $\{\boldsymbol{\Phi}(i',j'): (i',j') \neq (i,j)\}$.*

*b) All entries $\boldsymbol{\Phi}^{-1}(i,j)$ of the de-standardization inverse $\boldsymbol{\Phi}^{-1}$ keep the Wishart marginal probability density function: complex-valued inverse Gamma probability density function with parameter of shape $(T - q + 1)$ and scale $\boldsymbol{\Sigma}^{-1}(i,j)$.*

***Proof of Lemma 4***

**Proposition 1**

If $\boldsymbol{\Psi}$ is a $(q \times q)$ complex-valued random matrix with Wishart probability density function (equation VII-1) of $T$ degrees of freedom and positive definite scale matrix $\boldsymbol{\Sigma}$. Then the random matrix obtained from the consecutive rows and columns permutation operations, has also Wishart probability density function of $T$ degrees of freedom and positive definite scale matrix (equation VII-3).

$$\left(\mathbf{P}_{j\leftrightarrow j'}^{col}\mathbf{P}_{i\leftrightarrow i'}^{row}\boldsymbol{\Sigma}^{-1}\right)^{-1} \tag{VII-3}$$

**Proof of Proposition 1:**

For the Wishart probability density function of the random matrix $\boldsymbol{\Psi}$ that can be expressed as (equation VII-4) it holds that the determinants and trace terms are invariant to consecutive rows and columns permutation operations (equation VII-3), represented as the identities (equation VII-5), (equation VII-6), and (equation VII-7).

$$W_{\mathrm{q}}^{\mathbb{C}}(\boldsymbol{\Psi}|\boldsymbol{\Sigma}, T) \propto |\boldsymbol{\Psi}|^{(T-\mathrm{q})}|\boldsymbol{\Sigma}|^{-T}e^{-tr(\boldsymbol{\Sigma}^{-1}\boldsymbol{\Psi})} \tag{VII-4}$$

$$|\boldsymbol{\Sigma}^{-1}| = \left|\mathbf{P}_{j\leftrightarrow j'}^{col}\mathbf{P}_{i\leftrightarrow i'}^{row}\boldsymbol{\Sigma}^{-1}\right| \tag{VII-5}$$

$$|\boldsymbol{\Psi}| = \left|\mathbf{P}_{j\leftrightarrow j'}^{col}\mathbf{P}_{i\leftrightarrow i'}^{row}\boldsymbol{\Psi}\right| \tag{VII-6}$$

$$tr(\boldsymbol{\Sigma}^{-1}\boldsymbol{\Psi}) = tr\left(\mathbf{P}_{j\leftrightarrow j'}^{col}\mathbf{P}_{i\leftrightarrow i'}^{row}\boldsymbol{\Sigma}^{-1}\mathbf{P}_{j\leftrightarrow j'}^{col}\mathbf{P}_{i\leftrightarrow i'}^{row}\boldsymbol{\Psi}\right) \tag{VII-7}$$

Due to identities (equation VII-5), (equation VII-6), and (equation VII-7) the Wishart $W_q^{\mathbb{C}}$ probability density function values for the original and permuted random matrices are identic (equation VII-8).

$$W_{\mathrm{q}}^{\mathbb{C}}(\boldsymbol{\Psi}|\boldsymbol{\Sigma},\mathrm{m}) = W_{\mathrm{q}}^{\mathbb{C}}\left(\mathbf{P}_{j\leftrightarrow j'}^{col}\mathbf{P}_{i\leftrightarrow i'}^{row}\boldsymbol{\Psi}\middle|\left(\mathbf{P}_{j\leftrightarrow j'}^{col}\mathbf{P}_{i\leftrightarrow i'}^{row}\boldsymbol{\Sigma}^{-1}\right)^{-1},T\right) \tag{VII-8}$$

By definition, the Wishart is a joint probability density function of the set of random matrix entries in $\boldsymbol{\Psi}$ (equation VII-9). For permutations this joint probability density function would be invariant, as represented by same Wishart (equation VII-10).

$$p(\{\boldsymbol{\Psi}(i,j): i,j = 1 \dots \mathrm{q}\}) = p(\boldsymbol{\Psi}) = W_{\mathrm{q}}^{\mathbb{C}}(\boldsymbol{\Psi}|\boldsymbol{\Sigma},T) \tag{VII-9}$$

$$p\left(\left\{\mathbf{P}_{j\leftrightarrow j'}^{col}\mathbf{P}_{i\leftrightarrow i'}^{row}\boldsymbol{\Psi}(i,j): i,j = 1 \dots \mathrm{q}\right\}\right) = p\left(\mathbf{P}_{j\leftrightarrow j'}^{col}\mathbf{P}_{i\leftrightarrow i'}^{row}\boldsymbol{\Psi}\right) = W_{\mathrm{q}}^{\mathbb{C}}(\boldsymbol{\Psi}|\boldsymbol{\Sigma},T) \tag{VII-10}$$

From the identity between Wishart probability density functions (equation VII-8), it is clear that the permuted random matrix follows also a Wishart probability density function with permuted scale matrix (equation VII-11).

$$p\left(\mathbf{P}_{j\leftrightarrow j'}^{col}\mathbf{P}_{i\leftrightarrow i'}^{row}\boldsymbol{\Psi}\right) = W_{\mathrm{q}}^{\mathbb{C}}\left(\mathbf{P}_{j\leftrightarrow j'}^{col}\mathbf{P}_{i\leftrightarrow i'}^{row}\boldsymbol{\Psi}\middle|\left(\mathbf{P}_{j\leftrightarrow j'}^{col}\mathbf{P}_{i\leftrightarrow i'}^{row}\boldsymbol{\Sigma}^{-1}\right)^{-1},T\right) \blacksquare \tag{VII-11}$$

An element $\boldsymbol{\Psi}(i,j)$ of a random matrix $\boldsymbol{\Psi}$ with Wishart probability density function (equation VII-4), the conditional probability density function regarding the remaining elements $\left\{\overline{\boldsymbol{\Psi}(i,j)}\right\} = \{\boldsymbol{\Psi}(i',j'): (i',j') \neq (i,j)\}$ may be expressed as (equation VII-12).

$$p\left(\boldsymbol{\Psi}(i,j)\middle|\left\{\overline{\boldsymbol{\Psi}(i,j)}\right\}\right) \propto |\boldsymbol{\Psi}|^{(T-\mathrm{q})}e^{-tr(\boldsymbol{\Sigma}^{-1}\boldsymbol{\Psi})} \tag{VII-12}$$

An explicit form of this conditional probability density function for $\boldsymbol{\Psi}(i,j)$ (equation VII-12) may be obtained by applying the consecutive permutation operations (equation VII-13) and then consider the conditional conditional probability density function of the first element (equation VII-14).

$$\mathbf{P}_{j\leftrightarrow 1}^{col}\mathbf{P}_{i\leftrightarrow 1}^{row}\boldsymbol{\Psi} \tag{VII-13}$$

$$p\left(\mathbf{P}_{j\leftrightarrow 1}^{col}\mathbf{P}_{i\leftrightarrow 1}^{row}\boldsymbol{\Psi}(1,1)\middle|\left\{\overline{\mathbf{P}_{j\leftrightarrow 1}^{col}\mathbf{P}_{i\leftrightarrow 1}^{row}\boldsymbol{\Psi}(1,1)}\right\}\right) \propto \left|\mathbf{P}_{j\leftrightarrow 1}^{col}\mathbf{P}_{i\leftrightarrow 1}^{row}\boldsymbol{\Psi}\right|^{(T-\mathrm{q})}e^{-tr\left(\mathbf{P}_{j\leftrightarrow 1}^{col}\mathbf{P}_{i\leftrightarrow 1}^{row}\boldsymbol{\Sigma}^{-1}\mathbf{P}_{j\leftrightarrow 1}^{col}\mathbf{P}_{i\leftrightarrow 1}^{row}\boldsymbol{\Psi}\right)} \tag{VII-14}$$

Due to Proposition 1 the probability density function of the permuted random matrix (equation VII-13) is also Wishart, and therefore, for the element (equation VII-14) the conditional probability density function can be expressed as (equation VII-11). Without losing generality we can consider first the element in the original random matrix $\boldsymbol{\Psi}(1,1)$ (equaion VII-15) and then this result will also apply for all $\boldsymbol{\Psi}(i,j)$ (equation VII-12).

$$p\left(\boldsymbol{\Psi}(1,1)\middle|\left\{\overline{\boldsymbol{\Psi}(1,1)}\right\}\right) \propto |\boldsymbol{\Psi}|^{(T-\mathrm{q})}e^{-tr(\boldsymbol{\Sigma}^{-1}\boldsymbol{\Psi})} \tag{VII-15}$$

Partitioning the random matrix $\boldsymbol{\Psi}$ and $\boldsymbol{\Sigma}^{-1}$ into the following block structures (equation VII-16) and (equation VII-17) we can find a simplified expression of the conditional probability density function (equation VII-15).

$$\boldsymbol{\Psi} = \begin{pmatrix} \boldsymbol{\Psi}(1,1) & \boldsymbol{\Psi}(1,2) \\ \boldsymbol{\Psi}(2,1) & \boldsymbol{\Psi}(2,2) \end{pmatrix} \tag{VII-16}$$

$$\boldsymbol{\Sigma}^{-1} = \begin{pmatrix} \boldsymbol{\Sigma}^{-1}(1,1) & \boldsymbol{\Sigma}^{-1}(1,2) \\ \boldsymbol{\Sigma}^{-1}(2,1) & \boldsymbol{\Sigma}^{-1}(2,2) \end{pmatrix} \tag{VII-17}$$

From this block structures (equation VII-16) and (equation VII-17) we can deduce the form of the term in the exponent $tr(\boldsymbol{\Sigma}^{-1}\boldsymbol{\Psi})$ (equation VII-15) due to the product $\boldsymbol{\Sigma}^{-1}\boldsymbol{\Psi}$ (equation VII-18) that yields an expression sparable for $\boldsymbol{\Psi}(1,1)$ (equation VII-19). From the block structure (equation VII-16) the detrminant $|\boldsymbol{\Psi}|$ (equation VII-15) also yields an expression sparable for $\boldsymbol{\Psi}(1,1)$ (equation VII-20).

$$\boldsymbol{\Sigma}^{-1}\boldsymbol{\Psi} = \begin{pmatrix} \boldsymbol{\Sigma}^{-1}(1,1)\boldsymbol{\Psi}(1,1) + \boldsymbol{\Sigma}^{-1}(1,2)\boldsymbol{\Psi}(2,1) & \boldsymbol{\Sigma}^{-1}(1,1)\boldsymbol{\Psi}(1,2) + \boldsymbol{\Sigma}^{-1}(1,2)\boldsymbol{\Psi}(2,2) \\ \boldsymbol{\Sigma}^{-1}(2,1)\boldsymbol{\Psi}(1,1) + \boldsymbol{\Sigma}^{-1}(2,2)\boldsymbol{\Psi}(2,1) & \boldsymbol{\Sigma}^{-1}(2,1)\boldsymbol{\Psi}(1,2) + \boldsymbol{\Sigma}^{-1}(2,2)\boldsymbol{\Psi}(2,2) \end{pmatrix} \tag{VII-18}$$

$$tr(\boldsymbol{\Sigma}^{-1}\boldsymbol{\Psi}) = \boldsymbol{\Sigma}^{-1}(1,1)\boldsymbol{\Psi}(1,1) + \boldsymbol{\Sigma}^{-1}(1,2)\boldsymbol{\Psi}(2,1) + tr\left(\boldsymbol{\Sigma}^{-1}(2,1)\boldsymbol{\Psi}(1,2) + \boldsymbol{\Sigma}^{-1}(2,2)\boldsymbol{\Psi}(2,2)\right) \tag{VII-19}$$

$$|\boldsymbol{\Psi}| = |\boldsymbol{\Psi}(2,2)|\left(\boldsymbol{\Psi}(1,1) - \boldsymbol{\Psi}(1,2)\left(\boldsymbol{\Psi}(2,2)\right)^{-1}\boldsymbol{\Psi}(2,1)\right) \tag{VII-20}$$

We therefore may express the conditional probability density function as (equation VII-21) by completing the exponent with the term $\boldsymbol{\Psi}(1,2)\left(\boldsymbol{\Psi}(2,2)\right)^{-1}\boldsymbol{\Psi}(2,1)$.

$$p\left(\boldsymbol{\Psi}(1,1)\middle|\left\{\overline{\boldsymbol{\Psi}(1,1)}\right\}\right) \propto \left(\boldsymbol{\Psi}(1,1) - \boldsymbol{\Psi}(1,2)\left(\boldsymbol{\Psi}(2,2)\right)^{-1}\boldsymbol{\Psi}(2,1)\right)^{(T-\mathrm{q})} e^{-\boldsymbol{\Sigma}^{-1}(1,1)\left(\boldsymbol{\Psi}(1,1)-\boldsymbol{\Psi}(1,2)\left(\boldsymbol{\Psi}(2,2)\right)^{-1}\boldsymbol{\Psi}(2,1)\right)} \tag{VII-21}$$

Due to the structure of the block inverse for $\boldsymbol{\Psi}^{-1}$ (equation VII-22), the argument of the conditional probability density function above (equation VII-21) is the element $\boldsymbol{\Psi}^{-1}(1,1)$ of this block inverse $\boldsymbol{\Psi}^{-1}$ (equation VII-23).

$$\boldsymbol{\Psi}^{-1} = \begin{pmatrix} \frac{1}{\boldsymbol{\Psi}(1,1)-\boldsymbol{\Psi}(1,2)\left(\boldsymbol{\Psi}(2,2)\right)^{-1}\boldsymbol{\Psi}(2,1)} & -\frac{\left(\boldsymbol{\Psi}(2,2)\right)^{-1}\boldsymbol{\Psi}(2,1)}{\boldsymbol{\Psi}(1,1)-\boldsymbol{\Psi}(1,2)\left(\boldsymbol{\Psi}(2,2)\right)^{-1}\boldsymbol{\Psi}(2,1)} \\ -\frac{\left(\boldsymbol{\Psi}(2,2)\right)^{-1}\boldsymbol{\Psi}(2,1)}{\boldsymbol{\Psi}(1,1)-\boldsymbol{\Psi}(1,2)\left(\boldsymbol{\Psi}(2,2)\right)^{-1}\boldsymbol{\Psi}(2,1)} & \left(\boldsymbol{\Psi}(2,2)\right)^{-1} + \frac{\left(\boldsymbol{\Psi}(2,2)\right)^{-1}\boldsymbol{\Psi}(2,1)\boldsymbol{\Psi}(1,2)\left(\boldsymbol{\Psi}(2,2)\right)^{-1}}{\boldsymbol{\Psi}(1,1)-\boldsymbol{\Psi}(1,2)\left(\boldsymbol{\Psi}(2,2)\right)^{-1}\boldsymbol{\Psi}(2,1)} \end{pmatrix} \tag{VII-22}$$

$$\boldsymbol{\Psi}^{-1}(1,1) = \frac{1}{\boldsymbol{\Psi}(1,1)-\boldsymbol{\Psi}(1,2)\left(\boldsymbol{\Psi}(2,2)\right)^{-1}\boldsymbol{\Psi}(2,1)} \tag{VII-23}$$

From these relationships (equation VII-22) and (equation VII-23) it is deduced that any element in the random matrix $\boldsymbol{\Psi}^{-1}$, a random variable $\boldsymbol{\Psi}^{-1}(1,1)$ is conditionally independent from $\left\{\overline{\boldsymbol{\Psi}(1,1)}\right\}$ and has complex-valued inverse Gamma probability density function with shape parameter $(T-\mathrm{q})$ and scale parameter $\boldsymbol{\Sigma}^{-1}(1,1)$ (equation VII-24).

$$p\left(\boldsymbol{\Psi}^{-\mathbf{1}}(1,1)\right) \propto \left(\boldsymbol{\Psi}^{-\mathbf{1}}(1,1)\right)^{-(T-\mathrm{q})} e^{-\frac{\boldsymbol{\Sigma}^{-\mathbf{1}}(1,1)}{\boldsymbol{\Psi}^{-\mathbf{1}}(1,1)}} \quad \text{(VII-24)}$$

Now we define, 1) a random matrix $\widetilde{\boldsymbol{\Psi}}$ (referred as standardized sampled covariance matrix) with Wishart probability density function that can be expressed as (equation VII-25) with scale matrix $\widetilde{\boldsymbol{\Sigma}}$, defined by the tranformation of the element $\boldsymbol{\Sigma}^{-1}(1,1)$ from the scale matrix for the Wishart probability density function of the random matrix $\boldsymbol{\Psi}$ (equation VII-4), by a positive number $\boldsymbol{\Gamma}(1,1)$ (equation VII-26), and 2) another random matrix $\boldsymbol{\Phi}$ through the relationsship with the standardized sampled covariance matrix $\widetilde{\boldsymbol{\Psi}}$ (equation VII-27).

$$W_{\mathrm{q}}^{\mathbb{C}}\left(\widetilde{\boldsymbol{\Psi}}\middle|\,\widetilde{\boldsymbol{\Sigma}}, T\right) \propto \left|\widetilde{\boldsymbol{\Psi}}\right|^{(T-\mathrm{q})}\left|\widetilde{\boldsymbol{\Sigma}}\right|^{-T} e^{-tr\left(\widetilde{\boldsymbol{\Sigma}}^{-1}\widetilde{\boldsymbol{\Sigma}}\right)} \quad \text{(VII-25)}$$

$$\widetilde{\boldsymbol{\Sigma}} = \begin{pmatrix} \frac{\boldsymbol{\Sigma}^{-1}(1,1)}{\boldsymbol{\Gamma}(1,1)} & \boldsymbol{\Sigma}^{-1}(1,2) \\ \boldsymbol{\Sigma}^{-1}(2,1) & \boldsymbol{\Sigma}^{-1}(1,1) \end{pmatrix}^{-1} \quad \text{(VII-26)}$$

$$\widetilde{\boldsymbol{\Psi}} = \begin{pmatrix} \frac{\boldsymbol{\Phi}^{-1}(1,1)}{\boldsymbol{\Gamma}(1,1)} & \boldsymbol{\Phi}^{-1}(1,2) \\ \boldsymbol{\Phi}^{-1}(2,1) & \boldsymbol{\Phi}^{-1}(2,2) \end{pmatrix}^{-1} \quad \text{(VII-27)}$$

**Proposition 2**

The element $\boldsymbol{\Phi}^{-1}(1,1)$ of the matrix $\boldsymbol{\Phi}^{-1}$ (equation VII-27) has identical marginal pdf that the element $\boldsymbol{\Psi}^{-\mathbf{1}}(1,1)$ of the matrix $\boldsymbol{\Psi}$, when $\widetilde{\boldsymbol{\Psi}}$ has Wishart probability density function (equation VII-25) with $T$ degrees of freedom and scale matrix $\widetilde{\boldsymbol{\Sigma}}$ (equation VII-26).

**Proof of Proposition 2**

Using analogous representation to (equation VII-12) for the conditional probability density function of $\widetilde{\boldsymbol{\Psi}}(1,1)$ given the remaining elements $\left\{\overline{\widetilde{\boldsymbol{\Psi}}(1,1)}\right\} = \left\{\widetilde{\boldsymbol{\Psi}}(i',j') : (i',j') \neq (i,j)\right\}$ (equation VII-28).

$$p\left(\widetilde{\boldsymbol{\Psi}}(1,1)\middle|\left\{\overline{\widetilde{\boldsymbol{\Psi}}(1,1)}\right\}\right) \propto \left|\widetilde{\boldsymbol{\Psi}}\right|^{(T-\mathrm{q})} e^{-tr\left(\widetilde{\boldsymbol{\Sigma}}^{-1}\boldsymbol{\Psi}\right)} \quad \text{(VII-28)}$$

Given the property of the Wishart probability density function deduced before in **Proposition 1** (equation VII-24) it is clear that the random variable $\widetilde{\boldsymbol{\Psi}}^{-\mathbf{1}}(1,1)$ is independent from $\left\{\overline{\widetilde{\boldsymbol{\Psi}}(1,1)}\right\}$ and it has inverse Gamma probability density function (equation VII-29) with shape parameter $(T-\mathrm{q})$ and scale parameter $\widetilde{\boldsymbol{\Sigma}}^{-1}(1,1)$ (equation VII-26).

$$p\left(\widetilde{\boldsymbol{\Psi}}^{-\mathbf{1}}(1,1)\right) \propto \left(\widetilde{\boldsymbol{\Psi}}^{-\mathbf{1}}(1,1)\right)^{-(T-\mathrm{q})} e^{-\frac{\widetilde{\boldsymbol{\Sigma}}^{-1}(1,1)}{\widetilde{\boldsymbol{\Psi}}^{-\mathbf{1}}(1,1)}} \quad \text{(VII-29)}$$

By definition of $\boldsymbol{\Phi}^{-1}(1,1) = \boldsymbol{\Gamma}(1,1)\widetilde{\boldsymbol{\Psi}}^{-\mathbf{1}}(1,1)$ (equation VII-27), it holds that $\boldsymbol{\Phi}^{-1}(1,1)$ is also independent and it has inverse Gamma probability density function (equation VII-29) with shape parameter $(\mathrm{m}-\mathrm{q})$ and scale parameter $\boldsymbol{\Gamma}(1,1)\widetilde{\boldsymbol{\Sigma}}^{-1}(1,1) = \boldsymbol{\Sigma}^{-1}(1,1)$ (equation VII-26).

$$p\left(\boldsymbol{\Phi}^{-1}(1,1)\right) \propto \left(\boldsymbol{\Phi}^{-1}(1,1)\right)^{-(T-\mathrm{q})} e^{-\frac{\boldsymbol{\Sigma}^{-1}(1,1)}{\boldsymbol{\Phi}^{-1}(1,1)}} \quad \text{(VII-30)}$$

From the relationshsip of $\boldsymbol{\Phi}^{-1}(1,1)$ with $\boldsymbol{\Phi}(1,1)$ due to the block inverse, the probability density function (equation VII-29) can be employed to express the conditional probability density function for $\boldsymbol{\Phi}(1,1)$ (equation VII-30) regarding the remaining elements $\left\{\overline{\boldsymbol{\Phi}(1,1)}\right\}$, which yields the same Wishart marginal distribution as for $\boldsymbol{\Psi}(1,1)$ (equation VII-21).

$$p\left(\boldsymbol{\Phi}(1,1)\middle|\left\{\overline{\boldsymbol{\Phi}(1,1)}\right\}\right) \propto \left(\boldsymbol{\Phi}(1,1) - \boldsymbol{\Phi}(1,2)\left(\boldsymbol{\Phi}(2,2)\right)^{-1}\boldsymbol{\Phi}(2,1)\right)^{(T-\mathrm{q})} e^{-\boldsymbol{\Sigma}^{-1}(1,1)\left(\boldsymbol{\Phi}(1,1)-\boldsymbol{\Phi}(1,2)\left(\boldsymbol{\Phi}(2,2)\right)^{-1}\boldsymbol{\Phi}(2,1)\right)} \blacksquare \quad \text{(VII-31)}$$

Due to **Proposition 1** we can iteratively apply the scaling operation described in (equation VII-26) and (equation VII-27) for all elements $\widetilde{\boldsymbol{\Psi}} = (\boldsymbol{\Phi}^{-1} \oslash \boldsymbol{\Gamma})^{-1}$, and with $\widetilde{\boldsymbol{\Psi}}$ described by $p\left(\widetilde{\boldsymbol{\Psi}}\right) = W_{\mathrm{q}}^{\mathbb{C}}\left(\widetilde{\boldsymbol{\Psi}}\middle|\,\widetilde{\boldsymbol{\Sigma}}, T\right)$ Wishart probability density function of $T$ degrees of freedom and positive definite scale matrix $\widetilde{\boldsymbol{\Sigma}} = (\boldsymbol{\Sigma}^{-1} \oslash \boldsymbol{\Gamma})^{-1}$. In this case the random matrix $\boldsymbol{\Phi}^{-1}$ identical marginal stochastic properties to $\boldsymbol{\Psi}^{-\mathbf{1}}$. To complete the proof it is enough to show that the scale matrix $\widetilde{\boldsymbol{\Sigma}} = (\boldsymbol{\Sigma}^{-1} \oslash \boldsymbol{\Gamma})^{-1}$ of the Wishart probability density function keeps being positive definite, as we can check from the following proposition which is a Corollary of Schur product theorem (Schur 1911).

**Proposition 3**

If $\boldsymbol{\Sigma}$ and $\boldsymbol{\Gamma}$ are positive definite matrices then the matrix $\widetilde{\boldsymbol{\Sigma}} = (\boldsymbol{\Sigma}^{-1} \oslash \boldsymbol{\Gamma})^{-1}$ from the scaling operation is also positive definite.

**Proof of Proposition 3**

If $\boldsymbol{\Sigma}$ is positive definite so it is also the inverse $\boldsymbol{\Sigma}^{-1}$. Then, let's analyze the positive definiteness of the Hadamard scaling $\mathbf{1}_{\mathrm{q}} \oslash \boldsymbol{\Gamma}$. This Hadamard scaling can be expressed as an elementwise exponentiation $\mathbf{1}_{\mathrm{q}} \oslash \boldsymbol{\Gamma} = e^{\cdot -log(\boldsymbol{\Gamma})}$, $log(\boldsymbol{\Gamma})$ acts as an elementwise function. The logarithm can be expressed as an infinite Taylor series (equation VII-32).

$$log(\boldsymbol{\Gamma}) = \sum_{\ell=1}^{\infty} \frac{(-1)^{\ell+1}}{\ell}\left(\boldsymbol{\Gamma} - \mathbf{1}_{\mathrm{q}}\right)^{\cdot\ell} \quad \text{(VII-32)}$$

Without losing generality we can consider that the elements in $\boldsymbol{\Gamma}$ belong to the open interval $\boldsymbol{\Gamma}(i,j) \in (0,1), \forall i,j$. Given the assumption before, the odd terms in the Taylor series of $-log(\boldsymbol{\Gamma})$ become positive, the series (equation VII-32) can be rearranged into (equation VII-33).

$$-log(\boldsymbol{\Gamma}) = \sum_{\ell=1}^{\infty} \frac{(\mathbf{1}_{\mathrm{q}} - \boldsymbol{\Gamma})^{\cdot\ell}}{\ell} \tag{VII-33}$$

Finally, by substituting in the elementwise exponential function in (equation VII-33) we can express the Hadamard Scaling as the element wise product of exponentials (equation VII-34).

$$\mathbf{1}_{\mathrm{q}} \oslash \boldsymbol{\Gamma} = \prod_{\ell=1}^{\infty}{}_{\cdot}\, \boldsymbol{e}^{\cdot\frac{(\mathbf{1}_q - \boldsymbol{\Gamma})^{\ell}}{\ell}} \tag{VII-34}$$

Here (equation VII-34), given the positive definiteness of $\boldsymbol{\Gamma}$ and since $\boldsymbol{\Gamma}(i,j) \in (0,1), \forall i,j$ it holds that $\left(\mathbf{1}_{\mathrm{q}} - \boldsymbol{\Gamma}\right)$ is positive definite. Also, the elementwise exponentiation (equation VII-35) and (equation VII-36), and element wise product operation (equation VII-37) keep the positive definiteness.

$$\left(\mathbf{1}_{\mathrm{q}} - \boldsymbol{\Gamma}\right)^{\cdot\ell} \tag{VII-35}$$

$$\boldsymbol{e}^{\cdot\frac{(\mathbf{1}_{\mathrm{q}} - \boldsymbol{\Gamma})^{\ell}}{\ell}} \tag{VII-36}$$

$$\prod_{\ell=1}^{\infty}{}_{\cdot} \tag{VII-37}$$

Thus, it is clear that $\widetilde{\boldsymbol{\Sigma}}^{-1} = \boldsymbol{\Sigma}^{-1} \oslash \boldsymbol{\Gamma}$ is positive definite so also it is $\widetilde{\boldsymbol{\Sigma}}$.■

**Remark:** By Lemma 4 we build a statistically equivalent model of the random matrix $\boldsymbol{\Psi}$, the de-standardization $\boldsymbol{\Phi}$ defined through the from the standard sampled covariance matrix $\widetilde{\boldsymbol{\Psi}}$, in the sense that it keeps the same multivariate structure through the Wishart probability density function, and its inverse $\boldsymbol{\Phi}^{-1}$ keeps the same stochastic properties than $\boldsymbol{\Psi}^{-1}$, i.e. all their elements are independent with identical marginal probability density function (Drton, Massam, and Olkin 2008). This result prescribes a statistical equivalence between a model of the sampled covariance matrix $\boldsymbol{\Psi}$ defined by a Wishart probability density function, of $T$ degrees of freedom and positive definite scale matrix $(T\boldsymbol{\Theta})^{-1}$, with precision matrix $\boldsymbol{\Theta}$, and model of the standard sampled covariance matrix $\widetilde{\boldsymbol{\Psi}}$ defined by a Wishart probability density function (equation VII-38) and (equation VII-39), of $T$ degrees of freedom and positive definite scale matrix $\left(T\widetilde{\boldsymbol{\Theta}}\right)^{-1}$, with standard Precision matrix $\widetilde{\boldsymbol{\Theta}}$ given.

$$p\left(\widetilde{\boldsymbol{\Psi}}\middle|\widetilde{\boldsymbol{\Theta}}\right) = W_{\mathrm{q}}^{\mathbb{C}}\left(\widetilde{\boldsymbol{\Psi}}\middle|\left(T\widetilde{\boldsymbol{\Theta}}\right)^{-1}, T\right) \tag{VII-38}$$

$$W_{\mathrm{q}}^{\mathbb{C}}\left(\widetilde{\boldsymbol{\Psi}}\middle|\left(T\widetilde{\boldsymbol{\Theta}}\right)^{-1}, T\right) \propto \left|\widetilde{\boldsymbol{\Psi}}\right|^{T-\mathrm{q}}\left|\widetilde{\boldsymbol{\Theta}}\right|^{T} \boldsymbol{e}^{-Ttr\left(\widetilde{\boldsymbol{\Theta}}\widetilde{\boldsymbol{\Psi}}\right)} \tag{VII-39}$$

This statistical standardization is also consistent when the sample number $T$ tends to infinite, given the inverse sampled covariance matrix tendency in probability to $\boldsymbol{\Theta}$, i.e. $P\left(\boldsymbol{\Psi}^{-\mathbf{1}} \in \boldsymbol{\mathcal{B}}(\boldsymbol{\Theta})\right) \to 1$ as $T \to \infty$ for any open set $\boldsymbol{\mathcal{B}}(\boldsymbol{\Theta})$ containing $\boldsymbol{\Theta}$ in the $\mathrm{q}^2$-dimensional complex-valued Euclidean space. A natural estimator of the standard precision matrix $\widetilde{\boldsymbol{\Theta}}$ can be computed by maximum Likelihood through direct differentiation of (equation VII-38) and (equation VII-39) $\widehat{\widetilde{\boldsymbol{\Theta}}} = \widetilde{\boldsymbol{\Psi}}^{-\mathbf{1}}$, then, in agreement with the conditions and equivalence shown in Lemma 4 it holds that (equation VII-40).

$$\widehat{\widetilde{\boldsymbol{\Theta}}} = \boldsymbol{\Psi} \oslash \boldsymbol{\Gamma} \tag{VII-40}$$

Therefore, given the tendency in probability of $\boldsymbol{\Psi}^{-\mathbf{1}}$ it can also be directly deduced the tendency in probability of the standard precision matrix estimator, i.e. $P\left(\widehat{\widetilde{\boldsymbol{\Theta}}} \in \boldsymbol{\mathcal{B}}(\boldsymbol{\Theta} \oslash \boldsymbol{\Gamma})\right) \to 1$ as $T \to \infty$ for any open set $\boldsymbol{\mathcal{B}}(\boldsymbol{\Theta} \oslash \boldsymbol{\Gamma})$ containing $\boldsymbol{\Theta} \oslash \boldsymbol{\Gamma}$ in the $\mathrm{q}^2$-dimensional complex Euclidean space. ■

# VIII. hermitian graphical Ridge (hgRidge) and hgLASSO LQA estimator (Lemma 5)

***Lemma 5 (hermitian graphical Ridge estimator):*** *Given the local quadratic approximation of Lemma 3 and standardization of Lemma 4 the hermitian graphical LASSO admits a Ridge local representation: the pair given by the Likelihood (equation VIII-1) and prior (equation VIII-2).*

$$p\left(\widetilde{\boldsymbol{\Psi}}\middle|\widetilde{\boldsymbol{\Theta}}\right) = W_{\mathrm{q}}^{\mathbb{C}}\left(\widetilde{\boldsymbol{\Psi}}\middle| T^{-1}\widetilde{\boldsymbol{\Theta}}^{-1}, T\right) \tag{VIII-1}$$

$$p\left(\widetilde{\boldsymbol{\Theta}}\right) = exp\left(\left\|\widetilde{\boldsymbol{\Theta}}\right\|_2^2\middle| T/2\right) \tag{VIII-2}$$

*Furthermore, the hermitian graphical Ridge estimator, that maximizes the posterior distribution due to the pair (equation VIII-1) and (equation VIII-2) is the solution of the Riccati matrix equation (equation VIII-3).*

$$\widetilde{\boldsymbol{\Theta}}^2 + \widetilde{\boldsymbol{\Psi}}\widetilde{\boldsymbol{\Theta}} - \mathbf{I}_{\mathrm{q}} = \mathbf{0}_{\mathrm{q}} \tag{VIII-3}$$

*There is the unique positive definite and hermitic solution to (equation VIII-3) that also shares the eigenspace with* $\widetilde{\boldsymbol{\Psi}}$ *expressed by the matrix square root formula (equation VIII-4).*

$$\widehat{\widetilde{\boldsymbol{\Theta}}} = -\tfrac{1}{2}\widetilde{\boldsymbol{\Psi}} + \tfrac{1}{2}\sqrt{\widetilde{\boldsymbol{\Psi}}^2 + 4\mathbf{I}_{\mathrm{q}}} \tag{VIII-4}$$

***Proof of Lemma 5***

We base on a model with Wishart likelihood for the sampled covariance matrix $\boldsymbol{\Psi}$ (equation VIII-5) and the local quadratic approximation of Lemma 3 for the hermitian graphical LASSO prior upon the precision matrix $\boldsymbol{\Theta}$ (equation VIII-6).

$$p(\boldsymbol{\Psi}|\boldsymbol{\Theta}) = W_{\mathrm{q}}^{\mathbb{C}}(\boldsymbol{\Psi}|\, T^{-1}\boldsymbol{\Theta}^{-1}, T) \tag{VIII-5}$$

$$p(\boldsymbol{\Theta}) = exp(\|\boldsymbol{\Theta} \oslash \boldsymbol{\Gamma}\|_2^2|\, T/2) \tag{VIII-6}$$

We demonstrated the statistical equivalence between this model (equation VIII-5) and (equation VIII-6) and standard formulation of Lemma 4 for the Wishart likelihood upon $\widetilde{\boldsymbol{\Psi}} = (\boldsymbol{\Psi}^{-1} \oslash \boldsymbol{\Gamma})^{-1}$ (equation VIII-7) with hermitian graphical Ridge prior upon $\widetilde{\boldsymbol{\Theta}} = \boldsymbol{\Theta} \oslash \boldsymbol{\Gamma}$ (equation VIII-8).

$$p\big(\widetilde{\boldsymbol{\Psi}}\big|\widetilde{\boldsymbol{\Theta}}\big) = W_{\mathrm{q}}^{\mathbb{C}}\big(\widetilde{\boldsymbol{\Psi}}\big|\, T^{-1}\widetilde{\boldsymbol{\Theta}}^{-1}, T\big) \tag{VIII-7}$$

$$p\big(\widetilde{\boldsymbol{\Theta}}\big) = exp\left(\big\|\widetilde{\boldsymbol{\Theta}}\big\|_2^2\Big|\, T/2\right) \tag{VIII-8}$$

The posterior distribution of based on the Bayes theorem for the likelihood (equation VIII-7) and prior (equation VIII-8) is the given by (equation VIII-9).

$$p\big(\widetilde{\boldsymbol{\Theta}}\big|\widetilde{\boldsymbol{\Psi}}\big) = \big|\widetilde{\boldsymbol{\Theta}}\big|^T \boldsymbol{e}^{-Ttr(\widetilde{\boldsymbol{\Theta}}\widetilde{\boldsymbol{\Psi}})} e^{-\frac{T}{2}\|\widetilde{\boldsymbol{\Theta}}\|_2^2} \tag{VIII-9}$$

Applying the $-log$ transformation to this posterior distribution (equation VIII-9) we obtain the standard target function $\mathcal{L}\big(\widetilde{\boldsymbol{\Theta}}\big)$ and estimator $\hat{\widetilde{\boldsymbol{\Theta}}}$ (equation VIII-10).

$$\begin{gathered} \mathcal{L}\big(\widetilde{\boldsymbol{\Theta}}\big) = -T\log\big|\widetilde{\boldsymbol{\Theta}}\big| + Ttr\big(\widetilde{\boldsymbol{\Theta}}\widetilde{\boldsymbol{\Psi}}\big) + \tfrac{T}{2}\big\|\widetilde{\boldsymbol{\Theta}}\big\|_2^2 \\ \hat{\widetilde{\boldsymbol{\Theta}}} = argmin_{\widetilde{\boldsymbol{\Theta}}}\left\{-\log\big|\widetilde{\boldsymbol{\Theta}}\big| + tr\big(\widetilde{\boldsymbol{\Theta}}\widetilde{\boldsymbol{\Psi}}\big) + \tfrac{1}{2}\big\|\widetilde{\boldsymbol{\Theta}}\big\|_2^2\right\} \end{gathered} \tag{VIII-10}$$

Since this target function is convex and differentiable the necessary and sufficient condition for the minimum given by the zero of the derivative $\partial\mathcal{L}\big(\widetilde{\boldsymbol{\Theta}}\big)/\partial\widetilde{\boldsymbol{\Theta}} = \mathbf{0}_{\mathrm{q}}$ represented by the following matrix equation (equation VIII-11) with solution for the zero $\hat{\widetilde{\boldsymbol{\Theta}}}$ given by the matrix squared root formula (equation VIII-12).

$$-\widetilde{\boldsymbol{\Theta}}^{-1} + \widetilde{\boldsymbol{\Psi}} + \widetilde{\boldsymbol{\Theta}} = \mathbf{0}_{\mathrm{q}} \tag{VIII-11}$$

$$\hat{\widetilde{\boldsymbol{\Theta}}} = -\tfrac{1}{2}\widetilde{\boldsymbol{\Psi}} + \tfrac{1}{2}\sqrt{\widetilde{\boldsymbol{\Psi}}^2 + 4\mathbf{I}_{\mathrm{q}}} \tag{VIII-12}$$

**Remark:** The expression (equation VIII-11) constitutes a special case of the Riccati matrix equation (Honorio and Jaakkola 2013). Positive definiteness and hermiticity of its solution $\hat{\widetilde{\boldsymbol{\Theta}}}$, as a natural property of this estimator, is required. Also, it can be checked that the solution $\hat{\widetilde{\boldsymbol{\Theta}}}$ (equation VIII-12) must share the same eigenspace with $\widetilde{\boldsymbol{\Psi}}$, since from (equation VIII-11) two auxiliary second order matrix equations hold, i.e. $\widetilde{\boldsymbol{\Theta}}^2 + \widetilde{\boldsymbol{\Theta}}\widetilde{\boldsymbol{\Psi}} - \mathbf{I}_{\mathrm{q}} = \mathbf{0}$ and $\widetilde{\boldsymbol{\Theta}}^2 + \widetilde{\boldsymbol{\Psi}}\widetilde{\boldsymbol{\Theta}} - \mathbf{I}_{\mathrm{q}} = \mathbf{0}_{\mathrm{q}}$, given the left and right multiplication by the Standard Precision matrix. This imply that any solution $\hat{\widetilde{\boldsymbol{\Theta}}}$ commute with $\widetilde{\boldsymbol{\Psi}}$, and thus they share the eigenspace. ■

Let's show first that the proposed solution is positive definite. Given the positive definiteness and hermiticity of the complex-valued matrix $\widetilde{\boldsymbol{\Psi}}$ it admits a singular value decomposition $\widetilde{\boldsymbol{\Psi}} = \widetilde{\mathbf{U}}\widetilde{\mathbf{D}}\widetilde{\mathbf{U}}^\dagger$ with real and positive singular values $\widetilde{\mathbf{D}}$. Thus, the argument in the matrix square root of formula (equation VIII-12) admits a singular value decomposition of the kind $\widetilde{\boldsymbol{\Psi}}^2 + 4\mathbf{I}_{\mathrm{q}} = \widetilde{\mathbf{U}}\big(\widetilde{\mathbf{D}}^{\mathbf{2}} + 4\mathbf{I}_{\mathrm{q}}\big)\widetilde{\mathbf{U}}^\dagger$, where its singular values, given by $\widetilde{\mathbf{D}}^{\mathbf{2}} + 4\mathbf{I}_{\mathrm{q}}$, are also real and positive. In consequence the square root term (equation VIII-12) has also real and positive singular values given by the following singular value decomposition:

$$\sqrt{\widetilde{\boldsymbol{\Psi}}^2 + 4\mathbf{I}_{\mathrm{q}}} = \widetilde{\mathbf{U}}\sqrt{\widetilde{\mathbf{D}}^{\mathbf{2}} + 4\mathbf{I}_{\mathrm{q}}}\,\widetilde{\mathbf{U}}^\dagger \tag{VIII-13}$$

Finally, the singular value decomposition of the standard precision matrix estimator $\hat{\widetilde{\boldsymbol{\Theta}}}$ (equation VIII-12) can be expressed as (equation VIII-14).

$$\hat{\widetilde{\boldsymbol{\Theta}}} = \widetilde{\mathbf{U}}\left(\tfrac{1}{2}\sqrt{\widetilde{\mathbf{D}}^{\mathbf{2}} + 4\mathbf{I}_{\mathrm{q}}} - \tfrac{1}{2}\boldsymbol{I}_q\right)\widetilde{\mathbf{U}}^\dagger \tag{VIII-14}$$

It is clear, by formula (equation VIII-14), that the singular values of the standard precision matrix estimator are real and positive numbers. The Standard Precision matrix estimator commutes with the standard sampled covariance matrix $\widetilde{\boldsymbol{\Psi}}$ since they share the eigenspace, i.e. $\widetilde{\boldsymbol{\Psi}}\hat{\widetilde{\boldsymbol{\Theta}}} = \hat{\widetilde{\boldsymbol{\Theta}}}\widetilde{\boldsymbol{\Psi}}$. Given the singular value decomposition in equation (equation VIII-14) we can get that the singular value decomposition of $\widetilde{\boldsymbol{\Psi}}\hat{\widetilde{\boldsymbol{\Theta}}}$ can be expressed as (equation VIII-15).

$$\widetilde{\boldsymbol{\Psi}}\hat{\widetilde{\boldsymbol{\Theta}}} = \widetilde{\mathbf{U}}\left(\widetilde{\mathbf{D}}\left(\tfrac{1}{2}\sqrt{\widetilde{\mathbf{D}}^2 + 4\mathbf{I}_{\mathrm{q}}} - \tfrac{1}{2}\mathbf{I}_{\mathrm{q}}\right)\right)\widetilde{\mathbf{U}}^\dagger \tag{VIII-15}$$

The product of diagonal matrices always commutes, so we can directly check that (equation VIII-15) is also the singular value decomposition of $\hat{\widetilde{\boldsymbol{\Theta}}}\widetilde{\boldsymbol{\Psi}}$. To check that the proposed estimator (equation VIII-12) satisfies the equation (equation VIII-11) it is enough to check that it also satisfies the pair of equations $\hat{\widetilde{\boldsymbol{\Theta}}}\widetilde{\boldsymbol{\Psi}} = \mathbf{I}_{\mathrm{q}} - \hat{\widetilde{\boldsymbol{\Theta}}}^2$ and $\widetilde{\boldsymbol{\Psi}}\hat{\widetilde{\boldsymbol{\Theta}}} = \mathbf{I}_{\mathrm{q}} - \hat{\widetilde{\boldsymbol{\Theta}}}^2$. The left side in both equations are equal, i.e. $\hat{\widetilde{\boldsymbol{\Theta}}}\widetilde{\boldsymbol{\Psi}} = \widetilde{\boldsymbol{\Psi}}\hat{\widetilde{\boldsymbol{\Theta}}}$, and given by formula (equation VIII-15), so, lets evaluate the right side.

$$\mathbf{I}_{\mathrm{q}} - \tilde{\tilde{\boldsymbol{\Theta}}}^2 = \mathbf{I}_{\mathrm{q}} - \left(-\tfrac{1}{2}\tilde{\boldsymbol{\Psi}} + \tfrac{1}{2}\sqrt{\tilde{\boldsymbol{\Psi}}^2 + 4\lambda\mathbf{I}_{\mathrm{q}}}\right)^2 \quad \text{(VIII-16)}$$

Effectuating the matrix square operation in (equation VIII-16) we obtain (equation VIII-17).

$$\mathbf{I}_{\mathrm{q}} - \tilde{\tilde{\boldsymbol{\Theta}}}^2 = \mathbf{I}_q - \left(\tfrac{1}{2}\right)^{\mathbf{2}}\left(\tilde{\boldsymbol{\Psi}}^2 - \tilde{\boldsymbol{\Psi}}\sqrt{\tilde{\boldsymbol{\Psi}}^{\mathbf{2}} + 4\mathbf{I}_{\mathrm{q}}} - \sqrt{\tilde{\boldsymbol{\Psi}}^{\mathbf{2}} + 4\mathbf{I}_q}\,\tilde{\boldsymbol{\Psi}} + \tilde{\boldsymbol{\Psi}}^2 + 4\mathbf{I}_{\mathrm{q}}\right) \quad \text{(VIII-17)}$$

From the singular value decomposition analysis above the matrices sampled covariance matrix $\tilde{\boldsymbol{\Psi}}$ and the square root term in (equation VIII-13) share the same eigenspace and thus commute, so, rearranging (equation VIII-17) it can be obtained (equation VIII-18).

$$\mathbf{I}_{\mathrm{q}} - \lambda\tilde{\tilde{\boldsymbol{\Theta}}}^2 = \mathbf{I}_{\mathrm{q}} - 2\left(\tfrac{1}{2}\right)^{\mathbf{2}}\tilde{\boldsymbol{\Psi}}^2 + 2\left(\tfrac{1}{2}\right)^{\mathbf{2}}\tilde{\boldsymbol{\Psi}}\sqrt{\tilde{\boldsymbol{\Psi}}^{\mathbf{2}} + 4\mathbf{I}_{\mathrm{q}}} - 4\left(\tfrac{1}{2}\right)^{\mathbf{2}}\mathbf{I}_{\mathrm{q}} \quad \text{(VIII-18)}$$

From (equation VIII-18) and considering (equation VIII-17) it is direct that following identity holds (equation VIII-19).

$$\mathbf{I}_{\mathrm{q}} - \tilde{\tilde{\boldsymbol{\Theta}}}^2 = -\tfrac{1}{2}\tilde{\boldsymbol{\Psi}}^2 + \tfrac{1}{2}\tilde{\boldsymbol{\Psi}}\sqrt{\tilde{\boldsymbol{\Psi}}^{\mathbf{2}} + 4\mathbf{I}_{\mathrm{q}}} = \tilde{\boldsymbol{\Psi}}\tilde{\tilde{\boldsymbol{\Theta}}} \quad \text{(VIII-19)}$$

Now we need to proof that the proposed estimator (equation VIII-12) is the unique solution of equation (equation VIII-11) that commute with $\tilde{\boldsymbol{\Psi}}$. Let $\tilde{\tilde{\boldsymbol{\Phi}}}$ be another solution that also commutes with $\tilde{\boldsymbol{\Psi}}$. Since the matrix $\tilde{\tilde{\boldsymbol{\Phi}}}$ commutes with $\tilde{\boldsymbol{\Psi}}$ it has the same eigenspace, i.e. it admits a singular value decomposition of the kind $\tilde{\tilde{\boldsymbol{\Phi}}} = \tilde{\mathbf{U}}\tilde{\mathbf{E}}\tilde{\mathbf{U}}^{\dagger}$. Also, since $\tilde{\tilde{\boldsymbol{\Phi}}}$ satisfies equation (equation VIII-11) it can be checked that $-\tilde{\mathbf{E}}^{-1} + \tilde{\mathbf{D}} + \tilde{\mathbf{E}} = \mathbf{0}_{\mathrm{q}}$. The solution of this equation is straightforward given diagonal matrices $\tilde{\mathbf{E}}$ and $\tilde{\mathbf{D}}$ (equation VIII-20).

$$\tilde{\mathbf{E}} = \tfrac{1}{2}\sqrt{\tilde{\mathbf{D}}^{\mathbf{2}} + 4\mathbf{I}_{\mathrm{q}}} - \tfrac{1}{2}\mathbf{I}_{\mathrm{q}} \quad \text{(VIII-20)}$$

This (equation VIII-20) therefore yields that $\tilde{\tilde{\boldsymbol{\Phi}}} = \tilde{\tilde{\boldsymbol{\Theta}}}$ since they have identical eigenspace and eigenvalues.■

# 7 References

Andersen, Heidi H, Malene Hojbjerre, Dorte Sorensen, and Poul S Eriksen. 1995. *Linear and Graphical Models: For the Multivariate Complex Normal Distribution*. Vol. 101. Springer Science & Business Media.

Andrews. 1974. “Scale Mixtures of Normal Distributions.” *Journal of the Royal Statistical Society. Series B (Methodological)* 148: 148–62. https://doi.org/https://www.jstor.org/stable/2984774.

Attias, Hagai. 2000. “A Variational {B}ayesian Framework for Graphical Models.” *In Advances in Neural Information Processing Systems* 12: 209–15.

Babiloni, Fabio, Febo Cincotti, Claudio Babiloni, Filippo Carducci, Donatella Mattia, Laura Astolfi, Alessandra Basilisco, Paolo Maria Rossini, Lei Ding, and Ying Ni. 2005. “Estimation of the Cortical Functional Connectivity with the Multimodal Integration of High-Resolution EEG and FMRI Data by Directed Transfer Function.” *Neuroimage* 24 (1): 118–31.

Baccalá, Luiz A, and Koichi Sameshima. 2001. “Partial Directed Coherence: A New Concept in Neural Structure Determination.” *Biological Cybernetics* 84 (6): 463–74.

Baillet, Sylvain, John C. Mosher, and Richard M. Leahy. 2001. “Electromagnetic Brain Mapping.” *IEEE Signal Processing Magazine* 18 (6): 14–30. https://doi.org/10.1109/79.962275.

Bastos, André M, and Jan-Mathijs Schoffelen. 2016. “A Tutorial Review of Functional Connectivity Analysis Methods and Their Interpretational Pitfalls.” *Frontiers in Systems Neuroscience* 9 (January): 175. https://doi.org/10.3389/fnsys.2015.00175.

Bien, Jacob, and Robert J. Tibshirani. 2011. “Sparse Estimation of a Covariance Matrix.” *Biometrika* 98 (4): 807–20. https://doi.org/10.1093/biomet/asr054.

Bishop, Christopher M. 1995. *Neural Networks for Pattern Recognition*. Oxford university press.

Blinowska, Katarzyna J. 2011. “Review of the Methods of Determination of Directed Connectivity from Multichannel Data.” *Medical & Biological Engineering & Computing* 49 (5): 521–29. https://doi.org/10.1007/s11517-011-0735-x.

Bollhöfer, Matthias, Aryan Eftekhari, Simon Scheidegger, and Olaf Schenk. 2019. “Large-Scale Sparse Inverse Covariance Matrix Estimation.” *SIAM Journal on Scientific Computing* 41 (1): A380–401.

Brillinger, David R. 1975. “The Identification of Point Process Systems.” *The Annals of Probability*, 909–24.

Brookes, M. J., M. W. Woolrich, and G. R. Barnes. 2012. “Measuring Functional Connectivity in MEG: A Multivariate Approach Insensitive to Linear Source Leakage.” *NeuroImage*. https://doi.org/10.1016/j.neuroimage.2012.03.048.

Brookes, Matthew J., Joanne R. Hale, Johanna M. Zumer, Claire M. Stevenson, Susan T. Francis, Gareth R. Barnes, Julia P. Owen, Peter G. Morris, and Srikantan S. Nagarajan. 2011. “Measuring Functional Connectivity Using MEG: Methodology and Comparison with FcMRI.” *NeuroImage* 56 (3): 1082–1104. https://doi.org/10.1016/j.neuroimage.2011.02.054.

Brookes, Matthew J., Claire M. Stevenson, Gareth R. Barnes, Arjan Hillebrand, Michael I G Simpson, Susan T. Francis, and Peter G. Morris. 2007. “Beamformer Reconstruction of Correlated Sources Using a Modified Source Model.” *NeuroImage* 34 (4): 1454–65. https://doi.org/10.1016/j.neuroimage.2006.11.012.

Brookes, Matthew J., Mark Woolrich, Henry Luckhoo, Darren Price, Joanne R. Hale, Mary C. Stephenson, Gareth R. Barnes, Stephen M. Smith, and Peter G. Morris. 2011a. "Investigating the Electrophysiological Basis of Resting State Networks Using Magnetoencephalography." *Proceedings of the National Academy of Sciences of the United States of America* 108 (40): 16783–88. https://doi.org/10.1073/pnas.1112685108.

———. 2011b. "Investigating the Electrophysiological Basis of Resting State Networks Using Magnetoencephalography." *Proceedings of the National Academy of Sciences of the United States of America* 108 (40): 16783–88. https://doi.org/10.1073/pnas.1112685108.

Bruns, Andreas. 2004. "Fourier-, Hilbert- and Wavelet-Based Signal Analysis: Are They Really Different Approaches?" *Journal of Neuroscience Methods* 137 (2): 237. https://doi.org/10.1016/j.jneumeth.2004.03.002.

Bullmore, Ed, and Olaf Sporns. 2009. "Complex Brain Networks: Graph Theoretical Analysis of Structural and Functional Systems." *Nature Reviews Neuroscience* 10 (3): 186–98.

Bunea, Florentina, Yiyuan She, Hernando Ombao, Assawin Gongvatana, Kate Devlin, and Ronald Cohen. 2011. "Penalized Least Squares Regression Methods and Applications to Neuroimaging." *NeuroImage* 55 (4): 1519–27. https://doi.org/10.1016/j.neuroimage.2010.12.028.

Cabras, Stefano, Maria Eugenia Castellanos Nueda, and Erlis Ruli. 2015. "Approximate Bayesian Computation by Modelling Summary Statistics in a Quasi-Likelihood Framework." *Bayesian Analysis* 10 (2): 411–39.

Carter, G C. 1987. "Coherence and Time Delay Estimation." *Proceedings of the IEEE* 75 (2): 236–55. https://doi.org/10.1109/PROC.1987.13723.

Casella, George, Malay Ghosh, Jeff Gill, and Minjung Kyung. 2010. "Penalized Regression, Standard Errors, and Bayesian Lassos." *Bayesian Analysis* 5 (2): 369–411.

Colclough, G. L., M. J. Brookes, S. M. Smith, and M. W. Woolrich. 2015. "A Symmetric Multivariate Leakage Correction for MEG Connectomes." *NeuroImage* 117: 439–48. https://doi.org/10.1016/j.neuroimage.2015.03.071.

Colclough, Giles L, Matthew J Brookes, Stephen M Smith, and Mark W Woolrich. 2015. "A Symmetric Multivariate Leakage Correction for MEG Connectomes." *Neuroimage* 117: 439–48.

Csilléry, Katalin, Michael G B Blum, Oscar E Gaggiotti, and Olivier François. 2010. "Approximate Bayesian Computation (ABC) in Practice." *Trends in Ecology & Evolution* 25 (7): 410–18.

David, Olivier, and Karl J. Friston. 2003. "A Neural Mass Model for MEG/EEG: Coupling and Neuronal Dynamics." *NeuroImage* 20 (3): 1743–55. https://doi.org/10.1016/j.neuroimage.2003.07.015.

Dempster, Arthur P, Nan M Laird, and Donald B Rubin. 1977. "Maximum Likelihood from Incomplete Data via the EM Algorithm." *Journal of the Royal Statistical Society: Series B (Methodological)* 39 (1): 1–22.

Drton, Mathias, Hélène Massam, and Ingram Olkin. 2008. "Moments of Minors of Wishart Matrices." *The Annals of Statistics* 36 (5): 2261–83.

Engel, Andreas K, Pascal Fries, and Wolf Singer. 2001. "Dynamic Predictions: Oscillations and Synchrony in Top–down Processing." *Nature Reviews Neuroscience* 2 (10): 704–16.

Ephremidze, Lasha, Gigla Janashia, and Edem Lagvilava. 2007. "A New Efficient Matrix Spectral Factorization Algorithm." In *SICE Annual Conference 2007*, 20–23. IEEE.

Faes, Luca, Silvia Erla, and Giandomenico Nollo. 2012. "Measuring Connectivity in Linear Multivariate Processes: Definitions, Interpretation, and Practical Analysis." *Computational and Mathematical Methods in Medicine* 2012.

Faes, Luca, and Giandomenico Nollo. 2011. "Multivariate Frequency Domain Analysis of Causal Interactions in Physiological Time Series." *Biomedical Engineering, Trends in Electronics, Communications and Software* 8: 403–28.

Faes, Luca, Sebastiano Stramaglia, and Daniele Marinazzo. n.d. "On the Interpretability and Computational Reliability of Frequency-Domain Granger Causality," 1–4. https://arxiv.org/ftp/arxiv/papers/1708/1708.06990.pdf.

Fan Jianqing, Li Runze. 2001. "Variable Selection via Nonconcave Penalized." *Journal of the American Statistical Association* 96 (456): 1348–60. https://doi.org/10.1198/016214501753382273.

Fearnhead, Paul, and Dennis Prangle. 2012. "Constructing Summary Statistics for Approximate Bayesian Computation: Semi-automatic Approximate Bayesian Computation." *Journal of the Royal Statistical Society: Series B (Statistical Methodology)* 74 (3): 419–74.

Frauscher, Birgit, Nicolas Von Ellenrieder, Rina Zelmann, Irena Doležalová, Lorella Minotti, André Olivier, Jeffery Hall, Dominique Hoffmann, Dang Khoa Nguyen, and Philippe Kahane. 2018. "Atlas of the Normal Intracranial Electroencephalogram: Neurophysiological Awake Activity in Different Cortical Areas." *Brain* 141 (4): 1130–44.

Freeman, Walter J. 1975. *Mass Action in the Nervous System : Examination of the Neurophysiological Basis of Adaptive Behavior through the EEG*. New York ; London: Academic Press. https://librarysearch.cardiff.ac.uk/openurl/44WHELF_CAR/44WHELF_CAR:44WHELF_CAR_VU1? ?u.ignore_date_coverage=true&rft.mms_id=9911670637102420.

Friedman, Jerome, Trevor Hastie, and Robert Tibshirani. 2007. "Sparse Inverse Covariance Estimation with the Lasso," 1–14.

https://doi.org/10.1093/biostatistics/kxm045.
———. 2008. “Sparse Inverse Covariance Estimation with the Graphical Lasso.” *Biostatistics* 9 (3): 432–41. https://doi.org/10.1093/biostatistics/kxm045.
Friedman, N, and D Koller. 2003. “Being Bayesian about Network Structure. A Bayesian Approach to Structure Discovery in Bayesian Networks.” *Machine Learning* 50: 95–125. https://doi.org/10.1023/A:1020249912095.
Friston, K. J., A. Bastos, V. Litvak, K. E. Stephan, P. Fries, and R. J. Moran. 2012. “DCM for Complex-Valued Data: Cross-Spectra, Coherence and Phase-Delays.” *NeuroImage* 59 (1): 439–55. https://doi.org/10.1016/j.neuroimage.2011.07.048.
Friston, K. J., N. Trujillo-Barreto, and J. Daunizeau. 2008. “DEM: A Variational Treatment of Dynamic Systems.” *NeuroImage* 41 (3): 849–85. https://doi.org/10.1016/j.neuroimage.2008.02.054.
Friston, Karl. 2011. “Dynamic Causal Modeling and Granger Causality Comments on: The Identification of Interacting Networks in the Brain Using FMRI: Model Selection, Causality and Deconvolution.” *NeuroImage* 58 (2): 303–5. https://doi.org/10.1016/j.neuroimage.2009.09.031.
Friston, Karl, Lee Harrison, Jean Daunizeau, Stefan Kiebel, Christophe Phillips, Nelson Trujillo-Barreto, Richard Henson, Guillaume Flandin, and Jérémie Mattout. 2008. “Multiple Sparse Priors for the M/EEG Inverse Problem.” *NeuroImage* 39 (3): 1104–20. https://doi.org/10.1016/j.neuroimage.2007.09.048.
Friston, Karl, Richard Henson, Christophe Phillips, and J??r??mie Mattout. 2006. “Bayesian Estimation of Evoked and Induced Responses.” *Human Brain Mapping* 27 (9): 722–35. https://doi.org/10.1002/hbm.20214.
Friston, Karl, Jérémie Mattout, Nelson Trujillo-Barreto, John Ashburner, and Will Penny. 2007. “Variational Free Energy and the Laplace Approximation.” *Neuroimage* 34 (1): 220–34.
Galka, Andreas, Okito Yamashita, and Tohru Ozaki. 2004. “GARCH Modelling of Covariance in Dynamical Estimation of Inverse Solutions.” *Physics Letters, Section A: General, Atomic and Solid State Physics* 333 (3–4): 261–68. https://doi.org/10.1016/j.physleta.2004.10.045.
Galka, Andreas, Okito Yamashita, Tohru Ozaki, Rolando Biscay, and Pedro Valdés-Sosa. 2004. “A Solution to the Dynamical Inverse Problem of EEG Generation Using Spatiotemporal Kalman Filtering.” *NeuroImage* 23 (2): 435–53. https://doi.org/10.1016/j.neuroimage.2004.02.022.
Ghahramani, Z, and M J Beal. 2001. “Propagation Algorithms for Variational Bayesian Learning.” *Advances in Neural Information Processing Systems*, no. section 3: 507–13. https://doi.org/10.1.1.67.770.
Gonzalez-Moreira, Eduardo, Deirel Paz-Linares, Ariosky Areces-Gonzalez, Ying Wang, Min Li, Thalia Harmony, and Pedro A Valdes-Sosa. 2020. “Bottom-up Control of Leakage in Spectral Electrophysiological Source Imaging via Structured Sparse Bayesian Learning.” *BioRxiv*.
Gramfort, Alexandre, Daniel Strohmeier, Jens Haueisen, Matti S Hämäläinen, and Matthieu Kowalski. 2013. “Time-Frequency Mixed-Norm Estimates: Sparse M/EEG Imaging with Non-Stationary Source Activations.” *NeuroImage* 70: 410–22.
Grech, Roberta, Tracey Cassar, Joseph Muscat, Kenneth P Camilleri, Simon G Fabri, Michalis Zervakis, Petros Xanthopoulos, Vangelis Sakkalis, and Bart Vanrumste. 2008. “Review on Solving the Inverse Problem in EEG Source Analysis.” *Journal of Neuroengineering and Rehabilitation* 5 (1): 1–33.
Hadamard, Jacques, and Philip M. Morse. 1953. “Lectures on Cauchy’s Problem in Linear Partial Differential Equations.” *Physics Today* 6 (8): 18–18. https://doi.org/10.1063/1.3061337.
Hämäläinen, M. S., and R. J. Ilmoniemi. 1994. “Interpreting Magnetic Fields of the Brain: Minimum Norm Estimates.” *Medical & Biological Engineering & Computing* 32 (1): 35–42. https://doi.org/10.1007/BF02512476.
Haufe, Stefan, Vadim V. Nikulin, Andreas Ziehe, Klaus Robert Müller, and Guido Nolte. 2008. “Combining Sparsity and Rotational Invariance in EEG/MEG Source Reconstruction.” *NeuroImage* 42 (2): 726–38. https://doi.org/10.1016/j.neuroimage.2008.04.246.
Haufe, Stefan, Ryota Tomioka, Thorsten Dickhaus, Claudia Sannelli, Benjamin Blankertz, Guido Nolte, and Klaus-Robert Müller. 2011. “Large-Scale EEG/MEG Source Localization with Spatial Flexibility.” *NeuroImage* 54 (2): 851–59.
He, Bin, Laura Astolfi, Pedro Antonio Valdes-Sosa, Daniele Marinazzo, Satu O. Palva, Christian-George Benar, Christoph M. Michel, and Thomas Koenig. 2019. “Electrophysiological Brain Connectivity: Theory and Implementation.” *IEEE Transactions on Biomedical Engineering* 66 (7): 2115–37. https://doi.org/10.1109/TBME.2019.2913928.
Hillebrand, Arjan, Gareth R. Barnes, Johannes L. Bosboom, Henk W. Berendse, and Cornelis J. Stam. 2012. “Frequency-Dependent Functional Connectivity within Resting-State Networks: An Atlas-Based MEG Beamformer Solution.” *NeuroImage* 59 (4): 3909–21. https://doi.org/10.1016/j.neuroimage.2011.11.005.
Hipp, Joerg F, David J Hawellek, Maurizio Corbetta, Markus Siegel, and Andreas K Engel. 2012. “Large-Scale Cortical Correlation Structure of Spontaneous Oscillatory Activity.” *Nature Neuroscience* 15 (6): 884–90. https://doi.org/10.1038/nn.3101.
Honorio, Jean, and Ts Jaakkola. 2013. “Inverse Covariance Estimation for High-Dimensional Data in Linear Time and Space: Spectral Methods for Riccati and Sparse Models.” *ArXiv Preprint ArXiv:1309.6838*, 1329–36. http://arxiv.org/abs/1309.6838.
Hsieh, Cho-jui. 2014. “QUIC : Quadratic Approximation for Sparse Inverse Covariance Estimation.” *Journal of Machine Learning*

*Research* 15: 2911–47.
Hsieh, Cho-Jui, Arindam Banerjee, Inderjit Dhillon, and Pradeep Ravikumar. 2012. "A Divide-and-Conquer Method for Sparse Inverse Covariance Estimation." *Advances in Neural Information Processing Systems* 25: 2330–38.
Hsieh, Cho-Jui, Matyas A. Sustik, Inderjit S. Dhillon, and Pradeep Ravikumar. 2011. "Sparse Inverse Covariance Matrix Estimation Using Quadratic Approximation." *Advances in Neural* ..., 1–18. http://www.cs.utexas.edu/users/inderjit/public_papers/invcov_nips11.pdf.
Hsieh, Cho-Jui, Mátyás A Sustik, Inderjit S Dhillon, Pradeep Ravikumar, and Russell A Poldrack. 2013. "BIG & QUIC: Sparse Inverse Covariance Estimation for a Million Variables." In *NIPS*, 26:3165–73.
Hunter, David R., and Runze Li. 2005. "Variable Selection Using MM Algorithms." *Annals of Statistics* 33 (4): 1617–42. https://doi.org/10.1214/009053605000000200.
Jaakkola, T S, and M I Jordan. 2000. "Bayesian Parameter Estimation via Variational Methods." *Statistics And Computing* 10 (1): 25–37. https://doi.org/10.1023/A:1008932416310.
Jafarian, Amirhossein, and John G McWhirter. 2012. "A Novel Method for Multichannel Spectral Factorization." In *2012 Proceedings of the 20th European Signal Processing Conference (EUSIPCO)*, 1069–73. IEEE.
Janashia, Gigla, Edem Lagvilava, and Lasha Ephremidze. 2011. "A New Method of Matrix Spectral Factorization." *IEEE Transactions on Information Theory* 57 (4): 2318–26.
Janková, Jana, and Sara van de Geer. 2018. "Inference in High-Dimensional Graphical Models." In *Handbook of Graphical Models*, 325–50. CRC Press.
Jirsa, V.K., and H. Haken. 1997. "A Derivation of a Macroscopic Field Theory of the Brain from the Quasi-Microscopic Neural Dynamics." *Physica D: Nonlinear Phenomena* 99 (4): 503–26. https://doi.org/10.1016/S0167-2789(96)00166-2.
Jordan, Michael Irwin. 1998. *Learning in Graphical Models*. Vol. 89. Springer Science & Business Media.
Koller, Daphne, and Nir Friedman. 2009. *Probabilistic Graphical Models: Principles and Techniques*. MIT press.
Krishnaswamy, Pavitra, Gabriel Obregon-Henao, Jyrki Ahveninen, Sheraz Khan, Behtash Babadi, Juan Eugenio Iglesias, Matti S Hämäläinen, and Patrick L Purdon. 2017. "Sparsity Enables Estimation of Both Subcortical and Cortical Activity from MEG and EEG." *Proceedings of the National Academy of Sciences* 114 (48): E10465–74.
Kullback, S. 1968. *Information Theory and Statistics*.
Larson-Prior, L. J., R. Oostenveld, S. Della Penna, G. Michalareas, F. Prior, A. Babajani-Feremi, J. M. Schoffelen, et al. 2013. "Adding Dynamics to the Human Connectome Project with MEG." *NeuroImage* 80: 190–201. https://doi.org/10.1016/j.neuroimage.2013.05.056.
Ledoit, Olivier, and Michael Wolf. 2004. "A Well-Conditioned Estimator for Large-Dimensional Covariance Matrices." *Journal of Multivariate Analysis* 88 (2): 365–411.
Lee, Lucy, Karl Friston, and Barry Horwitz. 2006. "Large-Scale Neural Models and Dynamic Causal Modelling." *NeuroImage* 30 (4): 1243–54. https://doi.org/10.1016/j.neuroimage.2005.11.007.
Liu, Chuanhai, and Donald B Rubin. 1994. "The ECME Algorithm: A Simple Extension of EM and ECM with Faster Monotone Convergence." *Biometrika* 81 (4): 633–48.
Liu, Weidong, and Xi Luo. 2015. "Fast and Adaptive Sparse Precision Matrix Estimation in High Dimensions." *Journal of Multivariate Analysis* 135: 153–62. https://doi.org/10.1016/j.jmva.2014.11.005.
Lopes da Silva, Fernando. 2013. "EEG and MEG: Relevance to Neuroscience." *Neuron* 80 (5): 1112–28. https://doi.org/10.1016/j.neuron.2013.10.017.
MacKay, David J.C. 2003. *Information Theory, Inference and Learning Algorithms*. Cambridge: Cambridge University Press.
Mackay, David J C. 1999. "Hyperparameters" 1068: 1035–68.
Mahjoory, Keyvan, Vadim V Nikulin, Loïc Botrel, Klaus Linkenkaer-Hansen, Marco M Fato, and Stefan Haufe. 2017. "Consistency of EEG Source Localization and Connectivity Estimates." *Neuroimage* 152: 590–601.
Makeig, Scott. 2002. "Response: Event-Related Brain Dynamics - Unifying Brain Electrophysiology." *Trends in Neurosciences* 25 (8): 390. https://doi.org/10.1016/S0166-2236(02)02198-7.
Makeig, Scott, Stefan Debener, Julie Onton, and Arnaud Delorme. 2004. "Mining Event-Related Brain Dynamics." *Trends in Cognitive Sciences* 8 (5): 204–10. https://doi.org/10.1016/j.tics.2004.03.008.
Mantini, Dante, Mauro G Perrucci, Cosimo Del Gratta, Gian L Romani, and Maurizio Corbetta. 2007. "Electrophysiological Signatures of Resting State Networks in the Human Brain." *Proceedings of the National Academy of Sciences* 104 (32): 13170–75.
Marinazzo, Daniele, Jorge J. Riera, Laura Marzetti, Laura Astolfi, Dezhong Yao, and Pedro A. Valdés Sosa. 2019. "Controversies in EEG Source Imaging and Connectivity: Modeling, Validation, Benchmarking." *Brain Topography* 32 (4): 527–29. https://doi.org/10.1007/s10548-019-00709-9.
Marzetti, Laura, Cosimo Del Gratta, and Guido Nolte. 2008. "Understanding Brain Connectivity from EEG Data by Identifying Systems Composed of Interacting Sources." *NeuroImage* 42 (1): 87–98. https://doi.org/10.1016/j.neuroimage.2008.04.250.
Mattout, Jérémie, Christophe Phillips, William D Penny, Michael D Rugg, and Karl J Friston. 2006. "MEG Source Localization under Multiple Constraints: An Extended Bayesian Framework." *NeuroImage* 30 (3): 753–67.
McLachlan, Geoffrey J, and Thriyambakam Krishnan. 2007. *The EM Algorithm and Extensions*. Vol. 382. John Wiley & Sons.
Moffett, Steven X, Sean M O'Malley, Shushuang Man, Dawei Hong, and Joseph V Martin. 2017. "Dynamics of High Frequency Brain Activity." *Scientific Reports* 7 (1): 1–5.
Nagasaka, Yasuo, Kentaro Shimoda, and Naotaka Fujii. 2011. "Multidimensional Recording (MDR) and Data Sharing: An Ecological Open Research and Educational Platform for Neuroscience." *PloS One* 6 (7): e22561.
Niedermeyer, E, and F H L da Silva. 2005. *Electroencephalography: Basic Principles, Clinical Applications, and Related Fields*. LWW Doody's All Reviewed Collection. Lippincott Williams \& Wilkins. https://books.google.com.sg/books?id=tndqYGPHQdEC.
Nolte, Guido, Edgar Galindo-Leon, Zhenghan Li, Xun Liu, and Andreas K. Engel. 2020. "Mathematical Relations Between Measures of Brain Connectivity Estimated From Electrophysiological Recordings for Gaussian Distributed Data." *Frontiers in Neuroscience* 14. https://doi.org/10.3389/fnins.2020.577574.
Nunez, P. L., R. B. Silberstein, P. J. Cadusch, R. S. Wijesinghe, A. F. Westdorp, and R. Srinivasan. 1994. "A Theoretical and Experimental Study of High Resolution EEG Based on Surface Laplacians and Cortical Imaging." *Electroencephalography and Clinical Neurophysiology* 90 (1): 40–57. https://doi.org/10.1016/0013-4694(94)90112-0.
Nunez, Paul L. 1974. "The Brain Wave Equation: A Model for the EEG." *Mathematical Biosciences* 21 (3–4): 279–97. https://doi.org/10.1016/0025-5564(74)90020-0.
Nunez, Paul L, and Ramesh Srinivasan. 2006. *Electric Fields of the Brain: The Neurophysics of EEG*. 2nd ed. New York: Oxford University Press. https://doi.org/10.1093/acprof:oso/9780195050387.001.0001.
Palva, J. Matias, Sheng H. Wang, Satu Palva, Alexander Zhigalov, Simo Monto, Matthew J. Brookes, Jan Mathijs Schoffelen, and Karim Jerbi. 2018. "Ghost Interactions in MEG/EEG Source Space: A Note of Caution on Inter-Areal Coupling Measures." *NeuroImage* 173 (March 2017): 632–43. https://doi.org/10.1016/j.neuroimage.2018.02.032.
Papadopoulou, Margarita, Karl Friston, and Daniele Marinazzo. 2019. "Estimating Directed Connectivity from Cortical Recordings and Reconstructed Sources." *Brain Topography* 32 (4): 741–52.
Pascual-Marqui, Roberto D, Alberto D Pascual-Montano, Dietrich Lehmann, Kieko Kochi, Michaela Esslen, Lutz Jancke, Peter Anderer, Bernd Saletu, Hideaki Tanaka, and Koichi Hirata. 2006. "Exact Low Resolution Brain Electromagnetic Tomography (ELORETA)." *Neuroimage* 31 (Suppl 1).
Pascual-marqui, Roberto D, Pedro A Valdes-sosa, and Alfredo Alvarez-amador. 1988. "A Parametric Model for Multichannel EEG Spectra." *International Journal of Neuroscience* 40 (1–2): 89–99.
Paz-Linares, D., E. Gonzalez-Moreira, E. Martinez-Montes, P.A. Valdes-Hernandez, J. Bosch-Bayard, M.L. Bringas-Vega, and P.A. Valdés-Sosa. 2018. "Caulking the 'Leakage Effect' in MEEG Source Connectivity Analysis." *ArXiv*.
Paz-Linares, D., M. Vega-Hernández, P.A. Rojas-López, P.A. Valdés-Hernández, E. Martínez-Montes, and P.A. Valdés-Sosa. 2017. "Spatio Temporal EEG Source Imaging with the Hierarchical Bayesian Elastic Net and Elitist Lasso Models." *Frontiers in Neuroscience* 11 (NOV). https://doi.org/10.3389/fnins.2017.00635.
Qiao, Xinghao, Shaojun Guo, and Gareth M James. 2019. "Functional Graphical Models." *Journal of the American Statistical Association* 114 (525): 211–22.
Razi, Adeel, and Karl J. Friston. 2016. "The Connected Brain." *Journal of Experimental Psychology: Applied* 15 (3): 228–42. https://doi.org/10.1037/a0016995.
Reid, Andrew T, Drew B Headley, Ravi D Mill, Ruben Sanchez-Romero, Lucina Q Uddin, Daniele Marinazzo, Daniel J Lurie, Pedro A Valdés-Sosa, Stephen José Hanson, and Bharat B Biswal. 2019. "Advancing Functional Connectivity Research from Association to Causation." *Nature Neuroscience* 22 (11): 1751–60.
Riera, Jorge J., Jobu Watanabe, Iwata Kazuki, Miura Naoki, Eduardo Aubert, Tohru Ozaki, and Ryuta Kawashima. 2004. "A State-Space Model of the Hemodynamic Approach: Nonlinear Filtering of BOLD Signals." *NeuroImage*. https://doi.org/10.1016/j.neuroimage.2003.09.052.
Riera, Jörge Javier, and Maria Elena Fuentes. 1998. "Electric Lead Field for a Piecewise Homogeneous Volume Conductor Model of the Head." *IEEE Transactions on Biomedical Engineering* 45 (6): 746–53. https://doi.org/10.1109/10.678609.
Rosenblatt, Murray. 1956. "A Central Limit Theorem and a Strong Mixing Condition." *Proceedings of the National Academy of Sciences of the United States of America* 42 (1): 43.
Roweis, Sam, and Zoubin Ghahramani. 1999. "A Unifying Review of Linear Gaussian Models." *Neural Computation* 11 (2): 305–45.
Sayed, Ali H, and Thomas Kailath. 2001. "A Survey of Spectral Factorization Methods." *Numerical Linear Algebra with Applications* 8 (6-7): 467–96.
Schreier, Peter J, and Louis L Scharf. 2010. *Statistical Signal Processing of Complex-Valued Data: The Theory of Improper and Noncircular Signals*. Cambridge university press.
Schur, Jssai. 1911. "Bemerkungen Zur Theorie Der Beschränkten Bilinearformen Mit Unendlich Vielen Veränderlichen."
Smith, Stephen M, Diego Vidaurre, Christian F Beckmann, Matthew F Glasser, Mark Jenkinson, Karla L Miller, Thomas E Nichols, Emma C Robinson, Gholamreza Salimi-Khorshidi, and Mark W Woolrich. 2013. "Functional Connectomics from Resting-State

FMRI.” *Trends in Cognitive Sciences* 17 (12): 666–82.

Steen, Frederik Van De, Pedro A Valdes-Sosa, and Daniele Marinazzo. 2016. “Critical Comments on EEG Sensor Space Dynamical Connectivity Analysis.” *Brain Topography*. https://doi.org/10.1007/s10548-016-0538-7.

Tarantola, Albert. 2005. *Inverse Problem Theory and Methods for Model Parameter Estimation*. *Inverse Problem Theory and Methods for Model Parameter Estimation*. https://doi.org/10.1137/1.9780898717921.

Tewarie, Prejaas, Lucrezia Liuzzi, George C. O’Neill, Andrew J. Quinn, Alessandra Griffa, Mark W. Woolrich, Cornelis J. Stam, Arjan Hillebrand, and Matthew J. Brookes. 2019. “Tracking Dynamic Brain Networks Using High Temporal Resolution MEG Measures of Functional Connectivity.” *NeuroImage* 200: 38–50. https://doi.org/10.1016/j.neuroimage.2019.06.006.

Tibshirani, Robert. 1996. “Regression Shrinkage and Selection via the Lasso.” *Journal of the Royal Statistical Society: Series B (Methodological)* 58 (1): 267–88.

Tipping, Michael E. 2001. “Sparse Bayesian Learning and the Relevance Vector Machine.” *Journal of Machine Learning Research* 1 (Jun): 211–44.

Trujillo-Barreto, Nelson J, Eduardo Aubert-Vázquez, and Pedro A Valdés-Sosa. 2004. “Bayesian Model Averaging in EEG/MEG Imaging.” *NeuroImage* 21 (4): 1300–1319.

Tugnait, Jitendra K. 2019. “Graphical Lasso for High-Dimensional Complex Gaussian Graphical Model Selection.” In *ICASSP 2019-2019 IEEE International Conference on Acoustics, Speech and Signal Processing (ICASSP)*, 2952–56. IEEE.

Valdés-Sosa, Pedro A., Jose M. Sánchez-Bornot, Agustín Lage-Castellanos, Mayrim Vega-Hernández, Jorge Bosch-Bayard, Lester Melie-García, and Erick Canales-Rodríguez. 2005. “Estimating Brain Functional Connectivity with Sparse Multivariate Autoregression.” *Philosophical Transactions of the Royal Society B: Biological Sciences* 360 (1457): 969–81. https://doi.org/10.1098/rstb.2005.1654.

Valdes-Sosa, Pedro A., Jose Miguel Sanchez-Bornot, Roberto Carlos Sotero, Yasser Iturria-Medina, Yasser Aleman-Gomez, Jorge Bosch-Bayard, Felix Carbonell, and Tohru Ozaki. 2009. “Model Driven EEG/FMRI Fusion of Brain Oscillations.” *Human Brain Mapping* 30 (9): 2701–21. https://doi.org/10.1002/hbm.20704.

Valdes-Sosa, Pedro A, Alard Roebroeck, Jean Daunizeau, and Karl Friston. 2011. “Effective Connectivity: Influence, Causality and Biophysical Modeling.” *Neuroimage* 58 (2): 339–61.

Valdes-Sosa, Pedro A, Jose Miguel Sanchez-Bornot, Roberto Carlos Sotero, Yasser Iturria-Medina, Yasser Aleman-Gomez, Jorge Bosch-Bayard, Felix Carbonell, and Tohru Ozaki. 2009. “Model Driven EEG/FMRI Fusion of Brain Oscillations.” *Human Brain Mapping* 30 (9): 2701–21.

Valdes, P A, Juan C Jiménez, J Riera, R Biscay, and T Ozaki. 1999. “Nonlinear EEG Analysis Based on a Neural Mass Model.” *Biological Cybernetics* 81 (5): 415–24.

Valdés, Pedro, J Bosch, R Grave, J Hernandez, J Riera, R Pascual, and R Biscay. 1992. “Frequency Domain Models of the EEG.” *Brain Topography* 4 (4): 309–19.

Van Quyen, Michel Le, and Anatol Bragin. 2007. “Analysis of Dynamic Brain Oscillations: Methodological Advances.” *Trends in Neurosciences* 30 (7): 365–73. https://doi.org/10.1016/j.tins.2007.05.006.

Varela, Francisco, Jean-Philippe Lachaux, Eugenio Rodriguez, and Jacques Martinerie. 2001. “The Brainweb: Phase Synchronization and Large-Scale Integration.” *Nature Reviews Neuroscience* 2 (4): 229–39.

Veen, Barry D Van, Wim Van Drongelen, Moshe Yuchtman, and Akifumi Suzuki. 1997. “Localization of Brain Electrical Activity via Linearly Constrained Minimum Variance Spatial Filtering.” *IEEE Transactions on Biomedical Engineering* 44 (9): 867–80.

Vega-Hernández, Mayrim, Eduardo Martínez-Montes, José M Sánchez-Bornot, Agustín Lage-Castellanos, and Pedro A Valdés-Sosa. 2008. “Penalized Least Squares Methods for Solving the EEG Inverse Problem.” *Statistica Sinica*, 1535–51.

Vidaurre, Diego, Romesh Abeysuriya, Robert Becker, Andrew J. Quinn, Fidel Alfaro-Almagro, Stephen M. Smith, and Mark W. Woolrich. 2018. “Discovering Dynamic Brain Networks from Big Data in Rest and Task.” *NeuroImage* 180: 646–56. https://doi.org/10.1016/j.neuroimage.2017.06.077.

Vidaurre, Diego, Laurence T. Hunt, Andrew J. Quinn, Benjamin A.E. E. Hunt, Matthew J. Brookes, Anna C. Nobre, and Mark W. Woolrich. 2018. “Spontaneous Cortical Activity Transiently Organises into Frequency Specific Phase-Coupling Networks.” *Nature Communications* 9 (1): 2987. https://doi.org/10.1038/s41467-018-05316-z.

Wang, Hao. 2012. “Bayesian Graphical Lasso Models and Eficient Posterior Computation.” *Bayesian Analysis* 7 (4): 867–86. https://doi.org/10.1214/12-BA729.

Wang, Q., P.A. Valdés-Hernández, D. Paz-Linares, J. Bosch-Bayard, N. Oosugi, M. Komatsu, N. Fujii, and P.A. Valdés-Sosa. 2019. “EECoG-Comp: An Open Source Platform for Concurrent EEG/ECoG Comparisons—Applications to Connectivity Studies.” *Brain Topography* 32 (4). https://doi.org/10.1007/s10548-019-00708-w.

Wens, Vincent, Brice Marty, Alison Mary, Mathieu Bourguignon, Marc Op de Beeck, Serge Goldman, Patrick Van Bogaert, Philippe Peigneux, and Xavier De Ti??ge. 2015. “A Geometric Correction Scheme for Spatial Leakage Effects in MEG/EEG Seed-Based Functional Connectivity Mapping.” *Human Brain Mapping* 36 (11): 4604–21. https://doi.org/10.1002/hbm.22943.

Wieringen, Wessel N van. 2019. “The Generalized Ridge Estimator of the Inverse Covariance Matrix.” *Journal of Computational and*

*Graphical Statistics* 28 (4): 932–42.
Wills, Adrian, Brett Ninness, and Stuart Gibson. 2009. "Maximum Likelihood Estimation of State Space Models from Frequency Domain Data." *IEEE Transactions on Automatic Control* 54 (1): 19–33.
Wipf, D P, R R Ramırez, J A Palmer, S Makeig, and B D Rao. 2006. "Automatic Relevance Determination for Source Localization with MEG and EEG Data." Technical Report, University of California, San Diego.
Wipf, David, and Srikantan Nagarajan. 2009. "A Unified Bayesian Framework for MEG/EEG Source Imaging." *NeuroImage* 44 (3): 947–66.
Wodeyar, Anirudh. 2019. "Linking Structure to Function in Resting State Macroscale Neural Activity." UC Irvine.
Zhang, Teng, and Hui Zou. 2014. "Sparse Precision Matrix Estimation via Lasso Penalized D-Trace Loss." *Biometrika* 101 (1): 103–20. https://doi.org/10.1093/biomet/ast059.
Zou, Hui, and Trevor Hastie. 2005. "Regularization and Variable Selection via the Elastic Net." *Journal of the Royal Statistical Society: Series B (Statistical Methodology)* 67 (2): 301–20.